\documentclass[iop,numberedappendix]{emulateapj}

\usepackage[usenames,dvipsnames]{color}
\usepackage[normalem]{ulem}
\usepackage[colorlinks,urlcolor=blue,citecolor=black,linkcolor=blue]{hyperref}
\usepackage{subfigure}

\newcommand{\e}[1]{\times 10^{#1}}

\newcommand{\msun}{M$_\odot$}

\def\ni {$^{56}$Ni}

\def\kms {km~s$^{-1}$}
\def\ergs {erg s$^{-1}$}

\begin{document}

\title{Long-duration superluminous supernovae at late times}

\author{
A. Jerkstrand\altaffilmark{1,2},
S. J. Smartt\altaffilmark{1},
C. Inserra\altaffilmark{1},
M. Nicholl\altaffilmark{3},
T.-W. Chen\altaffilmark{4},
T. Kr{\"u}hler\altaffilmark{4},
J. Sollerman\altaffilmark{5},
S. Taubenberger\altaffilmark{2,6},
A. Gal-Yam\altaffilmark{7}, 
E. Kankare\altaffilmark{1},
K. Maguire\altaffilmark{1},
M. Fraser\altaffilmark{8},
S. Valenti\altaffilmark{9},
M. Sullivan\altaffilmark{10},
R. Cartier\altaffilmark{10}
D. R. Young\altaffilmark{1},\\
}

\altaffiltext{1}{Astrophysics Research Centre, School of Mathematics and Physics, Queen's University Belfast, Belfast BT7 1NN, UK}
\altaffiltext{2}{Max-Planck Institut f{\"u}r Astrophysik, Karl-Schwarzschild-Str. 1, 85748 Garching, Germany}
\altaffiltext{3}{Harvard-Smithsonian Center for Astrophysics, 60 Garden Street, Cambridge, Massachusetts 02138, USA}
\altaffiltext{4}{Max-Planck-Institut f\"ur Extraterrestrische Physik, Giessenbachstrasse 1, 85748, Garching, Germany}
\altaffiltext{5}{The Oskar Klein Centre, Department of Astronomy, Stockholm University, AlbaNova, SE-10691 Stockholm, Sweden}
\altaffiltext{6}{European Southern Observatory, Karl-Schwarzschild-Str. 2, 85748 Garching, Germany}
\altaffiltext{7}{Benoziyo Center for Astrophysics, Weizmann Institute of Science, 76100 Rehovot, Israel}
\altaffiltext{8}{Institute of Astronomy, University of Cambridge, Madingley Rd, Cambridge CB3 0HA, UK}
\altaffiltext{9}{Department of Physics, University of California, Davis, CA 95616, USA}
\altaffiltext{10}{Department of Physics and Astronomy, University of Southampton, Southampton SO17 1BJ, UK}

\email{anders@mpa-garching.mpg.de} 

\begin{abstract}
Nebular-phase observations and spectral models of Type Ic superluminous supernovae are presented. LSQ14an and SN 2015bn both display late-time spectra similar to galaxy-subtracted spectra of SN 2007bi, and the class shows strong similarity with broad-lined Type Ic SNe such as SN 1998bw. Near-infrared observations of SN 2015bn show a strong Ca II triplet, O I 9263, O I 1.13 $\mu$m and Mg I 1.50 $\mu$m, but no distinct He, Si, or S emission.  The high Ca II NIR/[Ca II] 7291,7323 ratio of $\sim$2 indicates a high electron density of $n_e \gtrsim 10^8$ cm$^{-3}$. Spectral models of oxygen-zone emission are investigated to put constraints on the emitting region.  Models require $M(\mbox{O-zone}) \gtrsim$10 \msun to produce enough [O I] 6300,6364 luminosity, irrespective of the powering situation and the density. The high oxygen-zone mass, supported by high estimated magnesium masses,  points to explosions of massive CO cores, requiring $M_{\rm ZAMS} \gtrsim 40 M_{\odot}$. Collisions of pair-instability pulsations do not provide enough mass to account for the emission. [O II] and [O III] lines emerge naturally in many models, which strengthens the identification of broad [O II] 7320,7330, [O III] 4363, and [O III] 4959,5007 in some spectra. A small filling factor $f\lesssim 0.01$ for the O/Mg zone is needed to produce enough luminosity in Mg I] 4571, Mg I 1.504 $\mu$m, and O I recombination lines, which shows that the ejecta is clumped.  We review the constraints from the nebular spectral modelling in the context of the various scenarios proposed for superluminous supernovae.
\end{abstract}


\section{Introduction}

The origin of the newly discovered class of superluminous supernovae (SLSNe) remains unclear.
Reaching peak brightness of $\leq-$21 mag and emitting $\sim$$10^{51}$ erg of energy \citep{GalYam2012},
these transients provide a challenge to explain within current paradigms in stellar
evolution and explosion theory \citep[see][for a recent overview]{Sukhbold2016}. Some SLSNe evolve fast on a time scale of weeks, whereas others evolve slowly on a time scale of months. There is diversity in spectral appearance, with some displaying
hydrogen \citep{Ofek2007,Smith2008,Gezari2009,Miller2009,Chatzo2011,Benetti2014,Inserra2016}, but most appearing like Type Ic SNe \citep{Gal-Yam2009,Pastorello2010,Quimby2011,Chomiuk2011,Chornock2013,Lunnan2013,Inserra2013,Howell2013,Nicholl2013,Nicholl2014,Vreeswijk2014,Nicholl2015,Lunnan2016}.

An important avenue for progress is to study and model SLSNe at late times, when their
inner ejecta become visible and constraints on the nucleosynthesis and core structure can be put.
The strengths and shapes of observed emission lines may be compared with models, and element masses,
distributions, and mixing may be inferred.
There is now a small but growing body of data that presents an opportunity to engage in
such analysis. Due to the intrinsically high late-time brightness of the slow-evolving (long-duration) transients with rise and decay times of several months, late-time observations are most easily obtained for these SNe.

The first late-time spectra of a Type Ic SLSN were presented by \citet{Gal-Yam2009}, for SN 2007bi ($z=0.1279$).
The spectra were taken at +367d and +471d post-peak rest-frame, and
displayed distinct lines of Ca II HK, Mg I] 4571, [Fe II] 5250, [O I] 6300, 6364, [Ca II] 7291, 7323 + [O II] 7320, 7330
and O I 7774. The expansion velocities of these lines varied, being 10,000 \kms~for
Mg I] 4571 and [Fe II] 5250, 5000 \kms~for [O I] 6300, 6364 and [Ca II] 7291, 7323, and 3000 \kms~for O I 7774.

A late-time spectrum of PTF12dam ($z=0.107$) at +509d post-peak rest frame was presented by \citet{Chen2015}. 
This spectrum had severe contamination by the bright host, and [O I] 6300, 6364 was
the only clearly detected feature (although hampered by noise,
Mg I] 4571 may also be seen). In contrast to SN 2007bi, the [O I] 6300, 6364 lines appeared to be significantly stronger than the detection limits for [Fe II] 5250 and [Ca II] 7291, 7323. Their expansion velocities were also of order 5,000 \kms.

\citet{Yan2015} reported a spectrum of iPTF13ehe ($z=0.3434$) at +251d post-peak rest frame. This spectrum
was dominated by a broad H$\alpha$ (4500 \kms), as well as a prominent He I 5876 or Na I D line.
The SN showed a plateau in the $r$-band from +50d to +200d, suggesting a phase of
circumstellar interaction to produce the H$\alpha$ line. 

\citet{Lunnan2016} presented spectra of PS1-14bj ($z=0.5215$) at +202d post-peak rest-frame, and of LSQ14an ($z=0.1637$) at +205d. They identified broad (2500 \kms) [O III] 4363 and [O III] 4959, 5007 lines in both SNe. 
Both SNe also showed Mg I] 4571 and hints of an emerging [O I] 6300 6364. 

Models developed to fit the light curves of SLSNe must also be able to reproduce these spectral features
seen at late times. The suggested model scenarios for SLSNe are \ni~powering in either pair-instability SNe (PISNe) \citep{Gal-Yam2009,Kozyreva2014,Kozyreva2015} or core-collapse SNe (CCSNe) \citep{Moriya2010,Yoshida2011,Yoshida2014}, central engine powering by a magnetar or accreting black hole \citep{Kasen2010, Woosley2010,Dexter2013,Metzger2015}, and strong circumstellar interaction \citep{Smith2007, Chevalier2011, Ginzburg2012, Moriya2013, Sorokina2016}, possibly between colliding shells ejected in pulsational pair-instability SNe (PPISNe) \citep{Woosley2007PP,Chatzo2012b}. Many of these model scenarios can produce light curves in good
agreement with observations \citep[e.g.][]{Chatzopoulos2012,Chatzopoulos2013a, Inserra2013,Nicholl2013}
but there is often degeneracy due to the many free parameters.

Very few nebular-phase models for SLSNe have so far been calculated. Initial single-zone modelling of SN 2007bi indicated large element masses and the possibility
of a pair-instability origin \citep{Gal-Yam2009}.
However, calculations of multi-zone PISN explosion models \citep{Dessart2013,Jerkstrand2016} showed that PISNe are cold and neutral at nebular times, and produce red spectra with emission lines of species such as Fe I and Si I, in poor
agreement with observed spectra of SN 2007bi and PTF12dam.
Models for photospheric-phase spectra have been calculated by \citet{Dessart2012}, \citet{Dessart2013}, and \citet{Mazzali2016}. These models generally have favoured a central-engine powered scenario over a radioactivity-powered one.

In this paper we aim to put further constraints on the nature of long-duration Type Ic
SLSNe by both observational and theoretical work on late-time spectra.
We present multi-epoch nebular observations of two SLSNe, LSQ14an and SN 2015bn, covering +250d to +410d. We make a new calibration of the nebular spectra of SN 2007bi, carefully considering host galaxy subtraction using post-explosion host photometry. 
This gives us a high-quality data set of three long-duration SLSNe, which allows an analysis
of similarities and differences within the class and with respect to other SNe.
We investigate spectral models of emission by oxygen-zone material to derive constraints
from the emission lines seen. 

The paper is structured as follows. In Section \ref{sec:data} we present the new data for LSQ14an and SN 2015bn, as well as
the new reductions of SN 2007bi. In Section \ref{sec:compobs} we compare the nebular spectra of these SLSNe with each other and with other non-superluminous Type Ic SNe. In Section \ref{sec:omodels} we study spectral formation in oxygen/magnesium material at powering levels relevant for SLSNe, providing a systematic analysis of dependencies on mass, clumping, and energy deposition. 
In Section \ref{sec:discussion} we discuss the results and its implications, as well as uncertainties. In Section \ref{sec:summary} we present our summary and conclusions.  

\section{Data} 
\label{sec:data}

Our dataset comprises two nebular-phase spectra of each of the three supernovae SN 2007bi, LSQ14an, and SN 2015bn. The LSQ14an and SN 2015bn spectra are new observations, whereas the SN 2007bi spectra are rereductions of the spectra presented in \citet{Gal-Yam2009}. Table \ref{table:data} summarizes the data. Below the observations and reductions are described in detail.

\begin{table*}
\caption{Summary of observational data.}
\centering
\begin{tabular}{ccccccccc}
\hline
\hline
Date            & MJD            & Phase from peak       & Phase from peak  & Instrument & Rest wavelength & Resolution & Source\\
                    &                    &   obs. frame (days) &     rest frame (days)         &                   &  coverage (\AA)                    & (\AA)            & \\
\hline
SN 2007bi\\
2008-04-10 & 54567      &  +414                          & +367                                 & VLT + FORS2 & 3320-8550 & 16 &  \citet{Gal-Yam2009}\\ 
2008-08-04 & 54683     & +530                                & +471                                 & Keck\,I + LIRIS & 3010-8250 & 4-7 &  \citet{Gal-Yam2009}\\ 
LSQ14an\\
2014-12-29  & 57020     & +414                                   & +365                                 & VLT + X-shooter & 3000-20000 & 1,1.1,3.3 & This paper\\ 
2015-02-07 & 57061      & +478                                   & +410                                  & VLT  + FORS2  & 3920-7960 & 16 & This paper\\
SN 2015bn\\     
2016-01-01 & 57389     & +280                                   & +250                                 & NTT + EFOSC2 & 3270-8290 & 18 &  This paper \\ 
2016-03-10  & 57458     & +350                                   & +315                                 & VLT + X-shooter & 3000-20000 & 1,1.1,3.3 &  This paper\\ 
\hline
\end{tabular}

\label{table:data}
\end{table*}

\subsection{New reduction of nebular SN 2007bi spectra}
\label{sec:2007bi}

SN 2007bi spectra were published by \citet{Gal-Yam2009}.
The redshift of $z=0.1279$ gives a luminosity distance of 583 Mpc with standard cosmology ($H_{0} = 72\ $\kms Mpc$^{-1}, \Omega_{\rm m} = 0.3,  \Omega_{\rm \Lambda} = 0.7$), which is used in later conversion to rest-frame spectral luminosities.

\subsubsection{SN 2007bi +367d spectrum}
The reduced VLT FORS2 spectrum at +367d post-peak (rest-frame) from \cite{Gal-Yam2009} is available on WISeREP \citep{Yaron2012}\footnote{http://wiserep.weizmann.ac.il}. 
This spectrum does not have any host galaxy subtracted. Inspection of the 2D spectral images (taken under seeing conditions of 0.8\arcsec) and the contemporary $BVR_{\rm c}I_{\rm c}$ images, shows that the flux of the SN is effectively unresolved from the host, preventing any direct subtraction. \citet{Chen2015} presented uncontaminated, deep, images of the host, taken on 2012 May 25, some 5 years after explosion. Those images show a compact host with a FWHM measured at 1.25\arcsec~in the $r-$band, under seeing conditions of 0.8\arcsec. 


The $griz$ magnitudes of the host determined by \citet{Chen2015} show that there must be significant host contamination in the VLT spectrum. For example, \cite{Chen2015} measure $g=22.84\pm0.1$ for the host, which corresponds to $V=22.56\pm0.15$ \citep[conversion to Vega mags done using the transformations of][]{Jester2005}. \citet{Young2010} measure $V=22.82\pm0.24$ for SN 2007bi at +367d after the host was subtracted (using a template taken at +669d), demonstrating that the host flux is similar to that of the SN at the VLT epoch in the $V$-band. In \cite{Gal-Yam2009}, the non-host-subtracted spectrum was scaled to the template-subtracted photometry of \cite{Young2010}, which gives a too low luminosity in the SN emission lines.

To attempt an improved calibration, we took the FORS2 spectrum which was wavelength and flux calibrated (with telluric lines removed) but had no other corrections applied. It had standard sky background subtraction applied, but no other attempts to remove the host. We obtained this from S. Valenti (priv. comm.). This spectrum is identical to the WISeREP spectrum, apart from the WISeREP spectrum having been scaled down by a factor of 1.8. We then scaled the original spectrum to match the $BVR_{\rm c}I_{\rm c}$ magnitudes in the FORS2 images from the same epoch (containing both SN + host). We calculated these using aperture photometry matched to a number of SDSS DR12 reference stars (converted to Vega mags), finding $B=22.55\pm0.08$, $V=21.73\pm0.09$, $R_{\rm c}=21.46\pm0.17$, $I_{\rm c}=21.26\pm0.15$, where the errors are the quadrature sum of the statistical error from aperture photometry and the standard deviation of the zeropoints from the reference stars. The photometry should ideally be S-corrected, however this was not attempted here. Compared to the VLT spectrum, the scaling factors for each band were 1.43($B$), 1.37($V$), 1.35($R_c$), and 1.26($I_c$) and we used the mean value 1.35 to scale the VLT spectrum to the photometry.

The next step is to subtract the galaxy component. The optimal method would be to use a spectrum of the host galaxy, which is not available. This leaves two options, either to subtract a simple interpolated function between the photometry points, or use a model which is calibrated to the host photometry. We employ here the second method, using a Starburst99 model \citep{Leitherer1999} with standard settings. The galaxy model was first reddened in the galaxy rest frame with $E(B-V)=0.3$ mag (see more below on this choice) and a \citet{Cardelli1989} law with $R_V=3.1$ (which we use throughout the paper), the wavelengths then transformed to observer frame, and SDSS photometry calculated. A scaling factor was determined that minimized the residuals of synthetic SDSS photometry of this spectrum compared to the \citet{Chen2015} host photometry. With just four data points to constrain the galaxy model \citep[the $griz$ magnitudes in][]{Chen2015}, it is not warranted to investigate various parameters in the Starburst99 models. We limit ourselves to the standard model and an age of 30 Myrs, and choose the $E(B-V)=0.3$ mag which gives a best fit. $E(B-V)\leq 0.2$ mag gives a too blue model, whereas $E(B-V) \ge 0.4$ mag gives a too red model. The host extinction of the SN line-of-sight is unknown. The Milky Way reddening is $E(B-V)=0.02$ mag \citep{Gal-Yam2009,Young2010}. The SN only samples one sightline through the host galaxy,  and one will therefore in general have two different extinctions for the SN and the "averaged" host.
Recently, \citet{Leloudas2015} estimated the host galaxy reddening to $E(B-V)=0.13$ mag, a factor $\sim$2 lower
than used here. This value gives
a too blue model to fit the photometry, possibly as our model is intrinsically too blue for this galaxy. It is
however preferential to subtract a model that fits all the photometry than to have the most
accurate value for the extinction.

\begin{figure}[htb]
\centering
\includegraphics[width=1\linewidth]{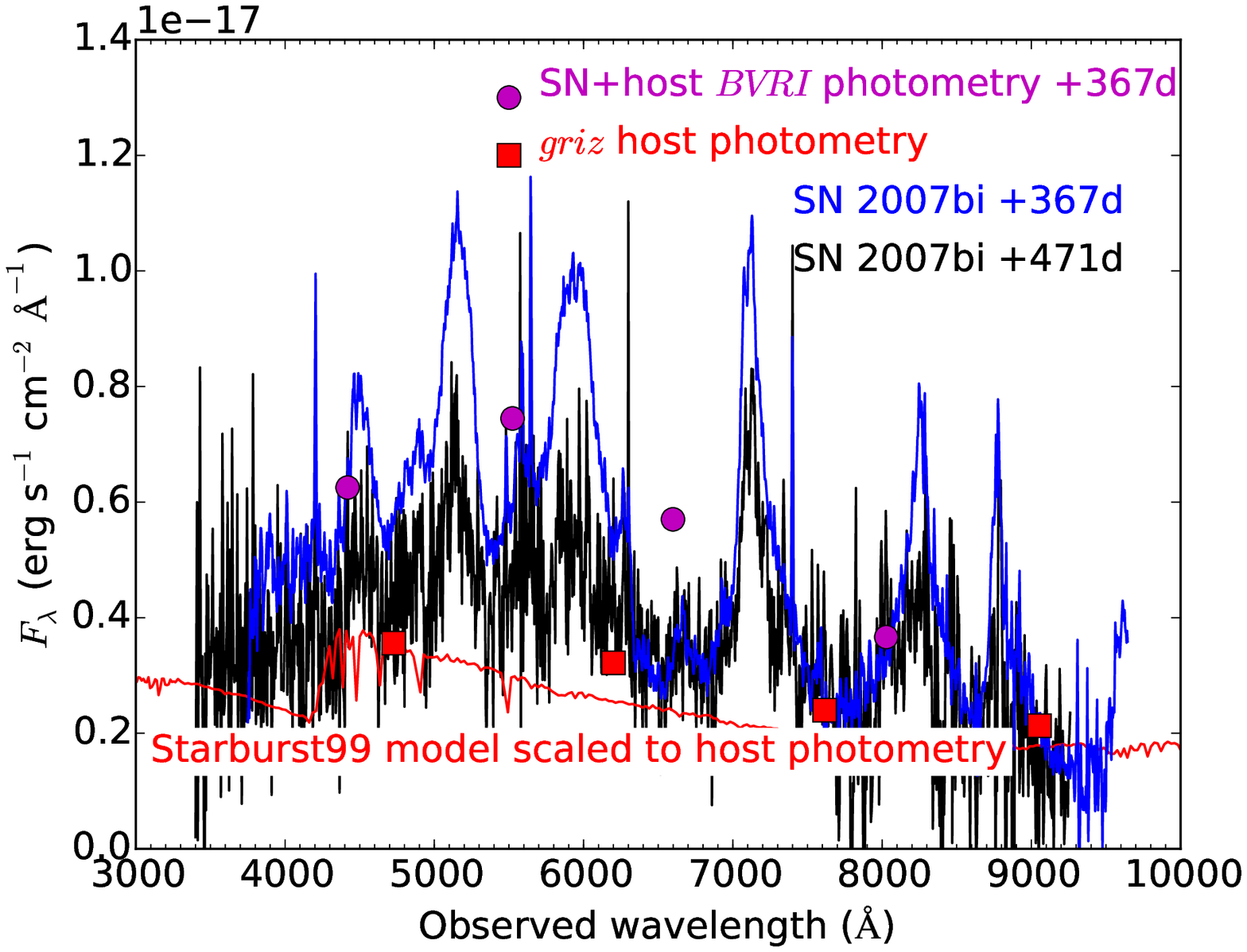} 
\includegraphics[width=1\linewidth]{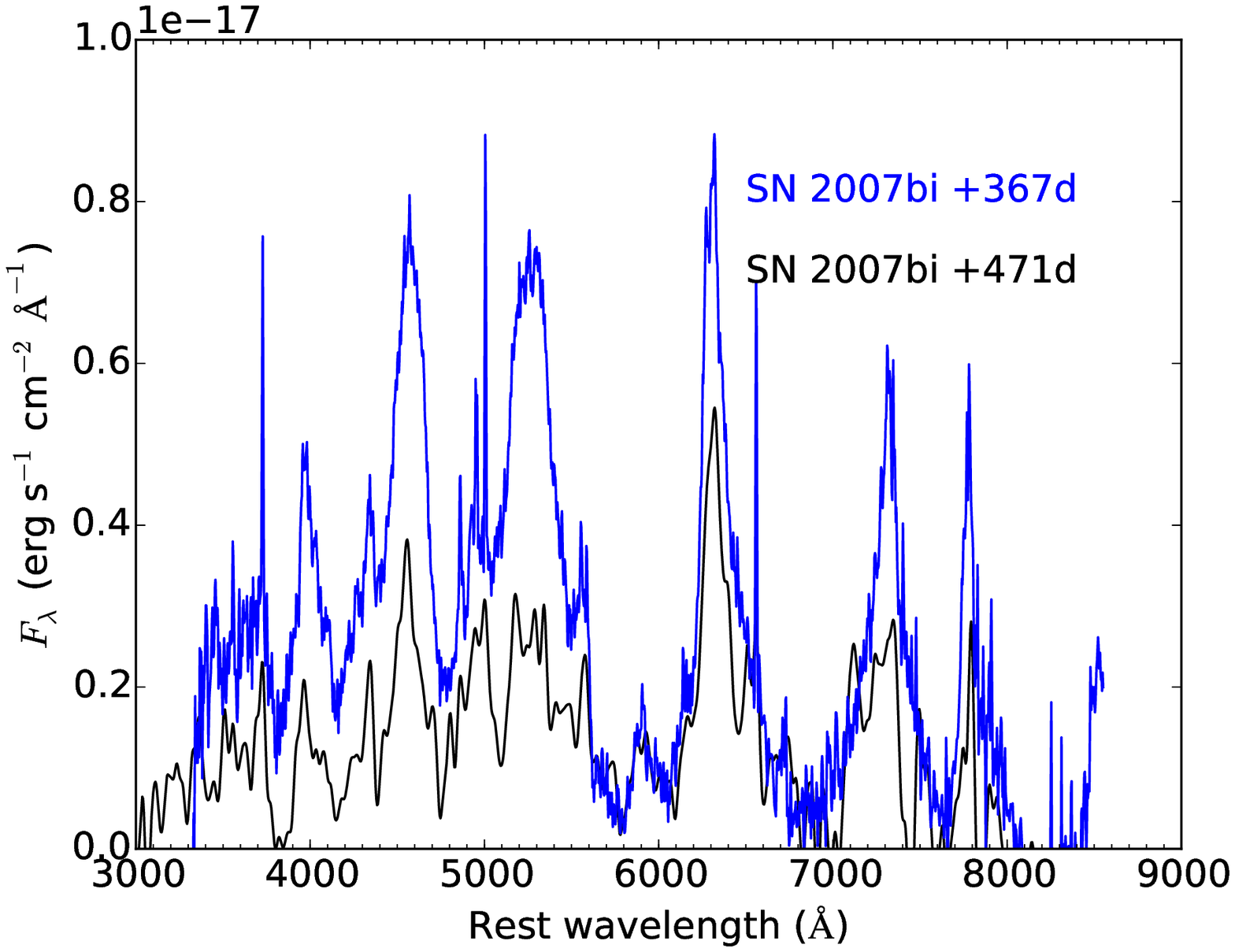}
\caption{Observed spectra of SN 2007bi. \emph{Top} : VLT spectrum scaled to SN + host photometry (blue line), contemporary SN+host photometry (magenta points), Keck spectrum scaled to estimated SN + host photometry (black line), host photometry from \citet{Chen2015} (red points), and a Starburst99 model ($E(B-V)=0.3$ mag) fitted to the \citet{Chen2015} photometry (red line). 
\emph{Bottom:} Final host-subtracted spectra in rest frame. The Keck spectrum has been convolved with a Gaussian of FWHM=2000 \kms.}
\label{fig:2007bi}
\end{figure}

The final SN spectrum was obtained by subtracting the host photometry-calibrated galaxy model from the SN+host-photometry-calibrated VLT spectrum. The resulting spectrum is shown in Fig. \ref{fig:2007bi} (bottom).
The narrow host lines result from UV reprocessing, an effect not included in the Starburst99 models. It is clear from the figure that they contribute little to the host photometry. We estimate a 5\% impact from synthetic photometry on exctracted lines, and make this small correction to the fit. For LSQ14an (Sec. \ref{sec:LSQ14an}), this effect is stronger, of order 30\%.

\subsubsection{SN 2007bi +471d spectrum}
There is another spectrum of SN 2007bi, taken with Keck 104 days later (rest-frame) and presented in  \cite{Gal-Yam2009} (also available on WISeREP).  There is no contemporary photometry to do flux calibration against, and
\cite{Gal-Yam2009} used spectral flux calibration using a standard taken 15 minutes before. However, synthetic photometry of this host+SN spectrum yields $r=23.3$ mag, 0.9 magnitudes fainter than even the host ($r_{host}=22.4$ mag). 
This means the flux calibration is likely off by a factor of several, and the night was indeed non-photometric (M. Sullivan, priv. comm), possibly explaining the discrepancy. 

The image quality in the 2D frames (from the Keck archive) is 1.2\arcsec, meaning that no host galaxy subtraction is possible in the 2D extraction. We can apply the model subtraction method of the preceding section to this spectrum as well, albeit with more uncertainty than for the VLT FORS2 spectrum.  We assume that SN 2007bi continued to decline linearly in the $R_{c}$-band with the slope given in \cite{Gal-Yam2009} and \citet{Young2010}, giving an expected magnitude of $R_{c}=23.2\pm0.2$ at the epoch of the Keck spectrum (this is a simple mean of the Gal-Yam et al. and Young et al. extrapolations). Thus the expected combined magnitude of host and SN is $R_{c}=21.9\pm0.2$. To match the Keck spectrum to this, it needs to be scaled by a factor 3.1. We then subtracted off the same Starburst99 model as used for the VLT FORS2 spectrum correction. We note that the factor 3.1 scaling gives a consistent position of the galaxy component with the VLT spectrum, which should dominate at wavelengths with weak SN lines. If we had chosen an approach to ignore the slope of previous $R$ points and just tried to align the galaxy flux, we would have got a very similar scaling. 

\subsubsection{Final spectra}
Figure \ref{fig:2007bi} shows the calibration process and final spectra. The top panel shows the raw VLT and Keck spectra
scaled to SN+host photometry. Also shown is the host photometry which demonstrates that host
light makes up about half the flux in the VLT spectrum and about 3/4 in the Keck spectrum.

Synthetic photometry on the galaxy-subtracted VLT spectrum agrees well with the template-subtracted
photometry, with errors of +0.31, $-$0.02, and $-$0.17 mags in $V$, $R_{\rm c}$, and $I_{\rm c}$ (\citet{Young2010} did not present a template-subtracted value in $B$), and means we have applied a self-consistent method to remove the host flux and match the image subtracted SN photometry.

The final spectra are shown in the bottom panel. Compared to the VLT spectrum presented by \cite{Gal-Yam2009}, the rising blue flux is removed and the emission lines are significantly stronger due. The spectrum appears more similar to typical Type Ib/c nebular SN spectra. 
The corrected Keck spectrum is similar to the VLT spectrum, but dimmer and with significantly lower signal to noise. Our final estimates for absolute R-band magnitudes are $M_R = -16.72$ at +367d and $M_R=-16.02$ at +471d.
The optical flux has declined by a factor 3.0 over 104d, close to the $^{56}$Co decay (2.55).
The Mg I] 4571 line and the Mg I 5180 + [Fe II] 5250 feature are noticeably weaker in the Keck spectrum. The redder emission lines of [O I] 6300, 6364 and [Ca II] 7291, 7323 have also weakened, but not as much.

\subsection{LSQ14an} 
\label{sec:LSQ14an}
LSQ14an was discovered by La Silla Quest Survey and spectroscopically classified as a SN 2007bi-like Type Ic SN at $z=0.163$ by the Public ESO Spectroscopic Survey of Transient Objects (PESSTO\footnote{www.pessto.org}). The classification spectrum of \cite{2014ATel.5718....1L} was taken on 2014 January 2 (MJD 56659) and 
suggested an epoch of $\sim$+55d post peak (rest-frame) from comparison with SN 2007bi and PTF12dam. This puts the peak around 2013 October 30 (MJD 56595). If we assume a rise time of 70d, the estimated explosion epoch is $\sim$ 2013 August 9 (MJD 56513). With our higher signal-to-noise spectra we refined the redshift estimate to $z=0.1637\pm 0.0001$, corresponding to a luminosity distance of 763 Mpc with standard cosmology.

We observed LSQ14an on two occasions in its nebular phase, with VLT X-shooter at +365d post-peak rest frame, and with VLT FORS2 at +410d. The observations and data reductions
are detailed in Section \ref{sec:14anobs} in the appendix. The full PESSTO data set on the early evolution of LSQ14an is 
presented by Inserra et al. (in prep).

\begin{figure*}
\centering
\includegraphics[width=0.93\linewidth]{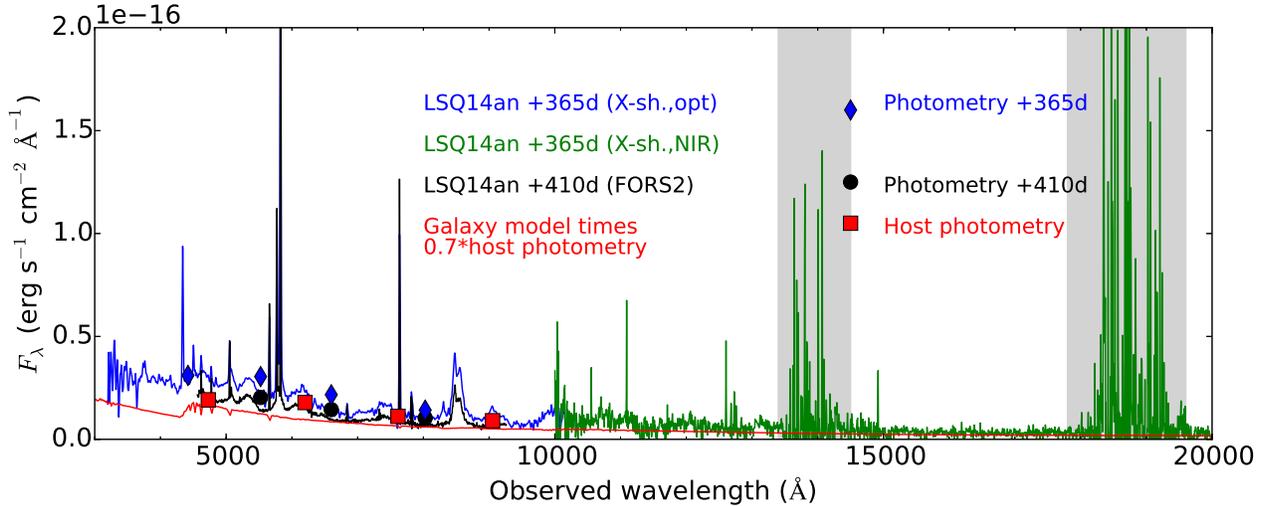} 
\caption{LSQ14an at +365d scaled to SN + host photometry (blue and green lines), at +410d (black lines), SN + host photometry at these epochs (blue and black points) and host photometry (red points). The galaxy model is shown as a red line. The X-shooter spectrum (+365d) has been smoothed.}
\label{fig:LSQ14an_calib1}
\end{figure*}

\begin{figure}
\centering
\includegraphics[width=1\linewidth]{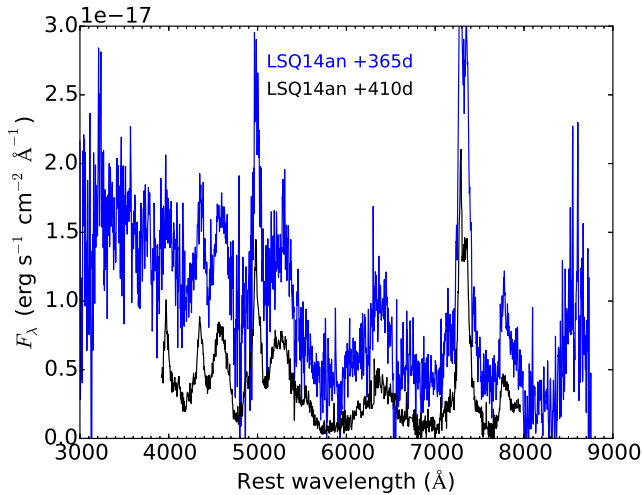} 
\caption{LSQ14an at +365d (blue) and +410d (black), both after subtraction of the galaxy model, and removal of host galaxy lines.}
\label{fig:LSQ14an_calib2}
\end{figure}

The spectra are shown in Fig. \ref{fig:LSQ14an_calib1}.
There is strong contamination by host galaxy lines and continuum, and as for SN 2007bi, the 
SN is unresolved from the host flux. Hence we need to employ a similar method of modelling the host 
flux and subtracting it off both the X-shooter and FORS2 spectra to leave a clean SN spectrum. 

The host galaxy was detected in the Pan-STARRS1 3$\pi$ survey (catalogued object PSO J193.4492-29.5243). 
Images in each of the filters $grizy_{\rm P1}$ were taken from the PS1 Processing
Version 2 stacks. All the individual images which make up
these stacked images were taken before MJD 56467 (2013 June 24), 1-2 months before the estimated
explosion date of early August 2013. Hence there is unlikely to be
any SN flux in the images. Magnitude measurements of the host were carried out using aperture photometry. We let the aperture size vary until we were confident that it encompassed the whole host flux and avoided to cover nearby object flux. The aperture radius we adopted is 2 arcsecs.
The zeropoint was determined with 15-18 PS1 reference stars in the field \citep{Schlafy2012,Magnier2013} resulting
in values of $g_{\rm P1}= 20.93\pm  0.11$, $r_{\rm P1}= 20.50\pm   0.12$, $i_{\rm P1}= 20.56\pm   0.10$, $z_{\rm P1}= 20.42\pm   0.14$, and $y_{\rm P1} = 20.43 \pm 0.19$ \citep[AB mags in the filter system described in][]{2012ApJ...750...99T}.
The position of LSQ14an was measured at
RA=193.44924    DEC=$-$29.52434 in
the astrometrically calibrated EFOSC2 images (with Sextractor). In comparison the centroid of the host galaxy was measured at RA=193.44921    DEC=$-$29.52429.
The difference between the two is
0.2 arcsecs, which illustrates that is coincident with (and unresolved from) the
dwarf galaxy host. 

Figure \ref{fig:LSQ14an_calib1} shows also the host photometry.  There is clearly significant contamination by galaxy light. However, distinct broad emission lines are seen on top of this. The situation is somewhat better than initial inspection suggests, because the narrow galaxy lines have an impact on the host photometry. 

Shown is also the 30 Myr Starburst99 model, reddened with $E(B-V)=0.1$ mag (in agreement with $E(B-V)$ estimates from the Balmer line ratio (Inserra et al., in prep.)). and scaled to give an average flux level of 0.7 times the observed photometry; this is to compensate for the narrow emission lines which are missing in the model. A factor smaller than 0.7 is not warranted as the lines do not add more than $\sim$1/3 to the photometry. A larger factor fails to give out much positive SN flux as the host photometry is even somewhat brighter than the host+SN photometry, presumably due to observational or calibration issues. One may attempt to estimate the contribution by the narrow lines more precisely, but this would not help too much here because of the discrepancy that host photometry is somewhat brighter than the SN+host photometry.

Figure \ref{fig:LSQ14an_calib2} shows the host-subtracted optical SN spectra. These spectra also have had the narrow galaxy lines clipped out (see Appendix).
The optical flux decreases by a factor 2 over 45d, or 1.6 mag per 100d. 
The bottom panel of Figure \ref{fig:PS15ae_optandnir} shows the combined optical and near-infrared spectrum. The NIR flux is weak compared to the optical. The NIR spectrum shows no clearly distinguished lines, apart from possibly O I 1.13 $\mu$m.

\subsection{SN 2015bn} 

SN 2015bn is alternatively known as PS15ae, CSS141223-113342+004332 and MLS150211-113342+004333. The earliest detected point is on 2014 December 23 by the Catalina Sky Survey and the evolution was then followed by PESSTO  after the initial classification as a SLSN in \cite{2015ATel.7102....1L}.  The SN was monitored from $-50$d to $+250$d (with respect to optical maximum) 
showing it to be a slowly evolving SLSN similar in observational characteristics to 
SN 2007bi and LSQ14an \citep{Nicholl2016}. Spectropolarimetry was presented by \citet{Inserra2016SP}. The redshift is $z=0.1136$ from the 
narrow host galaxy emission lines, giving a luminosity distance of 513 Mpc with
standard cosmology.  The light curve peaked around 2015 March 27 (MJD 56108)
so the SN had a rise time of $\gtrsim$ 80d rest-frame.

\begin{figure*}
\centering
\includegraphics[width=0.93\linewidth]{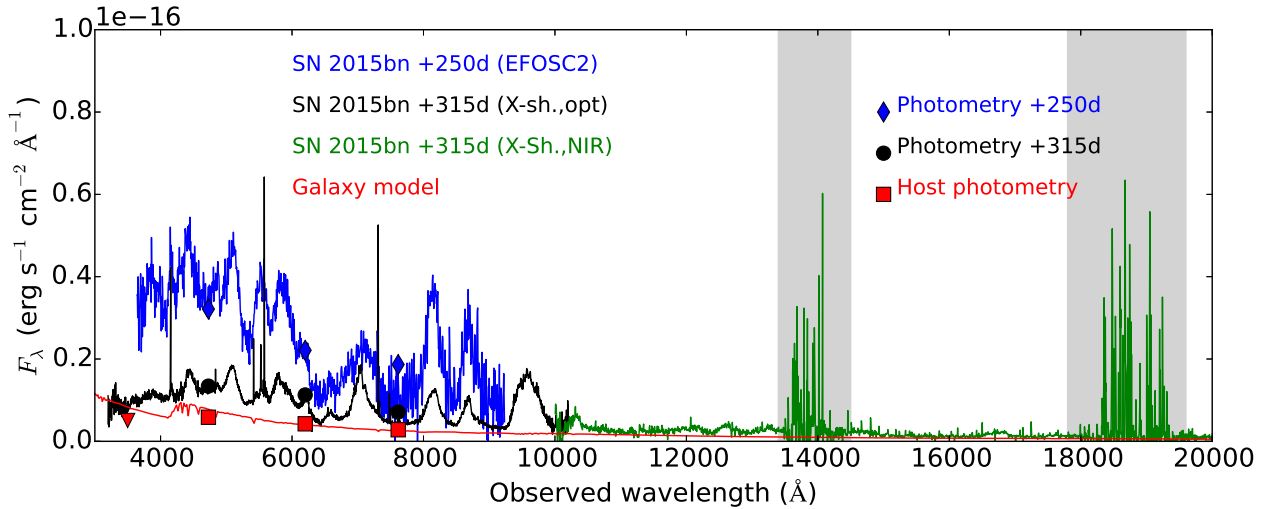} 
\caption{SN 2015bn at +250d (blue), photometry at this epoch (blue points), SN 2015bn at +315d (black), photometry at this epoch (black points), host photometry (red points), and Starburst99 model (red line).}
\label{fig:PS15ae_calib1}
\end{figure*}

\begin{figure}
\centering
\includegraphics[width=1\linewidth]{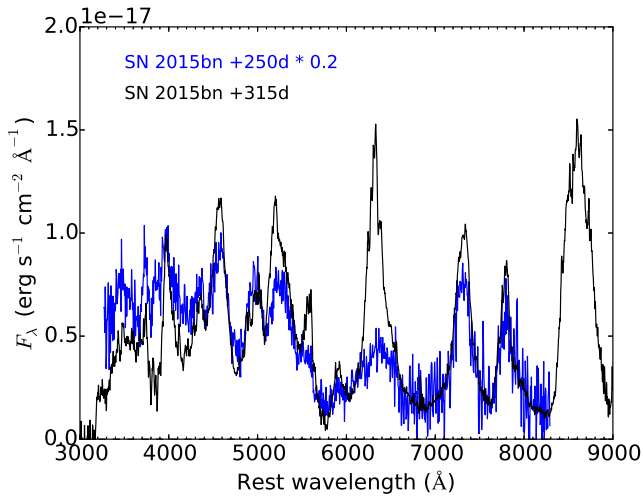} 
\caption{SN 2015bn at +250d (blue) and +315d (black), both after subtraction of the galaxy model, and removal of host galaxy lines. Note that the +250d spectrum has been scaled by 0.2.}
\label{fig:PS15ae_calib2}
\end{figure}

We observed SN 2015bn at two epochs in its nebular phase, at +250d post-peak rest frame with NTT EFOSC2 and at +315d with VLT X-shooter. The observations and data reductions are detailed in Sect. \ref{sec:2015bnobs} in the Appendix.
Figure \ref{fig:PS15ae_calib1} shows the spectra, the host photometry \citep{Nicholl2016}, and the same Starburst99 model as in previous section (using $E(B-V)=0$) scaled to this photometry. 
Figure \ref{fig:PS15ae_calib2} shows the host-subtracted spectra at rest-frame wavelengths (note that the +250d spectrum has been scaled by 0.2). From +250d to +315d, a more distinct [O I] 6300, 6364 develops in the spectrum, as well as higher contrast for Ca II HK, Mg I] 4571, and [Fe II] 5250. We see the same feature
at 5500 \AA~as in SN 2007bi, probably [O I] 5577. The optical flux has decreased by a factor 5 over 65d, or 2.2 mag per 100d.
At the end of the red arm we see the first detection of Ca II NIR in SLSNe at nebular times.
This line is strong and broad, with an expansion velocity of 8,000-10,000 \kms. 

\begin{figure*}
\centering
\hspace{0.13in}
\includegraphics[width=0.91\linewidth]{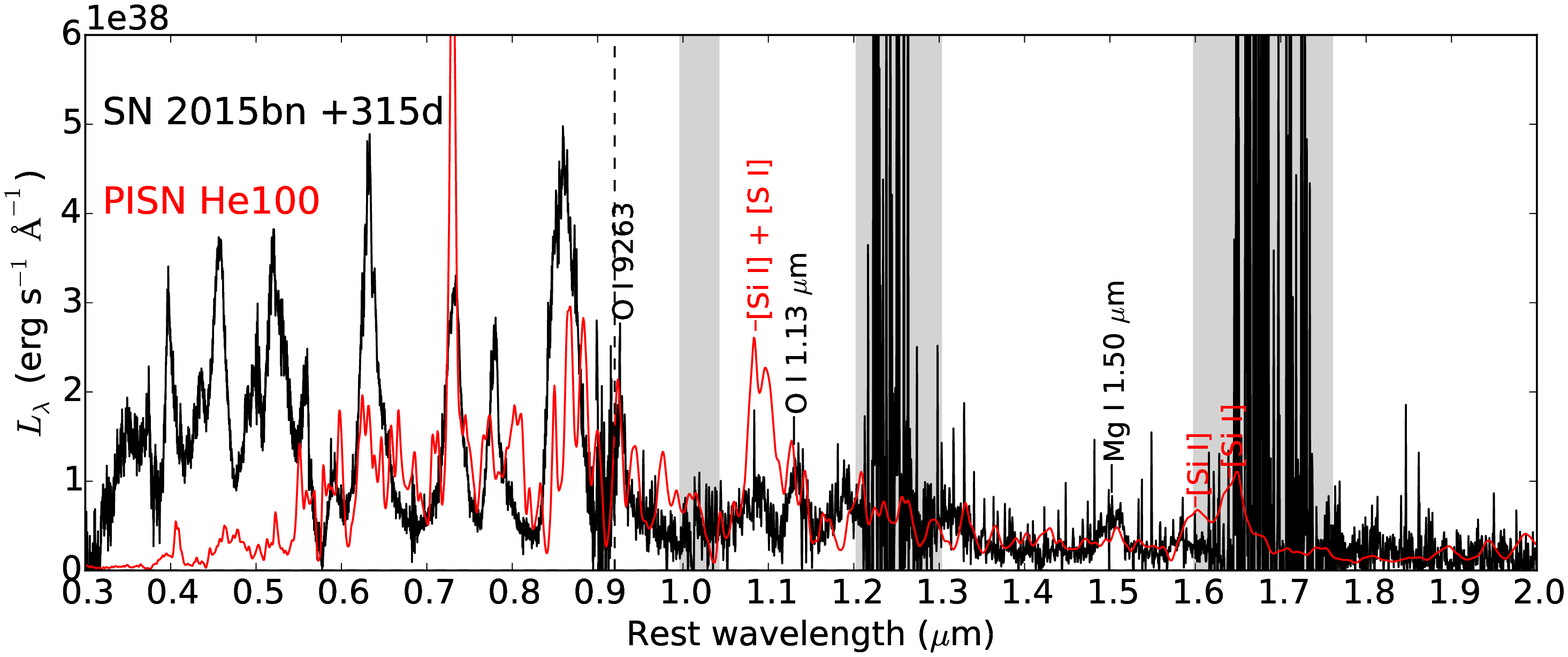} 
\includegraphics[width=0.93\linewidth]{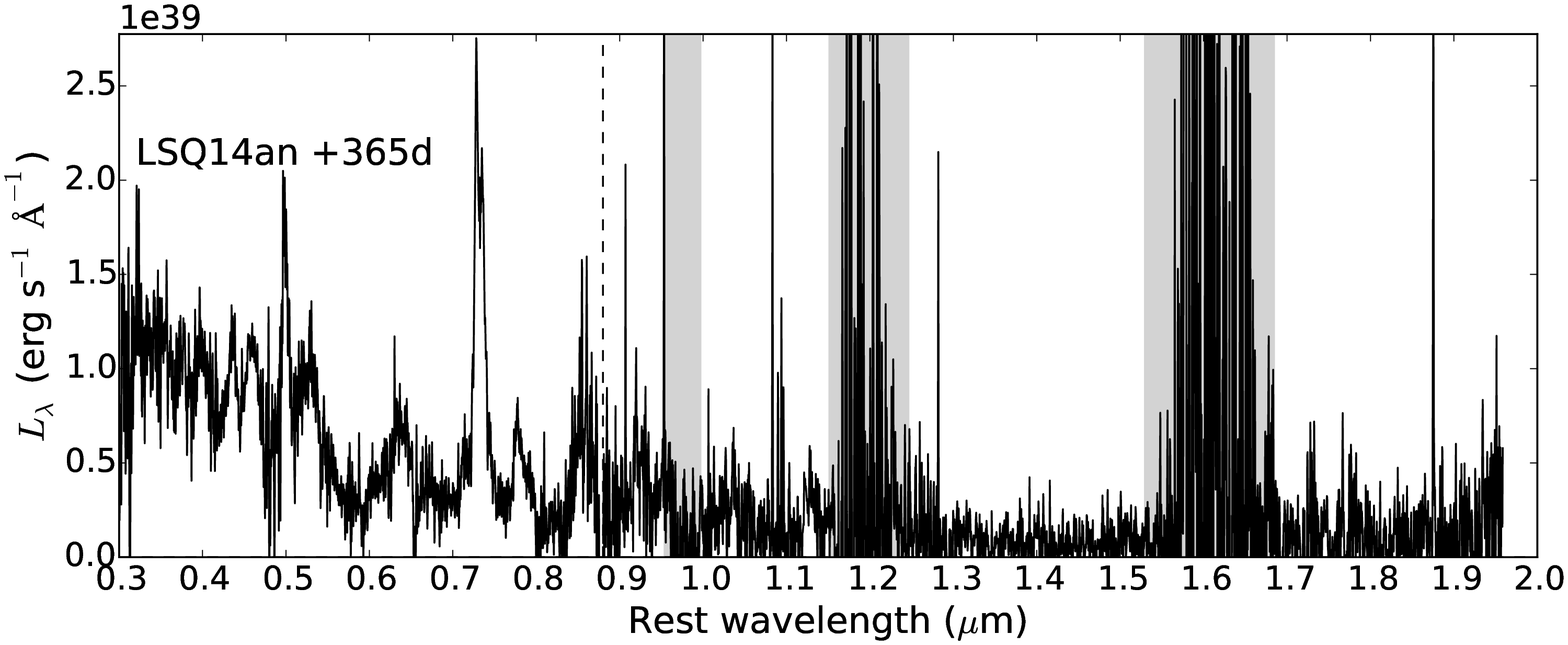} 
\caption{\emph{Top:} SN 2015bn from 3,000-20,000 \AA\ at +315d (black). The dashed vertical line shows the division between optical and NIR arms. The He100 PISN model of \citet{Jerkstrand2016} is also plotted (red). The PISN model has been scaled by a factor of 0.5 to give roughly right flux levels.
\emph{Bottom:} LSQ14an from 3,000-20,000 \AA\ at +365d. Telluric bands are marked.}
\label{fig:PS15ae_optandnir}
\end{figure*}

The top panel of Figure \ref{fig:PS15ae_optandnir} shows the combined optical and near-infrared spectrum. The NIR flux is weak compared to the optical. There are clear detections of O I 9263, O I 1.13 $\mu$m, and Mg I 1.50 $\mu$m. These lines are commonly seen in
stripped-envelope SNe with nebular NIR spectra, e.g. for the Type IIb SN 2008ax \citep{Taubenberger2011}, SN 2011dh \citep{Ergon2015,J15a}, the Type Ic SN 2007gr \citep{Hunter2009}, and in the broad-lined SN 2012au \citep{Mili2013}.

A comparison with the He100 PISN model of \citet{Jerkstrand2016} in the same figure shows no strong emission in the [Si I] + [S I] blend around 1.08 $\mu$m predicted to be strong. Also [Si I] 1.607, 1.645 $\mu$m are predicted to be strong, but this feature falls in the telluric band.

\section{Observational comparisons}
\label{sec:compobs}

In this section we compare the spectra of SN 2007bi, LSQ14an and SN 2015bn with each other as well as with other SNe. 
For the three SLSNe we attempt no dereddening as the host line-of-sight extinctions are not known, and the Milky Way extinctions are smaller ($E(B-V)=0.03$ mag for SN 2007bi \citep{Gal-Yam2009}, $E(B-V)=0.07$ mag for LSQ14an (NED\footnote{https://ned.ipac.caltech.edu}) and $E(B-V)=0.02$ mag for SN 2015bn \citep{Nicholl2016}) than the uncertainties in flux calibration
and host galaxy subtractions.

\begin{figure*}
\centering
\includegraphics[width=1\linewidth]{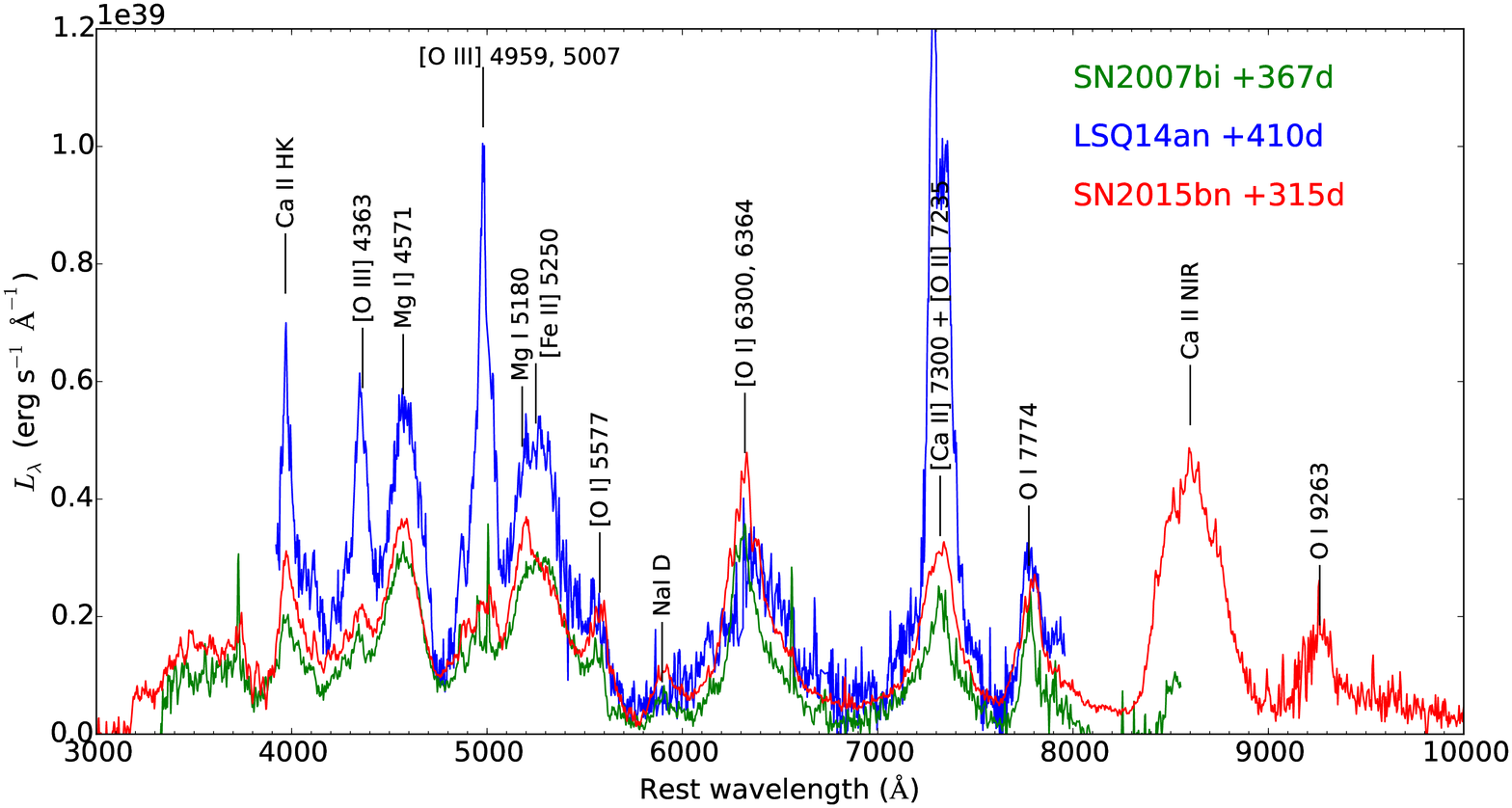} 
\caption{A comparison of rest-frame spectral luminosities of SN 2007bi (green), LSQ14an (blue), and SN 2015bn (red).}
\label{fig:allthree}
\end{figure*}

Figure \ref{fig:allthree} compares the spectra of SN 2007bi, LSQ14an, and SN 2015bn at around one year post-peak in the rest frame. All three SNe show qualitatively similar spectra, with Ca II HK,
Mg I] 4571, Mg I 5180 + [Fe II] 5250, [O I] 6300, 6364, [Ca II] 7291, 7323 and O I 7774 all distinct.
There are also clear detections  of [O I] 5577 and Na I D in SN 2015bn and SN 2007bi and hints of [O I] 5577 in LSQ14an.  The similarity in luminosity of both [O I] 6300, 6364 and O I 7774 in all three SNe is striking.  LSQ14an has, in addition, strong [O III] 4363, [O III] 4959, 5007, and the prominent feature usually identified as the forbidden [Ca II] 7291, 7323 doublet. Given the strength of the O III lines, we suggest this likely is dominated by [O II] 7320, 7330 in LSQ14an (see also Sect. \ref{sec:oIIIlines}). 
Figure \ref{fig:velspace} in Appendix \ref{sec:linezoom} shows comparisons of each of the main emission lines in velocity space.

\begin{figure}
\centering
\includegraphics[width=1\linewidth]{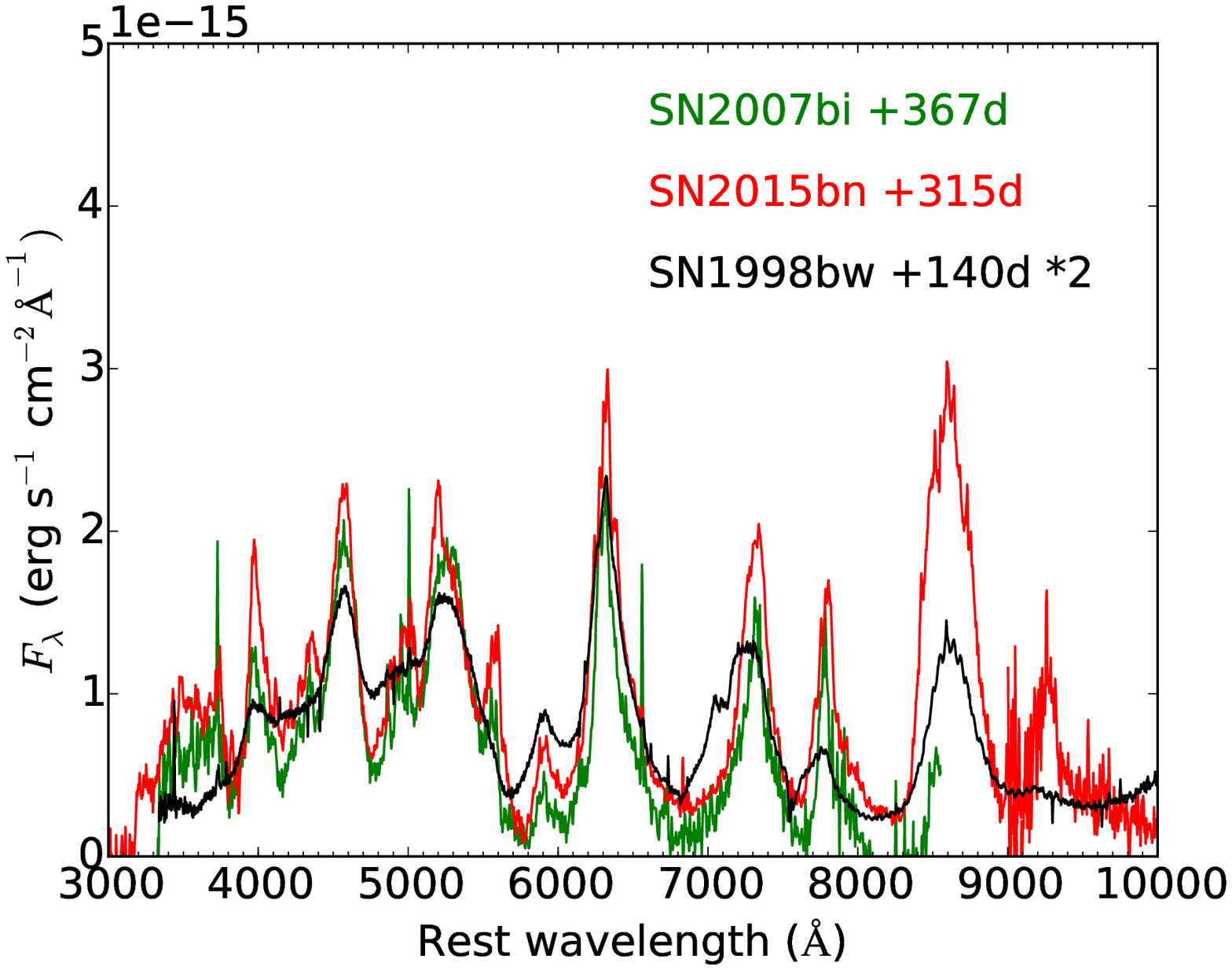}\\ 
\includegraphics[width=1\linewidth]{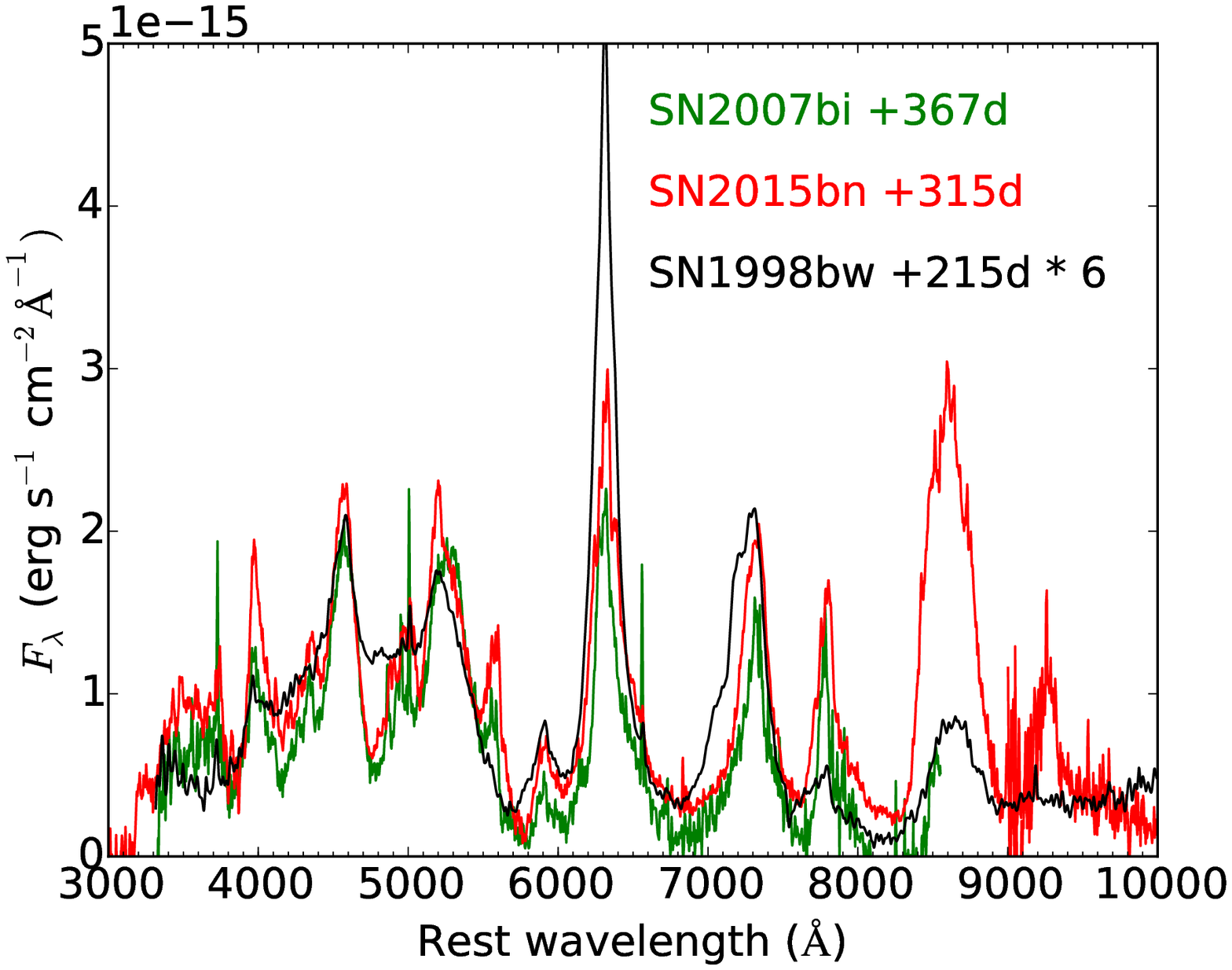}\\
\includegraphics[width=1\linewidth]{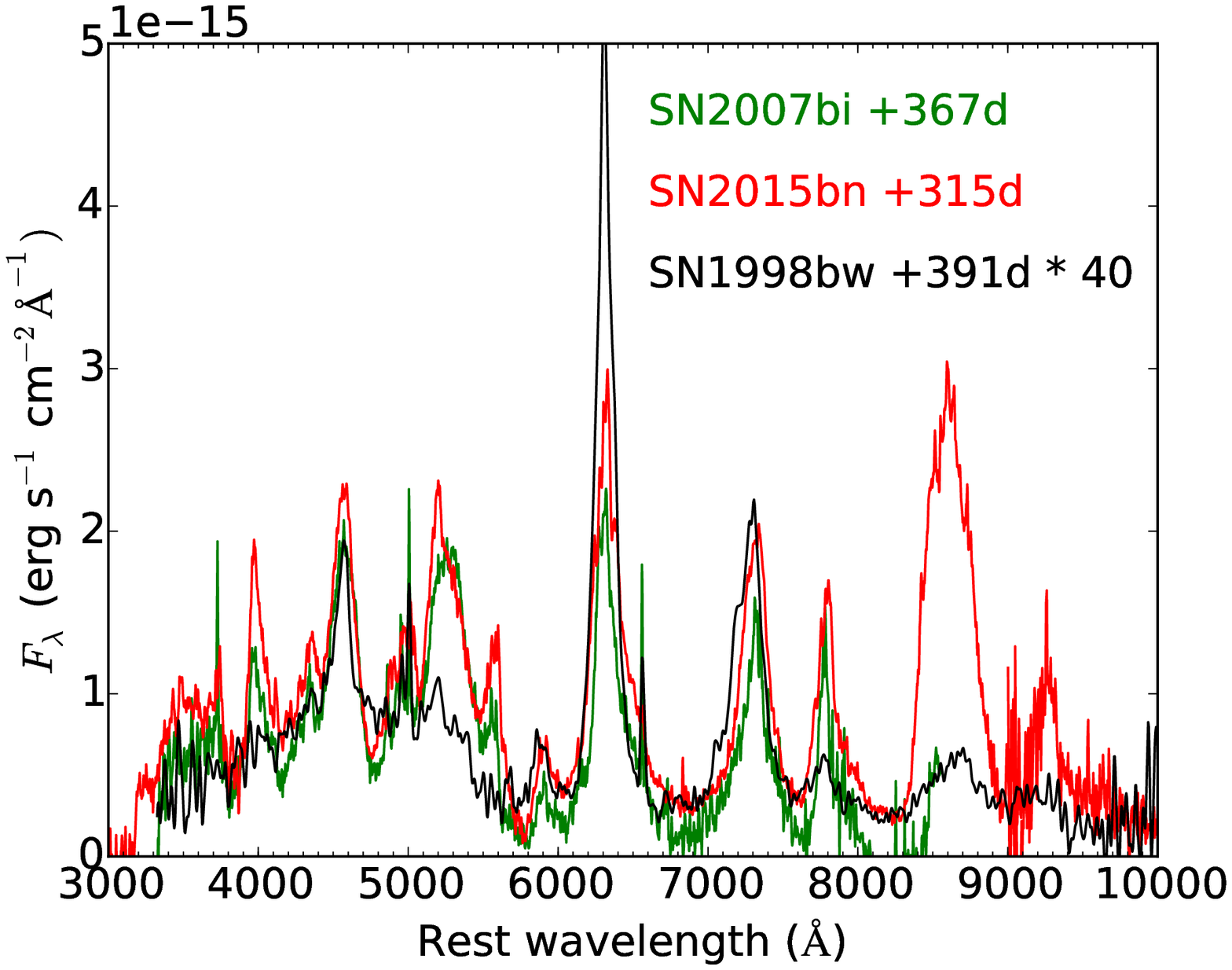} 
\caption{Comparison of SN 2007bi (green) and SN 2015bn (red) with SN 1998bw (black) at 140d (top), 215d (middle) and 391d (bottom). SN 2007bi and SN 2015bn have been scaled to the SN 1998bw distance of 36 Mpc, and SN 1998bw has been multiplied by a factor shown in each figure to bring about flux agreement.} 
\label{fig:comp98bw}
\end{figure}

In Fig. \ref{fig:comp98bw} we compare the spectra of SN 2007bi and SN 2015bn with SN 1998bw spectra at +140d, +215d, and +391d post-explosion \citep{Patat2001}, retrieved from WISeREP. For the comparisons we use $z=0.0085$ and $E(B-V)=0.06$ mag for SN 1998bw \citep{Patat2001}. SN 1998bw has a representative spectrum for the class of broad-lined Type Ic SNe,
which have relatively similar nebular phase spectra \citep[e.g.][]{Taubenberger2006, Mili2015,Modjaz2015}. At 200-300d after explosion, SN 1998bw, SN 2002ap  \citep{Foley2003}, SN 2003jd \citep{Valenti2008}, SN 2004aw \citep{Taubenberger2006}, and SN 2012ap \citep{Mili2015} all show [O I] 6300, 6364 as the strongest line, calcium lines with a Ca II NIR/[Ca II] 7291, 7323 ratio of $<$1, and distinct Mg I] 4571. Some show a distinct line at 5250 \AA\ (where all SLSNe have a line), and some do not.

The figure shows that there is strong  overall similarity between the SLSNe and SN 1998bw, but with distinct phase shifts. The spectra of SN 1998bw  at +140d and +215d show similar emission lines and relative strengths as the spectra of SN 2007bi and SN 2015bn at +315d to +367d after peak. Thus, SLSNe appear to behave like broad-lined Type Ic SNe but to evolve more slowly. It is particularly noteworthy that SN 1998bw also displays distinct Mg I] 4571 and Mg I 5180+ [Fe II] 5250.
The relative strength of these lines compared to the quasi-continuum is somewhat lower in SN 1998bw than in the SLSNe. 
Comparing with SN 1998bw at +391d shows that the blue quasi-continuum and the 5250 \AA~feature have faded compared to the red lines, and the similarity at blue wavelengths is reduced.
The [Ca II] 7291, 7323 lines are however more similar in shape at this epoch. At earlier epochs, SN 1998bw has a line at $\sim$7100 \AA~not seen in the SLSNe. This line may be [Fe II] 7155 or possibly He I 7065.

 At no epoch does SN 1998bw show a particularly strong O I 7774 line, which is a difference to the SLSNe. Neither do most other broad-lined Ic such as SN 1997ef \citep{Matheson2001}, SN 2002ap, SN 2003jd, SN 2006aj \citep{Maeda2007}, SN 2009bb \citep{Pignata2011}, or SN 2012ap.
 There are two known exceptions; SN 1997dq, which showed a strong narrow O I 7774 line at +262d post explosion \citep{Matheson2001,Mazzali2004}, and SN 2012au at +321d \citep{Mili2013}. \citet{Mili2013} showed an overall strong similarity between the nebular spectra of SN 2012au and SN 2007bi. Also O I 9263 is weaker in the broad-lined Ic SNe.

Another difference is [O I] 5577, which is distinct in the SLSNe but not seen in SN 1998bw. This probably indicates that the temperature is higher in the SLSNe, since the [O I] 5577 line has a high excitation energy.

SN 2015bn has additional coverage of Ca II NIR and O I 9263. The Ca II NIR/ [Ca II] ratio in SN 2015bn is higher than in SN 1998bw or any other Type Ic SN at any epoch. We study the implications of this in Sect. \ref{sec:caiilines}.

\section{Oxygen zone models} 
\label{sec:omodels}

The relative homogeneity in spectral appearance established in Section \ref{sec:compobs} warrants an investigation on what
physical conditions are needed to produce the observed line luminosities in O, Mg, and Ca lines. Here, we pursue single-zone
modelling of oxygen-rich compositions, exploring parameter space over mass, density, and
powering level. The aim is to understand the physical mechanisms behind the emission of distinct lines (e.g. [O I] 6300, 6364, [O I] 5577, O I 7774, Mg I] 4571, Mg I 5180, Mg I 1.50 $\mu$m, [Ca II] 7291, 7323 and Ca II NIR) and the diagnostic use of these. Generalized single-zone modelling is a complementary approach to the calculation of multi-zone explosion models, where constraints may be derived independent of the powering mechanism and parameter space can be explored to a larger extent. Indications of ejecta properties obtained from it can be used to select explosion models to test in follow-up multi-zone modelling.

For the modelling we use both the SUMO spectral synthesis code \citep{J11,Jerkstrand2012} and semi-analytic approaches. The SUMO code solves the equations for non-thermal energy degradation, Non-Local-Thermodynamic-Equilibrium excitation and ionization balance, temperature, and radiative transfer. It provides a self-consistently calculated
spectrum of any ejecta structure, without any free parameters.

Because the forward modelling is always limited to particular compositions and ejecta structures, 
and involves much complex physics, some of it treated in rough approximations, it is beneficial to 
complement it with backward modelling using analytic formulae, which we also pursue here.


\subsection{SUMO grid setup}
Each model in the SUMO grid was set up as $N=100$ randomly distributed spherical clumps within a volume specified by an expansion velocity $V$ and a time $t$ (assuming homology). Together with a specified total mass $M_{ej}$, the density of the clumps is determined by a filling factor $f$, with vacuum in the $1-f$ space. The radiative transport is computed self-consistently throughout the volume. We compute a model grid with
\begin{itemize}
\item $M_{ej}$ = 3, 10, and 30 \msun
\item Energy deposition (by gamma-rays or any other high-energy source) $2.0\e{41}$, $2.5\e{41}$, $5.0\e{41}$, $10\e{41}$, and $20\e{41}$ \ergs. 
\item Composition : Pure O, O and Mg (92\% and 8\% by mass), and full C burning ashes (see below). The pure O and OMg compositions were computed for selected values of other parameters only.
\item Filling factors : $f=0.001,0.01$, and 0.1. 
\item V = 8000 \kms (the characteristic observed line width.)
\item Time = 400d since explosion
\end{itemize}

The models are named by composition, mass, filling factor, and energy deposition, for example is O30-0.1-2.0 a pure oxygen gas of mass 30 \msun, with $f=0.1$ and deposition $2.0\e{41}$ \ergs. 

By allowing for a free energy deposition we decouple the problem of powering
and spectral formation. For energy input by high-energy particles, the model spectra are insensitive to the properties of the input spectrum as long as the primaries have an energy $E \gtrsim$ 1 keV  \citep{Axelrod1980,Kozma1992}. Thus, the results will be valid for any high-energy input such as gamma rays from
radioactive decay or hard X-rays from a central engine or from shocks.

The choice of C-burning ashes to represent the O-zone composition
needs some motivation. In SN explosion models there are three principal oxygen zones;
He burning ash (C/O dominated), C burning ash (O/Ne/Mg dominated), and Ne
burning ash (O/Si/S dominated). The O/Ne/Mg zone is in most CCSN models the most massive of these,
in particular at high $M_{\rm ZAMS}$ values \citep{Woosley2007}. Also in PISN models is O/Ne/Mg the most common composition \citep{Heger2002}. In addition, the O/C and O/Si/S zones are prone
to formation of CO and SiO molecules at nebular times \citep{Liu1995}, which can lead to strong cooling
and blackout of optical thermal emission. Therefore, C-burning composition is most
motivated for single-zone modelling. Clearly, the strong Mg I] 4571 line in SLSNe is
also stemming from this zone.

The next question is which composition to use for the C-burning ashes.
The composition in nucleosynthesis models depends on the particular progenitor, but the relative abundances of O, Ne, Mg, and Na are relatively insensitive to $M_{\rm ZAMS}$. The Mg/O mass ratio is typically around 0.09,
which we use in the pure OMg models (compare to the solar ratio of 0.12). For the C-burn models, we use the composition from the ONeMg zone in the $M_{\rm ZAMS}=25$ \msun~solar metallicity model of \citet{Woosley2007}, listed in Table \ref{table:Cash}. Elements heavier than calcium stem mainly from primordial (solar) abundances, although some modification is made by processes such as $n$ and $p$ capture. Because SLSNe have been shown to arise in regions of subsolar metallicity \citep{Chen2013,Lunnan2014,Chen2015,Leloudas2015,Perley2016,ChenTW2016}, a reduction of
iron-group abundance may be warranted. However, we did some experiments comparising models
with 10 times lower metallicity and found only moderate differences.

\begin{table}
\caption{Composition (mass fractions) of C burning ashes. The abundances are taken from the $M_{\rm ZAMS}=25$ \msun~model of \citet{Woosley2007}.}
\centering
\begin{tabular}{cccc}
\hline
H & 0 & S & $6.3\e{-4}$\\
He & $1.4\e{-6}$ & Ar & $9.9\e{-5}$\\
C & $7.9\e{-3}$ & Ca & $3.1\e{-5}$\\
N & $3.5\e{-5}$ & Sc & $1.6\e{-6}$\\
O & 0.74 & Ti & $8.0\e{-6}$\\
Ne & 0.15 & V & $6.1\e{-7}$\\
Na & $1.8\e{-3}$ & Mn & $2.8\e{-6}$\\
Mg & 0.072 & Fe & $5.6\e{-4}$\\
Al & $6.7\e{-3}$ & Ni & $6.6\e{-4}$\\
Si & $0.014$ & Co & $1.3\e{-4}$\\
\hline
\end{tabular}
\label{table:Cash}
\end{table}


All models were computed without charge transfer, as these reactions gave numeric instabilities for some
high-density models, and to allow comparison we prefer to use the same physics for all models.

\subsection{Varying composition and density}
\label{sec:varycomp}
\begin{figure*} 
\centering
\large{}{30 \msun~ejecta models}\\
\hspace{0.1in}
\includegraphics[width=0.32\linewidth]{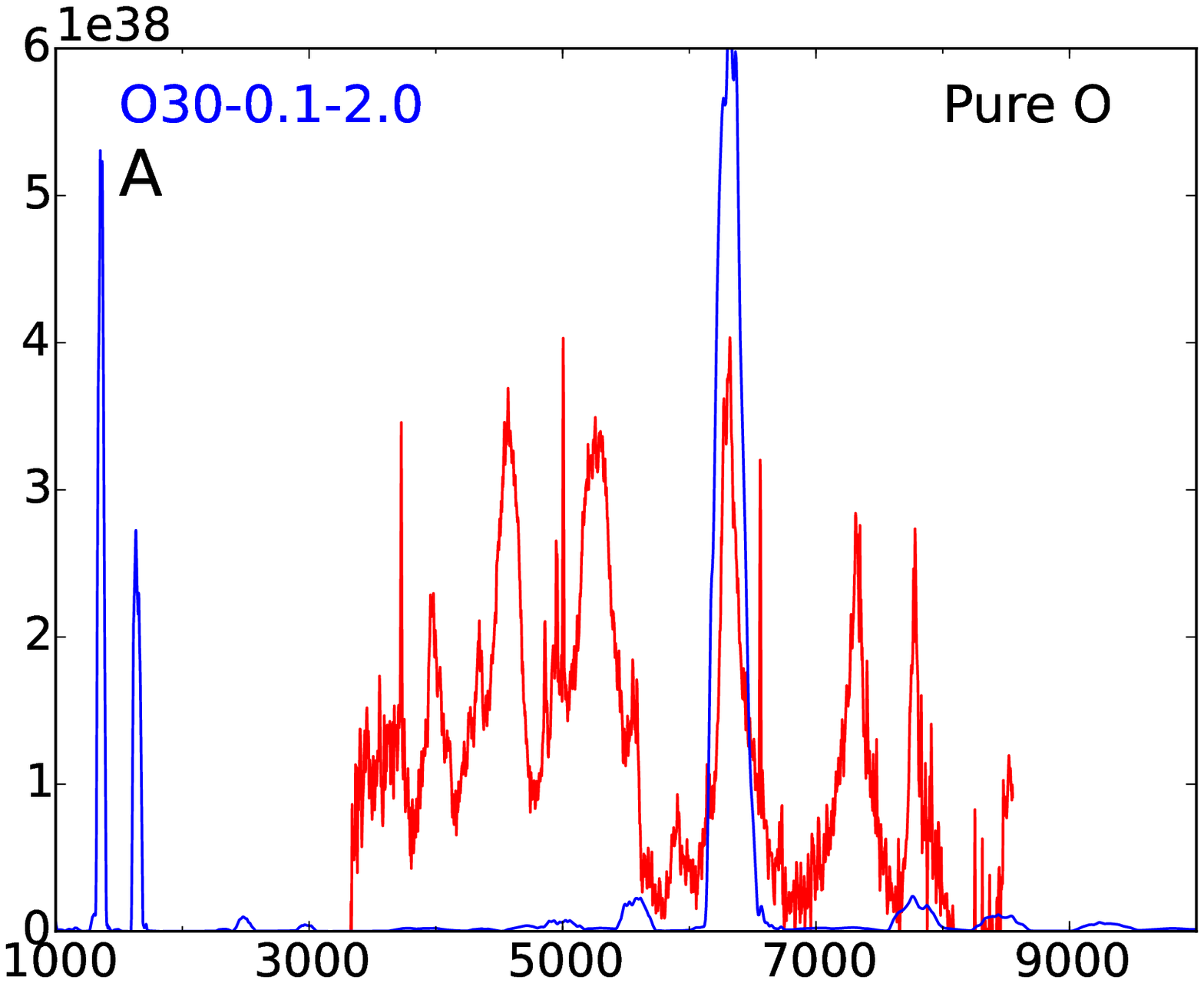} 
\includegraphics[width=0.32\linewidth]{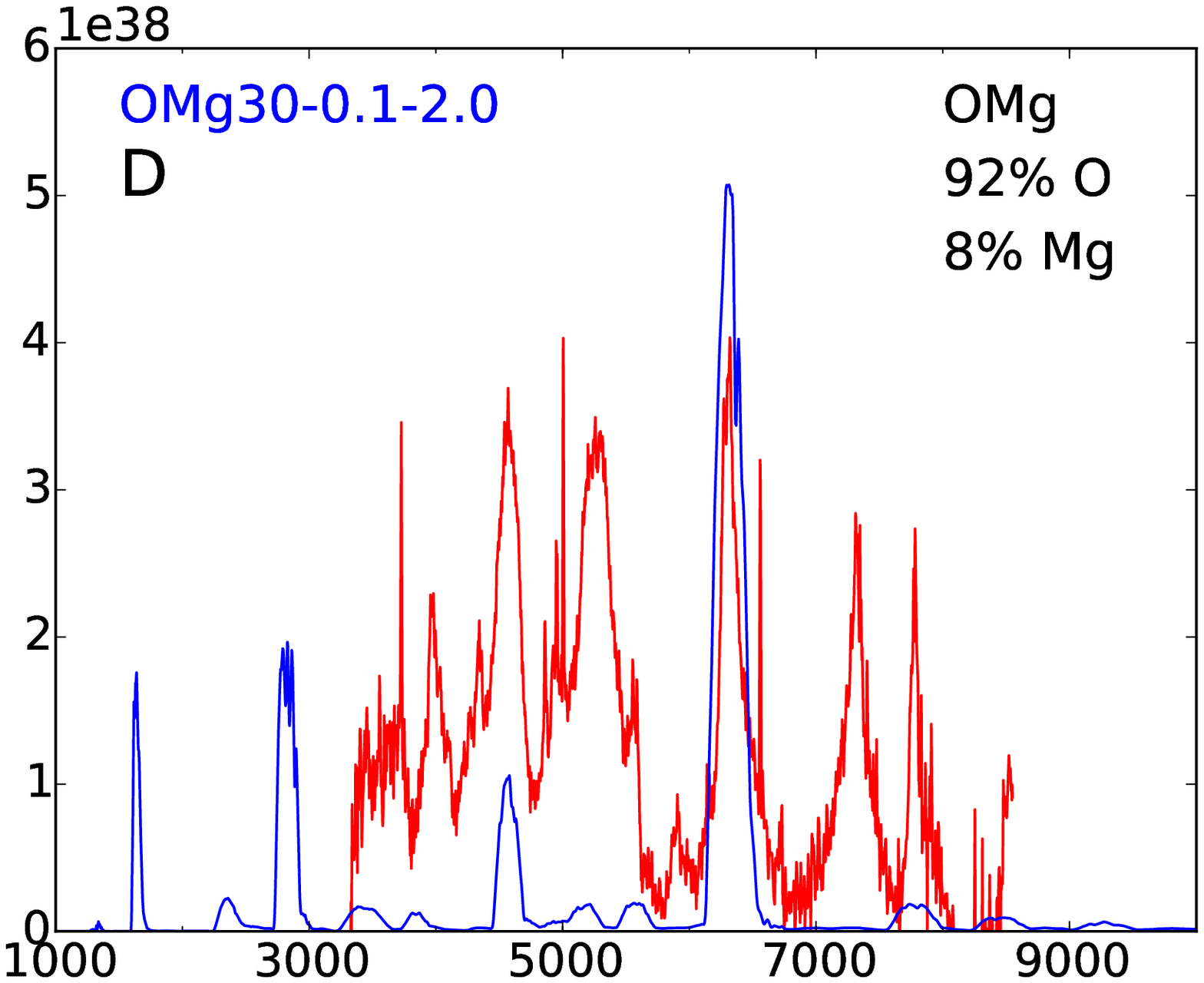} 
\includegraphics[width=0.32\linewidth]{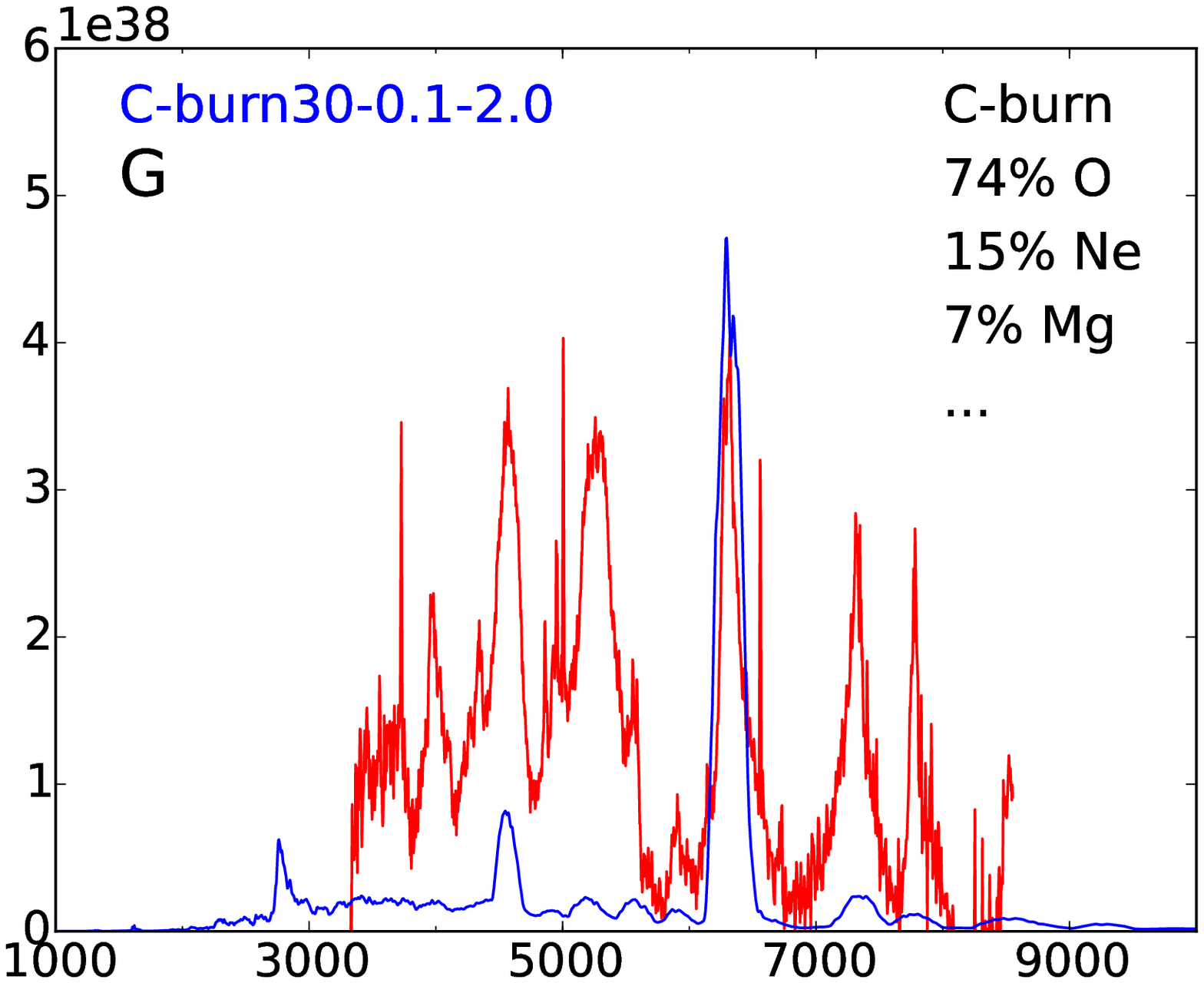}\\ 
\includegraphics[width=0.335\linewidth,height=1.9in]{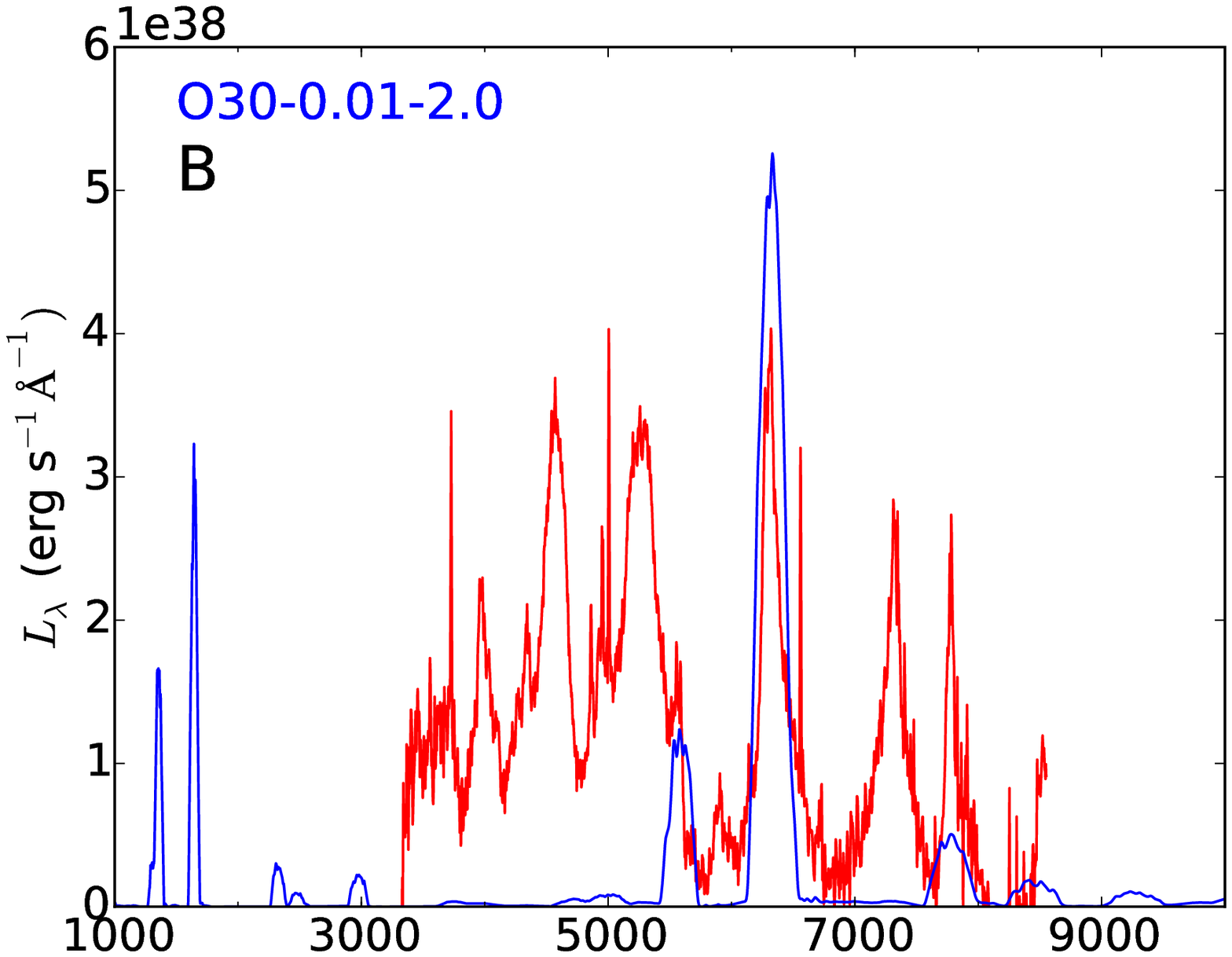} 
\includegraphics[width=0.32\linewidth]{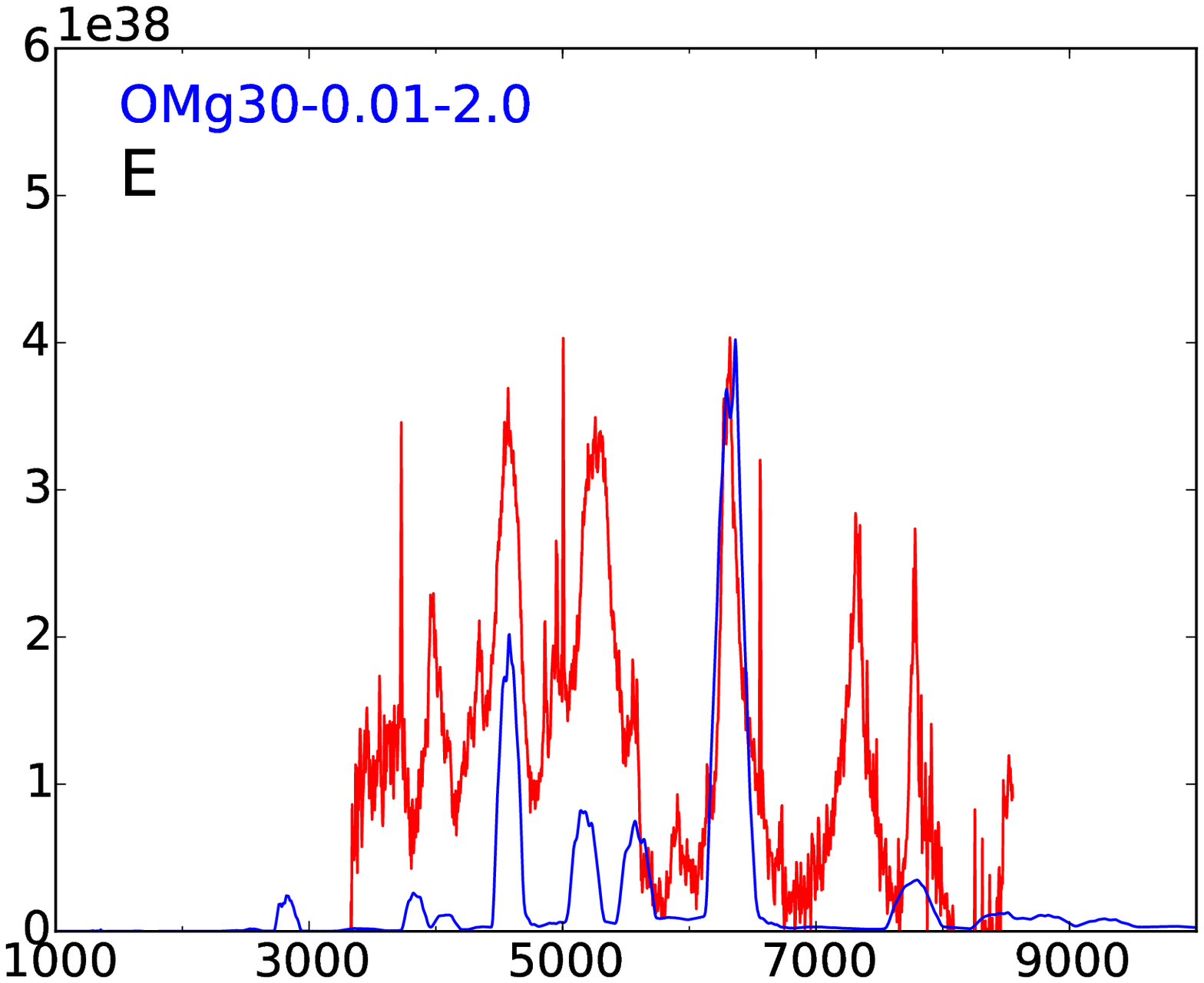} 
\includegraphics[width=0.32\linewidth]{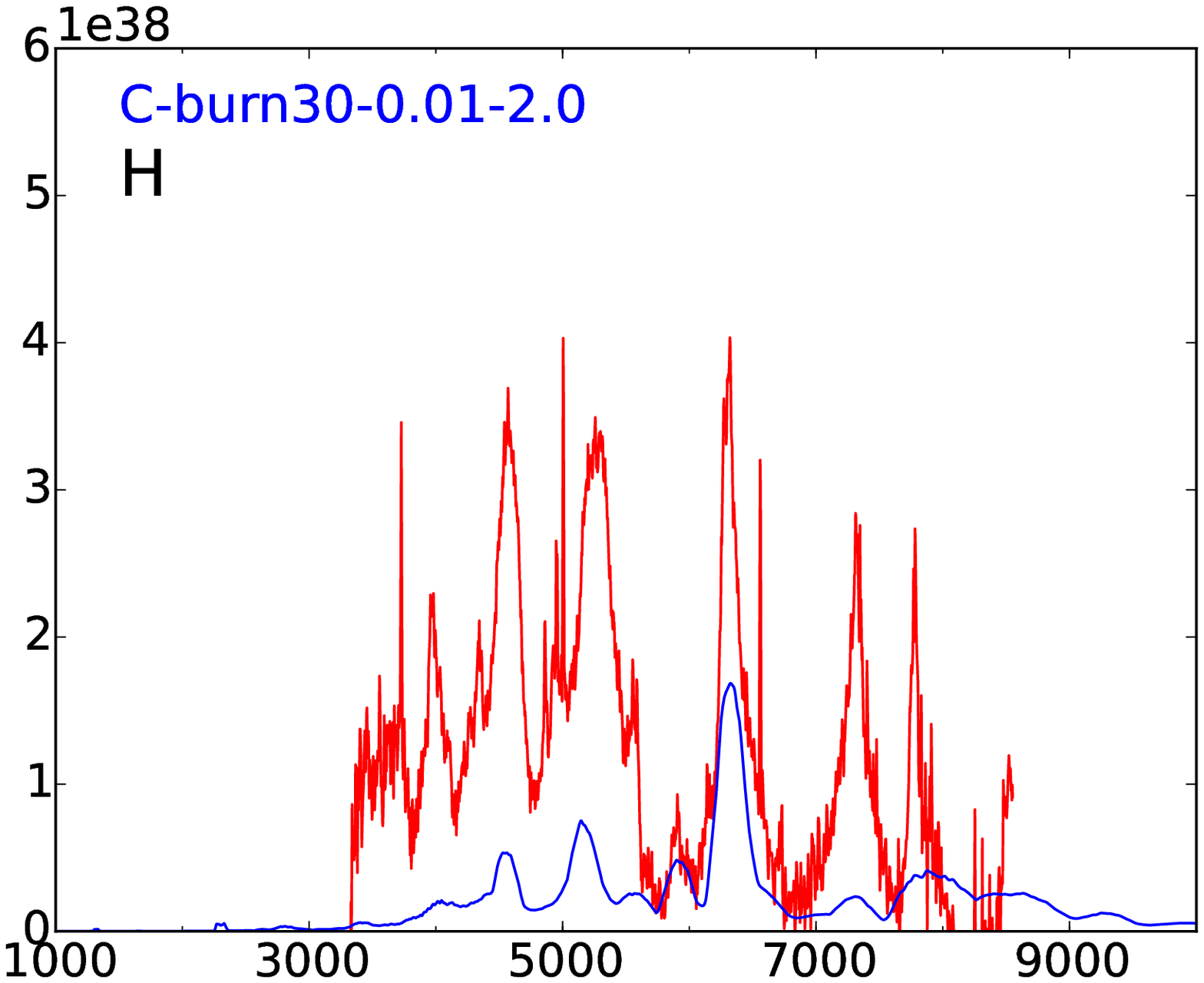}\\ 
\hspace{0.1in}
\includegraphics[width=0.32\linewidth,trim=0 0 0 0mm]{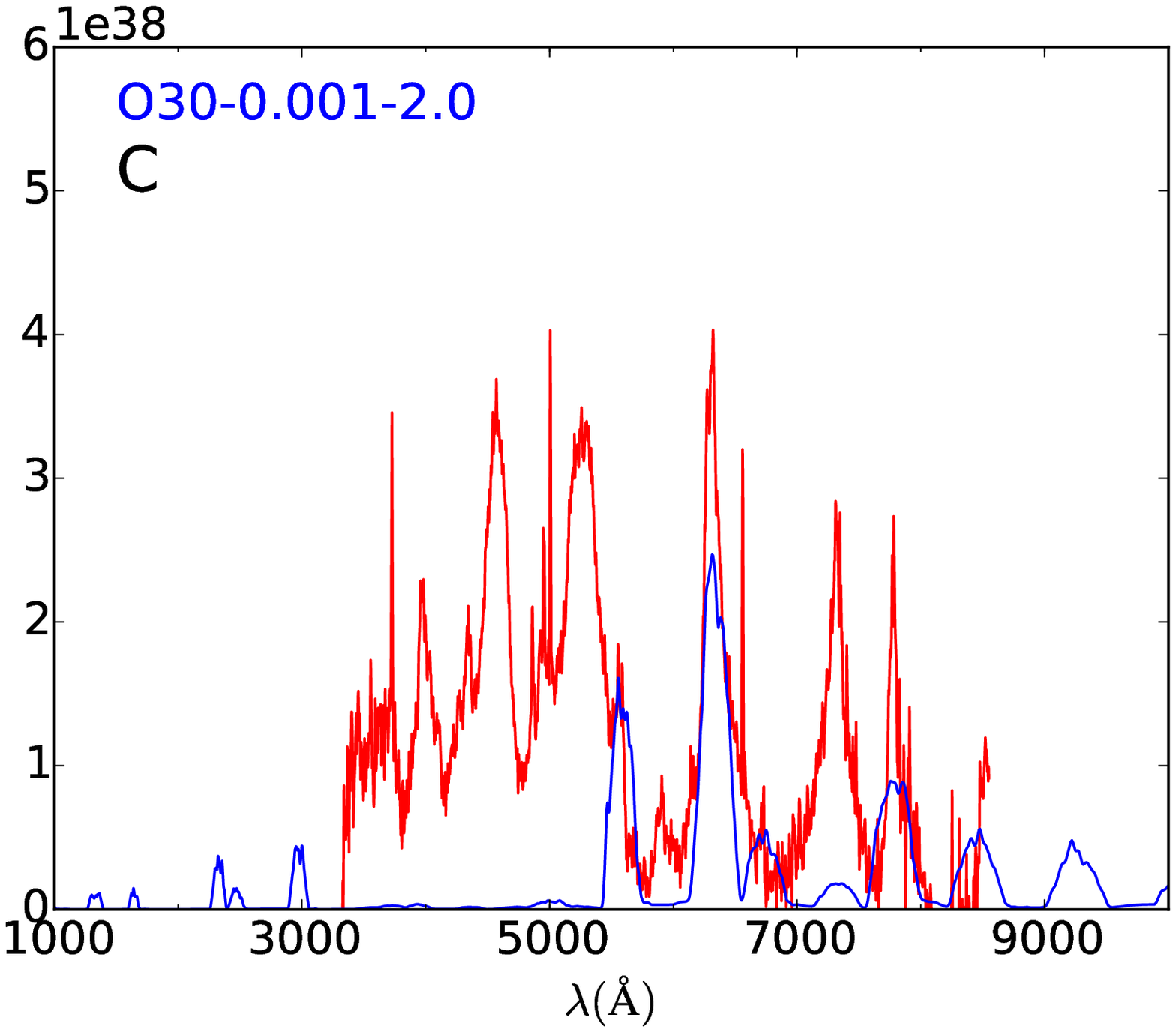} 
\includegraphics[width=0.32\linewidth,trim=0 0 0 0mm]{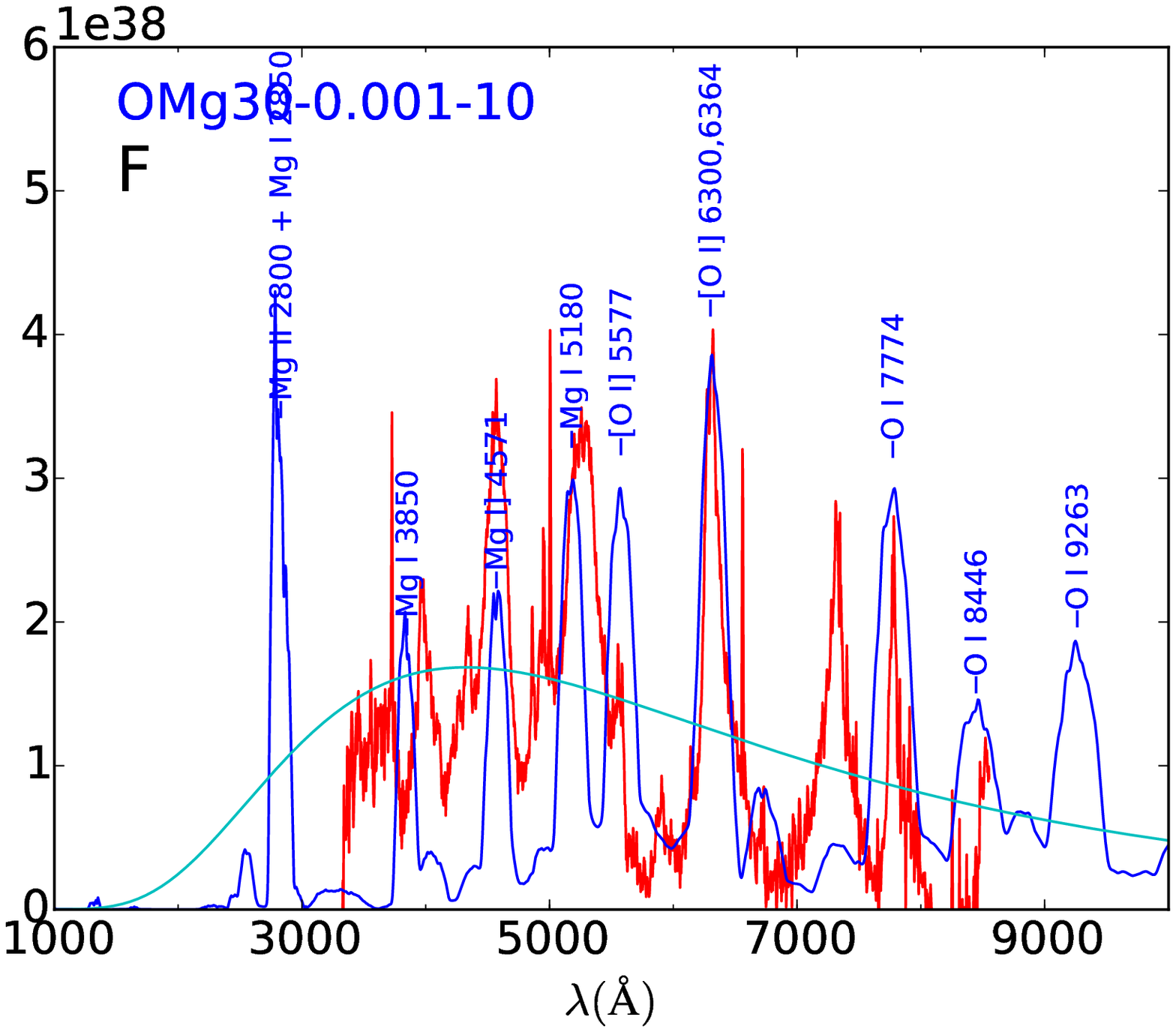} 
\includegraphics[width=0.32\linewidth]{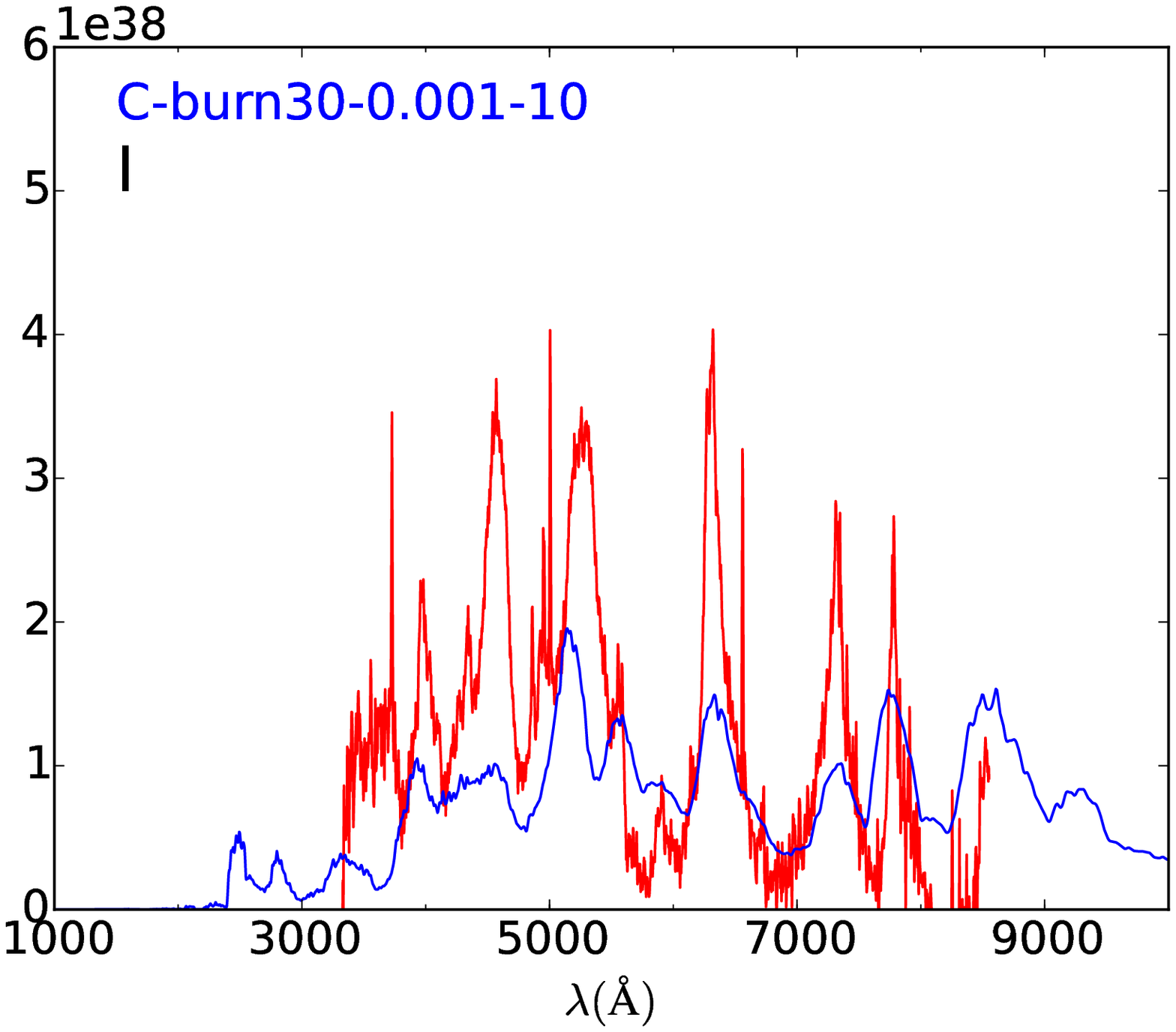} 
\caption{Comparison of SN 2007bi at +367d post-peak (red) and 30 \msun~O-zone models (blue). \emph{Left column} : pure O composition. \emph{Middle column} : OMg composition. \emph{Right column} : Full C burning composition. The rows are ordered by $f=0.1$ (top), $f=0.01$ (middle) and $f=0.001$ (bottom). In the bottom middle figure, a blackbody at the zone temperature of 6500 K is drawn.}
\label{fig:99d}
\end{figure*}

In Figure \ref{fig:99d} we illustrate the influence of composition and density, comparing $M_{ej}=30$ \msun~models with different compositions and filling factors. For each composition and $f$-value, we plot the model with energy deposition giving the best reproduction of [O I] 6300, 6364. 
The model spectra are compared with SN 2007bi at +367d. One should note that 
with our single-velocity approach we cannot directly address the issue of different line widths, but limit our discussion to line luminosities. Because we are only studying the emission from one of presumably many composition zones in the SN, the goal is not to reproduce the full
spectrum, but rather obtain an idea of what this zone contributes. Since the strongest and most prominent lines in the spectra are due to oxygen and magnesium (and calcium) significant parts of the spectrum should be recovered by an OMg composition. We review the main physical processes that produce the model spectra in the following sections.

\begin{figure*} 
\hspace{0.1in}
\includegraphics[width=0.32\linewidth]{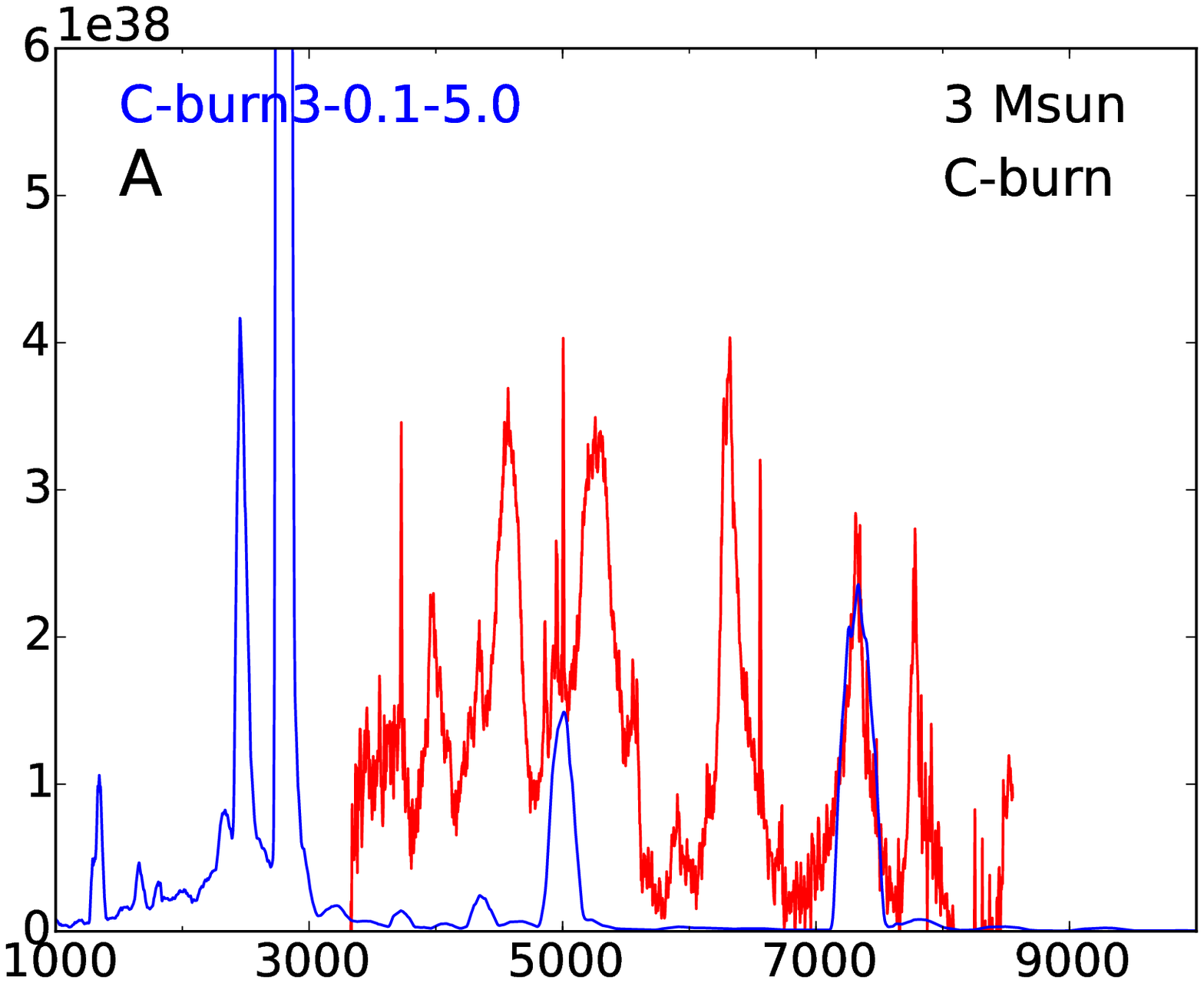} 
\includegraphics[width=0.32\linewidth]{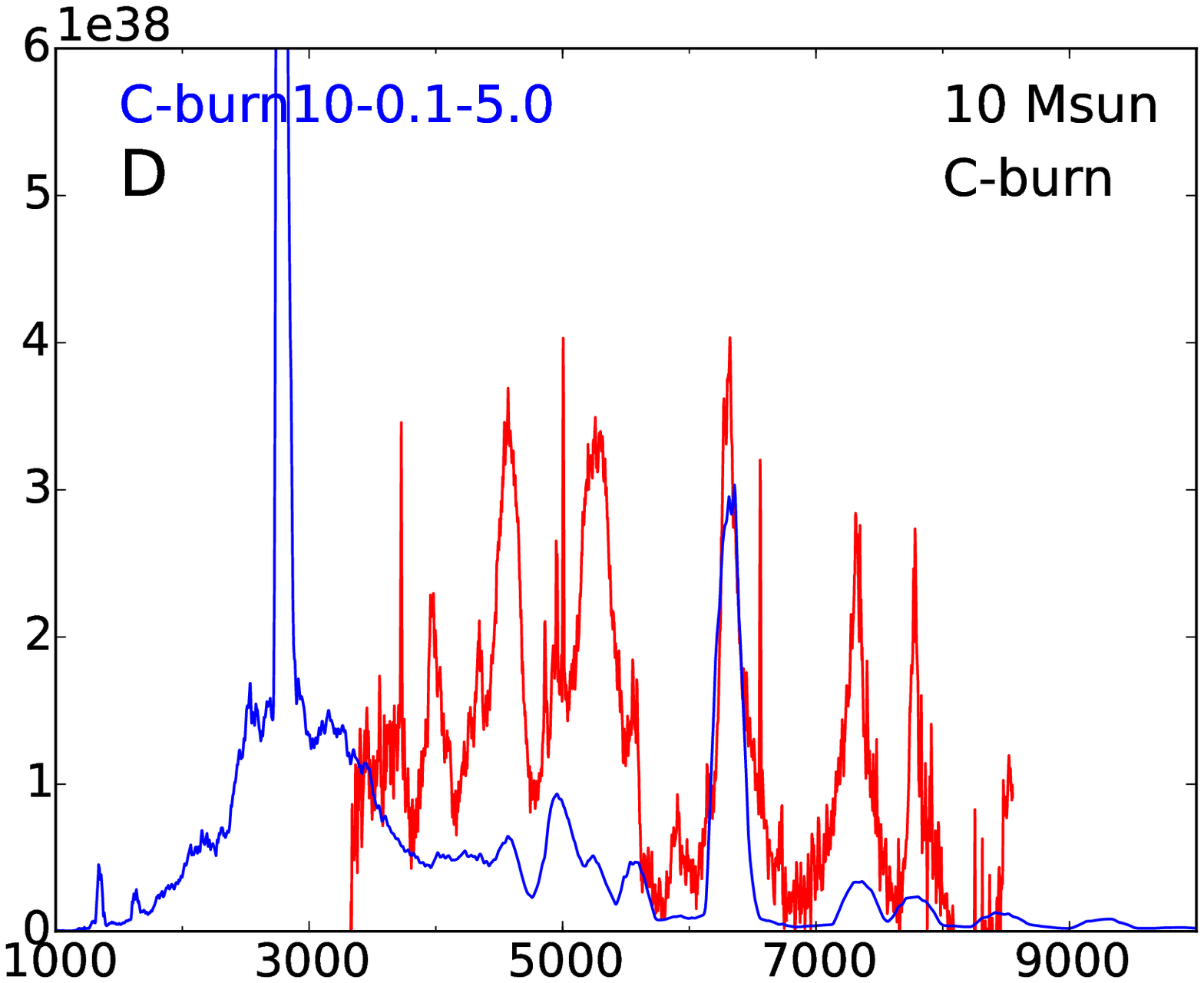} 
\includegraphics[width=0.32\linewidth]{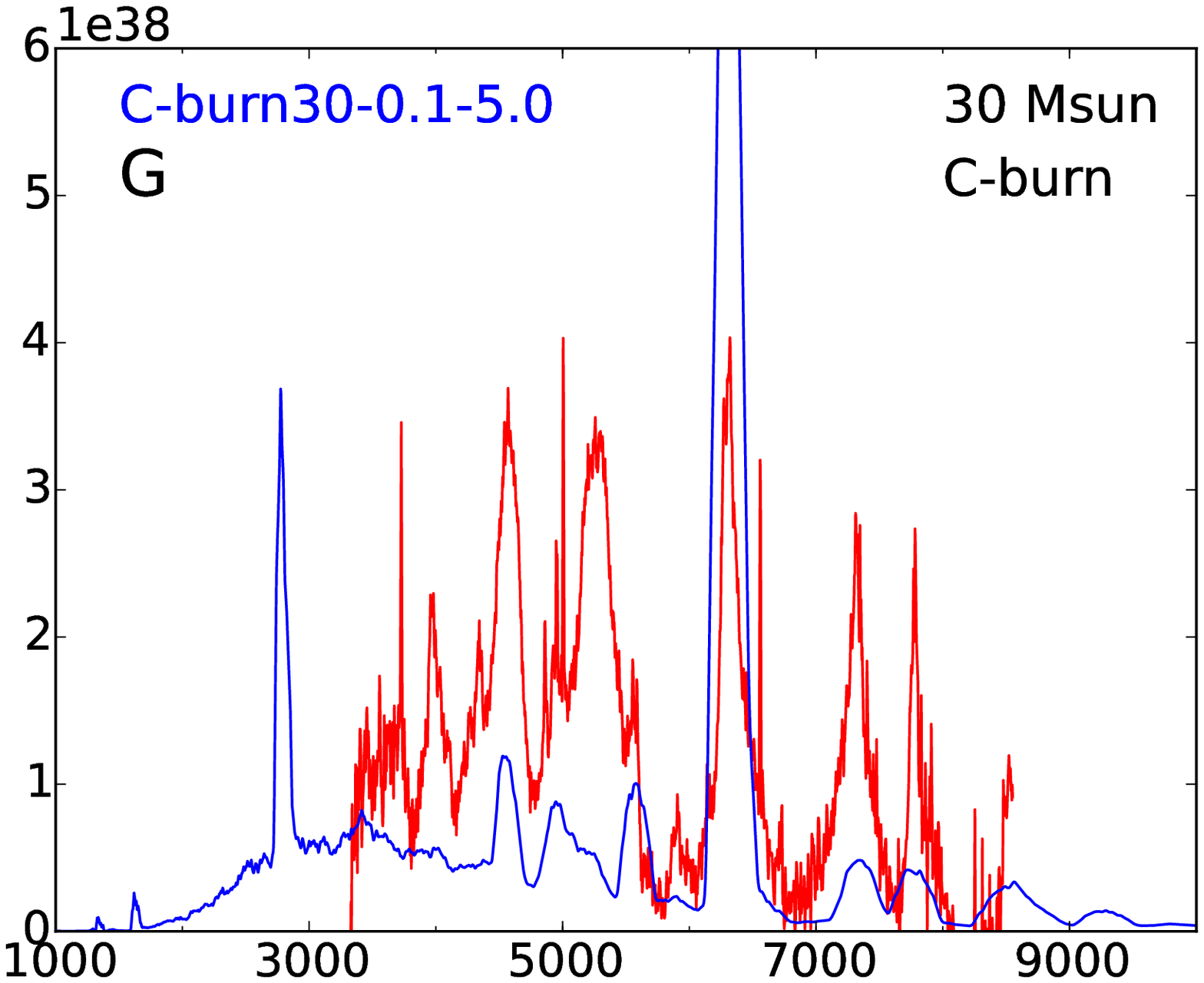}\\ 
\includegraphics[width=0.335\linewidth,height=1.9in]{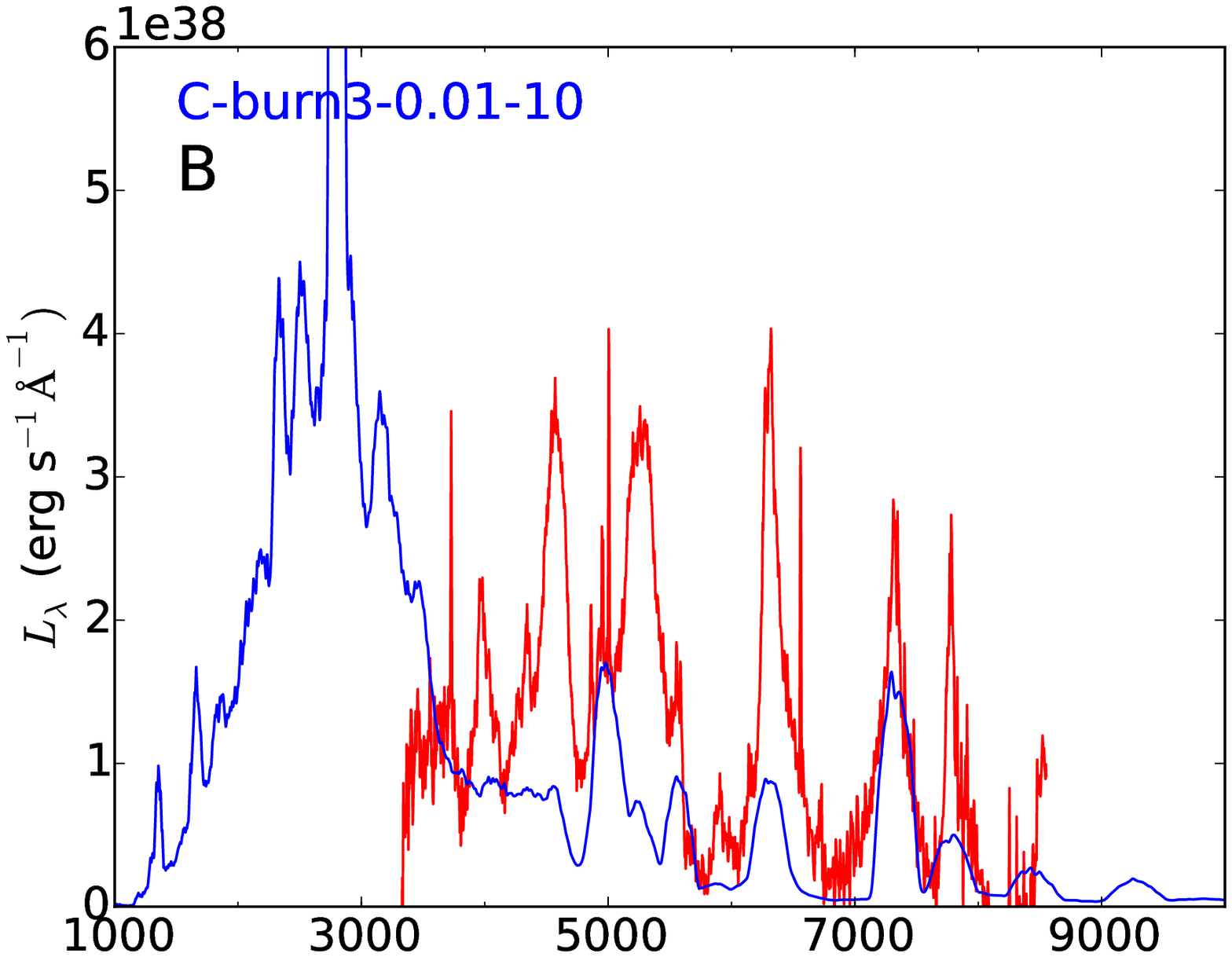} 
\includegraphics[width=0.32\linewidth]{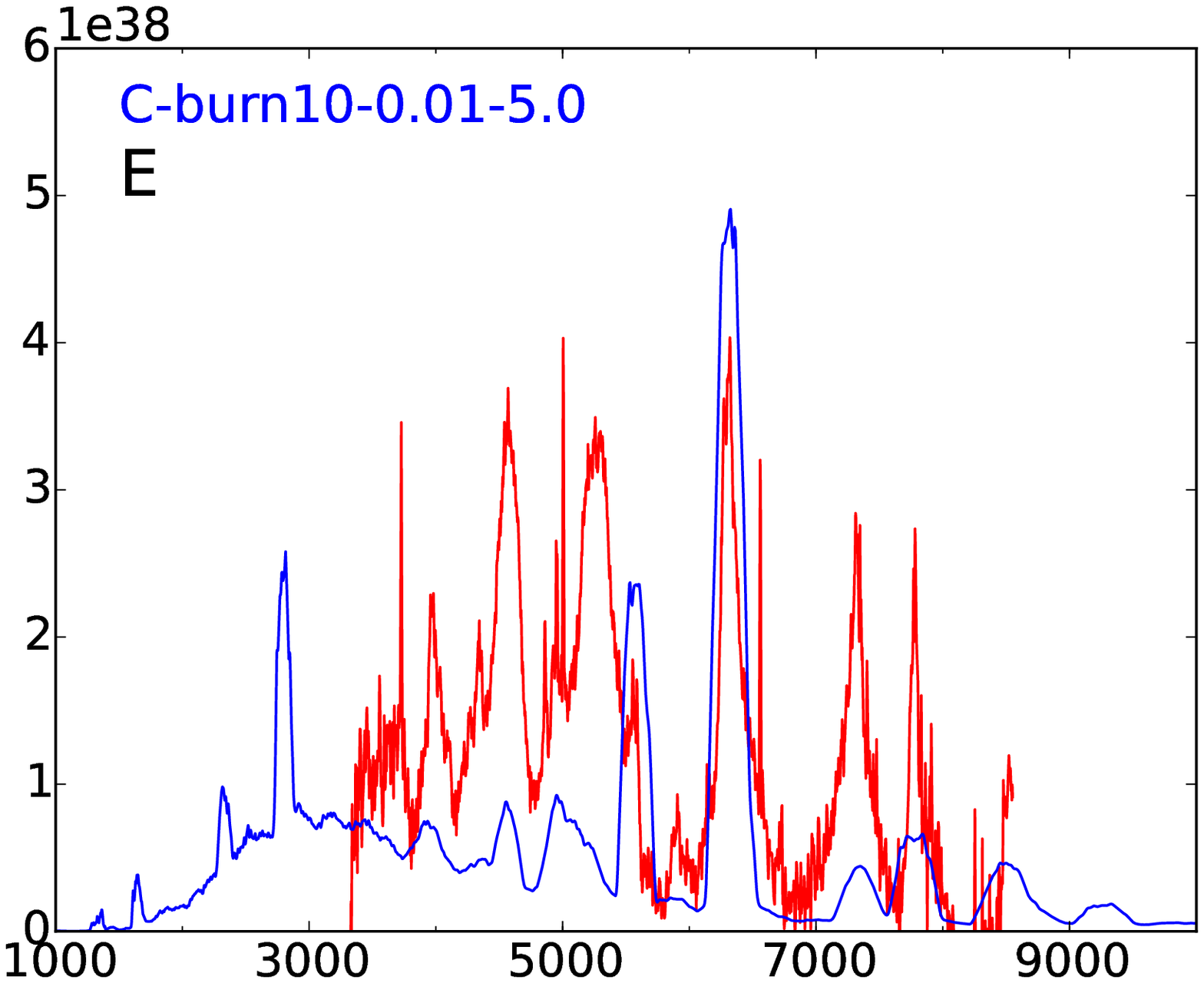} 
\includegraphics[width=0.32\linewidth]{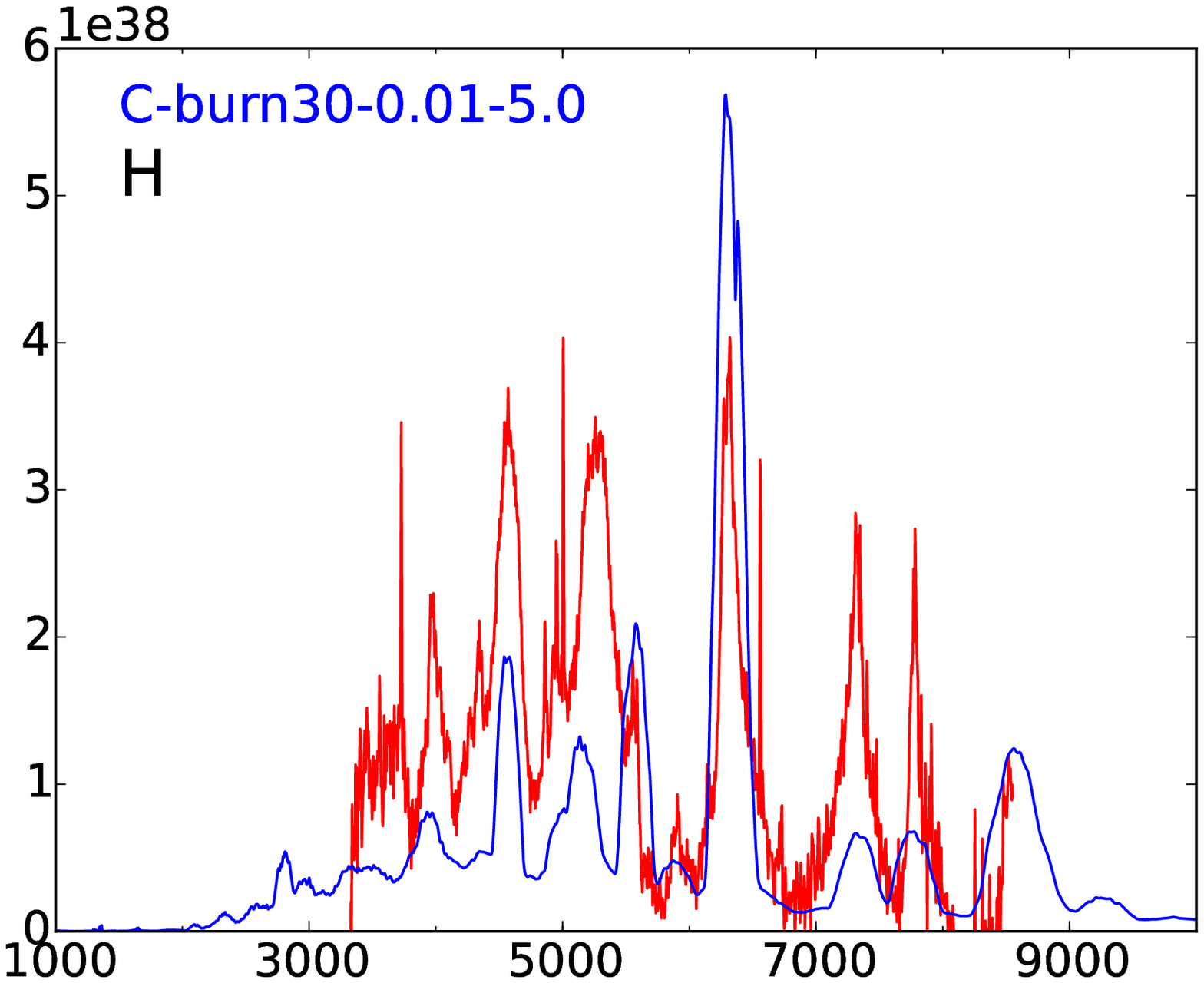}\\ 
\hspace{1in}
\includegraphics[width=0.32\linewidth]{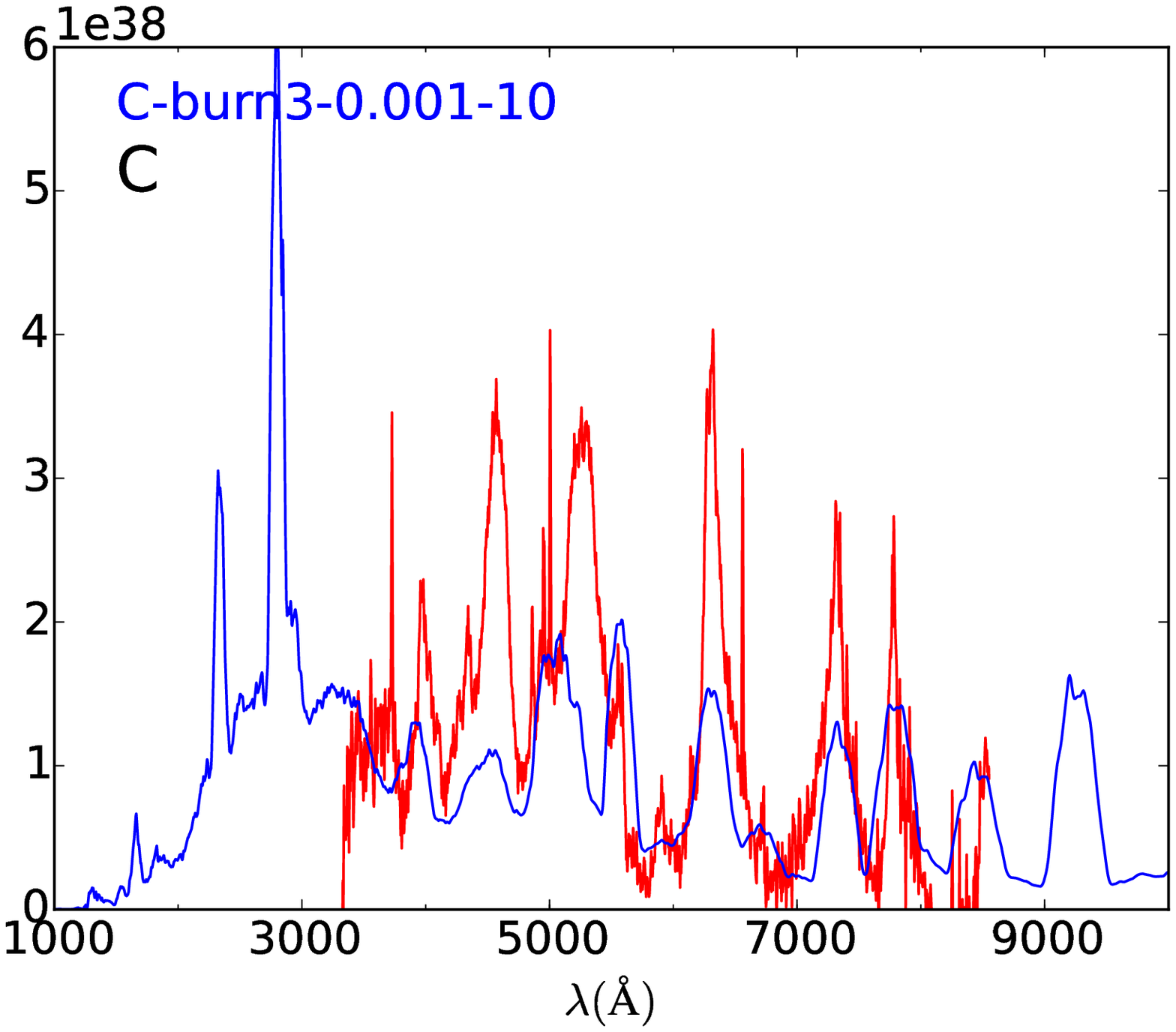} 
\includegraphics[width=0.32\linewidth]{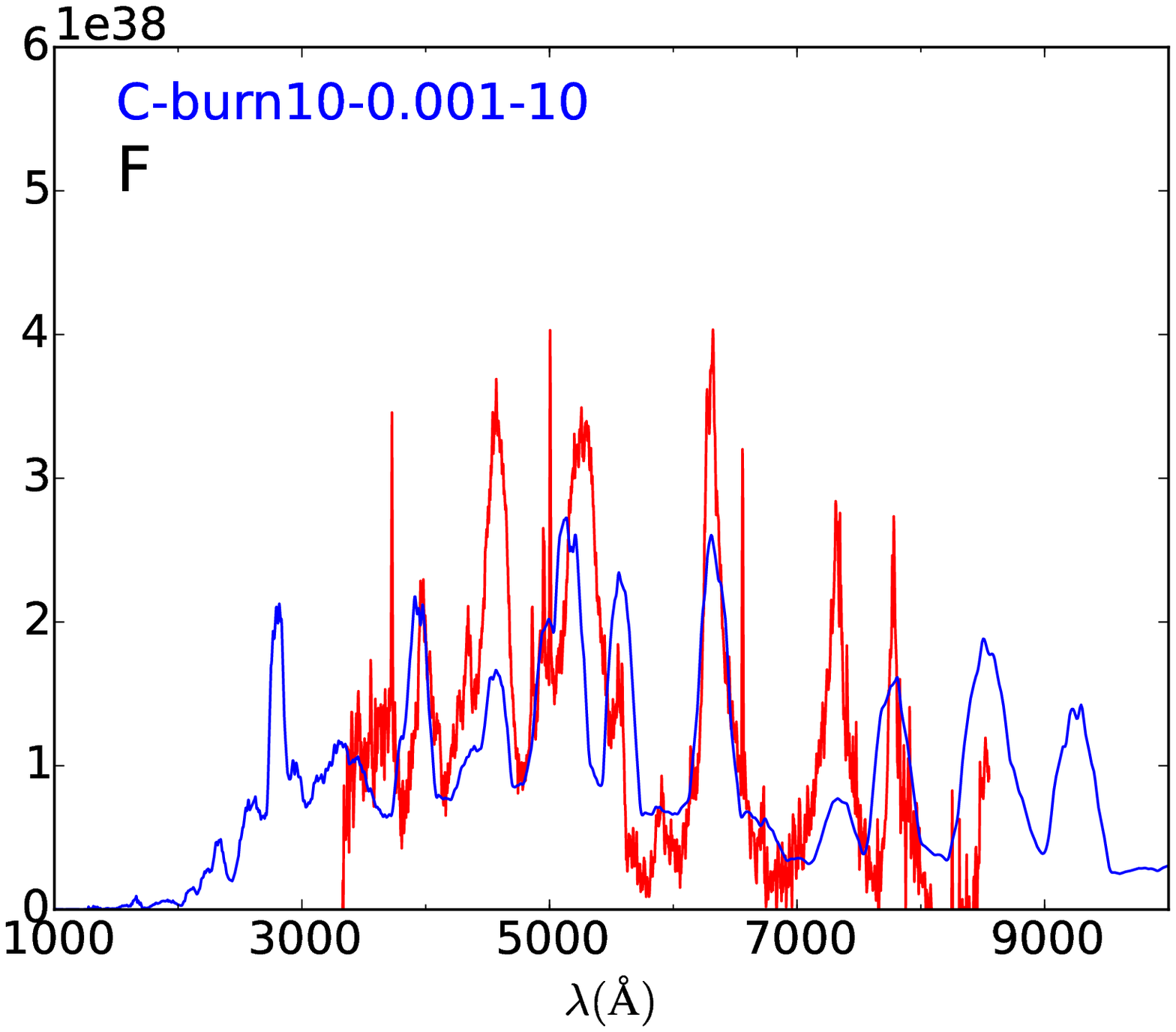} 
\includegraphics[width=0.32\linewidth]{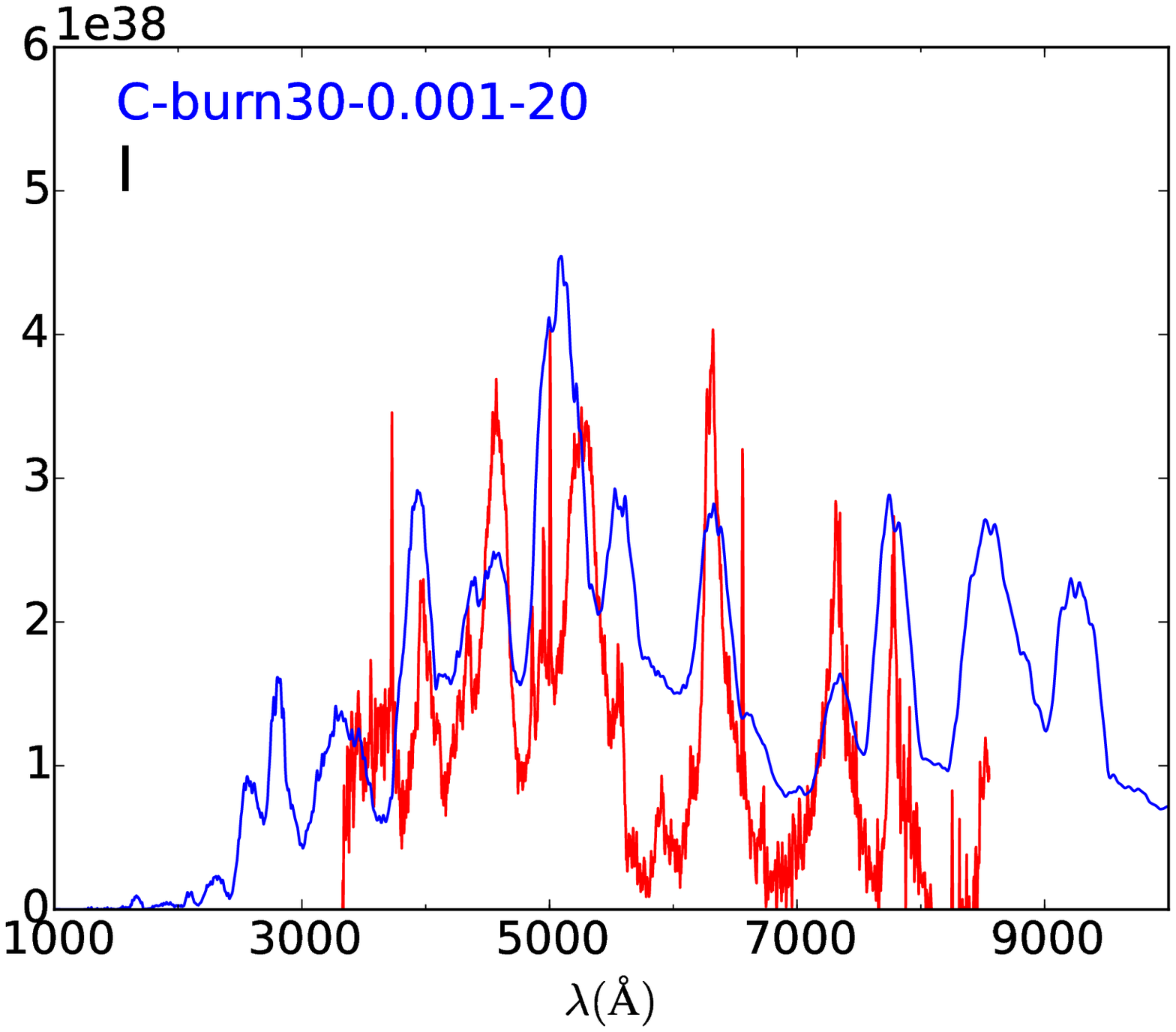} 
\caption{Comparison of SN 2007bi (red) and O-zone models (blue). \emph{Left column:} $M$=3 \msun~models. \emph{Middle column} : $M$ = 10 \msun~models. \emph{Right column:} $M$=30 \msun~models. The rows are ordered by $f=0.1$ (top), $f=0.01$ (middle) and $f=0.001$ (bottom). For each $M$ and $f$ combination, the model with deposition giving the best oxygen lines is plotted.}
\label{fig:varymass}
\end{figure*}


\subsubsection{Pure oxygen composition} 
At $f=0.1$ (model O30-0.1-2.0, Panel A in Fig. \ref{fig:99d}), there is little emission apart from [O I] 6300, 6364. The
temperature ($T=4400$ K) is too low for [O I] 5577. There is some power in O I 1350, O I 1640, O I 7774, O I 8446 and O I 9263, arising by recombination mainly. [O I] 6300, 6364 and [O I] 5577 are optically thin ($\tau\sim0.1$) but the recombination lines are optically thick.

O I 7774 is weak compared to the observed line. While about 30\% of the deposited energy goes into ionizing O I, the maximum energy fraction that can emerge in O I 7774 is $0.30\times 1064/7774 = 0.04$ (assuming Case B recombination so the effective ionization threshold is 1064 \AA\ for 2p($^1$D)). This is small compared with the cooling in [O I] 6300, 6364 which is about 50\% of the total deposition. Thermal population of the O I 7774 parent multiplet 3p($^5$P) from the ground state is not possible (the energy gap is 10.8 eV), and any such very hot situation would anyway lead to a very strong [O I] 5577 line (excitation energy of 4.2 eV), which is observed to be weaker than O I 7774. Only 2\% of the energy goes to non-thermal excitations in O I, ruling out this mechanism for powering O I 7774 or any other strong O I line. The NIR spectra of SN 2015bn and LSQ14an show a strong O I 9263 line (Fig. \ref{fig:PS15ae_optandnir}),  which is expected as this is also a recombination line with similar expected strength as O I 7774. This is also underproduced in the $f=0.1$ model.

Lowering $f$ to 0.01 (model O30-0.01-2.0, Panel B) leads to stronger [O I] 5577 because the temperature increases to $T = 4820$ K (from $T = 4400$ K). Collisional excitation of this line occurs mainly from the first excited state, 2p($^1$D) (i.e. in the [O I] 5577 transition itself), and the fact that there are a factor 1.6 more O I atoms in the 2p($^1$D) state in this model also boosts [O I] 5577.  The recombination lines strengthen somewhat as the higher
clumping leads to lower ionization (electron fraction $x_e=0.053$ compared to $x_e=0.13$), which reduces
the heating fraction and increases the ionization fraction from 0.36 to 0.44. In addition photoionization begins to contribute. The recombination lines are, however, still much weaker than [O I] 6300, 6364. [O I] 6300, 6364 and [O I] 5577 are marginally optically thick ($\tau=1-3$), whereas O I 7774, O I 8446 and O I 9263 have $\tau \gg 1$.

Lowering $f$ to 0.001 (model O30-0.001-2.0, panel C), [O I] 5577 strengthens further as the temperature continues to increase ($T = 6550$ K). The optical recombination lines strengthen again as the ionization fraction goes to 0.49 at $x_e=0.025$, and photoionization now dominates non-thermal rate, effectively transferring energy from the UV lines and continua to the optical recombination lines. In addition, collisional excitation now contributes a similar amount as recombinations to the O I 7774 luminosity. These excitations occur mainly from 3s($^5$S) which is effectively meta-stable having optically thick transitions to the ground state. That O I 7774 can become a cooling line at high density was pointed out by \citet{Maurer2010}, and the line holds promise as a tracer of high density.
 
In this model, all lines seen have $\tau \gg 1$.
The model has quite significant O I 7774, O I 8446, and O I 9263.
The UV lines are now quenched due to absorption by lines and continua in the high density clumps.
Note some emergence of [O II] 7320, 7330\footnote{Six lines, three at 7320 \AA\ and three at 7330 \AA.}, which requires high temperature. But overall, none of these 30 \msun~ejecta, constrained to fit the [O I] 6300, 6364 lines, produce any significant [O II] or [O III] lines.

By contrasting the three different $f$ values, we see that increasing the clumping increases [O I] 5577, O I 7774, O I 8446 and O I 9263 at the expense of [O I] 6300, 6364. The O I 7774, O I 8446 and O I 9263 lines gain from two effects at higher density; increased fraction of non-thermal ionization and photoionization, and more efficient cooling from excited states as their relative populations increase. The observed relation between these lines is better reproduced at high clumping. 

\subsubsection{O/Mg composition} 
We now explore models containing magnesium. Magnesium typically becomes much more ionized than oxygen because its low ionization thresholds 
at 1620 \AA\ (ground state) and 2510 \AA\ (first excited state) makes it prone to photoionization \citep[e.g.][]{Fransson1989}.
As discussed in \citet{J15a}, the exact solution for the neutral fraction determines whether Mg I] 4571
can contribute to the cooling or not. A high luminosity in this line as observed in SLSNe almost certainly requires cooling and not just recombination. How this is achieved is not easily understood,
since Mg I is easy to ionize when the energy deposition is high. The neutral fraction is typically increased by higher density, and this holds potential as a constraint to be explored.

Introducing magnesium at $f=0.1$ (model OMg30-0.1-2.0, panel D in Fig. \ref{fig:99d}) gives a temperature $T = 4330$ K (compared to $T =\ $4400 K for pure O composition), electron fraction $x_e = 0.15 (0.13)$,
and neutral magnesium fraction $x(\mbox{MgI}) = 0.004$, as expected much lower than the oxygen neutral fraction (0.90).

Mg II 2795, 2802 and Mg I] 4571 are distinct. Neutral magnesium, although rare, does about 10\% of the cooling. This makes a distinct line, although weaker than observed. The recombination emission in Mg I] 4571 is moderate, and the cooling dominates the line. Note that 10\% cooling corresponds to $dep \times f_{heating} \times f_{cool,line}=\sim 2\e{41}\times 0.6\times 0.10= 10^{40}$ erg s$^{-1}$, which with a typical line width of 100 \AA~is $10^{38}$ erg s$^{-1}$ \AA$^{-1}$, close to what is seen in the model. 
Mg II 2800 does 12\% of the cooling. Some of this emission fluoresces in O and Mg lines, but a big part escapes. 

At $f=0.01$, (model OMg30-0.01-2.0, panel E), the temperature is $T = 4530$ K (compare $T = 4820$ K pure O composition), $x_e = 0.076 (0.053)$, and
$x(\mbox{MgI}) = 0.072$.
Higher clumping boosts Mg I] 4571, because Mg has now crossed the "ionization runaway" threshold (see Appendix \ref{sec:icat}) and is neutral to 7.2\%, doing significant cooling (25\%). 
The line is in LTE and optically thick ($\tau=900$), as such its luminosity can be derived from
the volume $\mathcal{V}$ of the emitting region and the temperature (see e.g. Jerkstrand, in prep.\footnote{In ``Handbook of supernovae'', Springer}):

\begin{equation}
L = \frac{4\pi \mathcal{V}}{ct} \lambda B_\lambda(T)
\label{eq:LLTEthick}
\end{equation}

Once in this regime, the line luminosity cannot be easily increased further; the temperature would need to increase enough with compression so $B_\lambda(T)$ grows faster than $1/\mathcal{V}$. This will certainly not happen for lines with $\lambda \gtrsim hc/\left(4.96kT\right)$ (peak of the blackbody), and also shorter wavelength lines typically show decline or modest growth.
From the perspective of the single transition, increasing the Mg I fraction leads to increased optical depths, which offsets the increased number of atoms available. 
As a rough picture to understand the 25\% cooling in Mg I] 4571, most of the 12\% cooling done by Mg II in the previous model has now been transferred to Mg I] 4571. The Mg II lines are in the LTE and optically thick regime already at $f=0.1$, and thus from Eq. \ref{eq:LLTEthick}, reducing $\mathcal{V}$ by a factor of 10 while obtaining only a small change in $T$, their luminosity drops by a factor $\sim$10. Mg I] 4571, on the other hand, is in NLTE at $f=0.1$, and can increase its
luminosity until it saturates in LTE, compensating for the loss of Mg II cooling.

The Mg I 5180 multiplet has now also emerged, being formed mainly by recombinations but also
by some collisional excitation from 3p($^3$P). Because most Mg is still singly ionized, $\mathcal{V} \times n_{MgII}$ is constant and the Mg recombination line luminosities increase as $\mathcal{V} n_e n_{MgII} \propto n_e$. The Mg I 5180 lines are optically thick ($\tau \sim 10^5$), but the parent population is below LTE. 
At $f=0.001$ and an energy deposition of $2\e{41}$ \ergs~(as used in all models up until now), a much too dim optical spectrum (not shown) is produced, as Mg NIR lines dominate the cooling. To fit the [O I] 6300, 6364 lines, we need instead a deposition of $1\e{42}$ \ergs (model OMg30-0.001-10, Panel F). 
This model has temperature $T = 6470$ K, ionization $x_e = 0.11$, and Mg neutral fraction 
$x(\mbox{MgI}) = 0.001$. The higher energy deposition has offset the higher clumping to reduce the Mg I abundance again. But the Mg I] 4571 line is still in the LTE, optically thick regime, $B_{4571}(T) \mathcal{V}$ is roughly the same
as in the $f=0.01$ model, and the line luminosity therefore changes little.
All the lines of Mg I] 4571, Mg I 5180, [O I] 5577, [O I] 6300, 6364
are now close to the LTE and optically thick regime; their peak flux therefore follows a blackbody at the zone temperature of 6700 K, per Eq. \ref{eq:LLTEthick} ($F_\lambda \propto L/\lambda \propto B_\lambda$, since $\Delta \lambda \propto \lambda$). A multiplet with $N$ lines, all in LTE and optically thick, would give a line feature $N$ times above the blackbody curve. The O I 7774, 8446, and 9263 multiplets tend to have departure coefficients still a factor 2-3 below unity, which offsets this multiplicity effect and gives total line luminosities close to the blackbody curve. [O I] 6300, 6364 has more radiative transfer losses than the other lines, and emerges somewhat below its expected flux.

This model has a good overall optical brightness compared to the observed spectra, and reproduces Mg I] 4571, Mg I 5180 (although somewhat too blue compared to the observed line), and [O I] 6300, 6364. It produces also a strong O I 9263, in good agreement with the observed line in SN 2015bn (Fig. \ref{fig:allthree}). It has too strong [O I] 5577 and O I 7774. Mg I 3850 and O I 8446 will scatter in Ca II HK and Ca II NIR, respectively, in models including calcium, so their overproduction is not a big concern.

A hot model like this ubiquitously produces a strong [O I] 5577, which is not distinct in SN 2007bi. There is an unusual ``cut-off'' in the observed spectra around 5600 \AA~that may correspond to a line absorption threshold (e.g. in He I 5876 or Na I D), but it appears that a [O I] 5577 as strong as in this model should still reveal itself by its blue edge. 

This is an interesting model in the sense that it shows that a pure O/Mg gas can qualitatively reproduce many of the observed spectral features in long-duration SLSNe. However, several key discrepancies remain; blueshift of the 5200 \AA\ feature, and too strong [O I] 5577 and O I 7774.
It is unclear whether the observed 5200 \AA\ feature is dominated by Mg I 5180, [Fe II] 5250, or is a blend. A Mg I 5180 component is plausible because it emerges in the models where Mg I] 4571 approaches its observed strength.
 
\begin{figure*}
\centering
\includegraphics[width=0.93\linewidth]{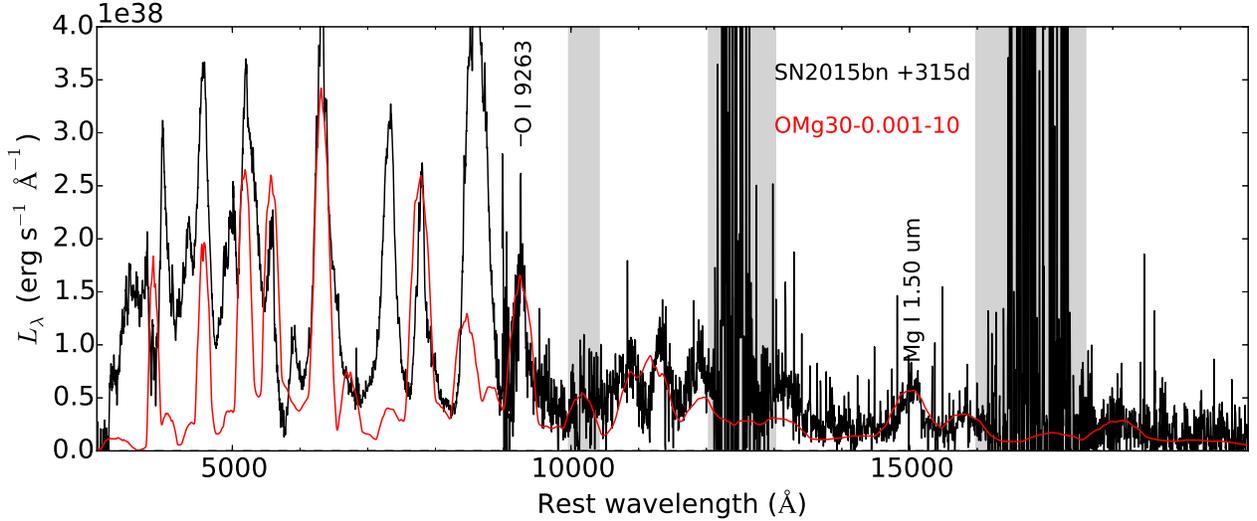} 
\caption{Comparison of SN 2015bn at +315d (black) and model OMg30-0.001-10 (red). Telluric bands are marked gray.}
\label{fig:OMg30PS15ae}
\end{figure*}

There is another constraint on the emission from magnesium from Mg I 1.50 $\mu$m. 
Figure \ref{fig:OMg30PS15ae} compares this model to the X-shooter spectrum of SN 2015bn which shows this line. 
The model reproduces O I 9263 and the Mg I 1.50 $\mu$m line, and overall the agreement across the NIR region is reasonably good. The data for LSQ14an are noisier but a feature at the 1.50 $\mu$m position may also be present (Fig. \ref{fig:PS15ae_optandnir}).

In summary, adding Mg to the composition (in its typical C-burning abundance) shows that Mg I] 4571 and Mg I 1.50 $\mu$m emerge naturally, and consolidates the identification of these lines. Mg I 5180 emerges at low $f$ values, and likely contributes to the 5200 \AA\ feature. The best reproductions of both O and Mg line ratios are at low filling factors. At high $f$ values magnesium tends to be too ionized
to make strong Mg I cooling lines, and the Mg I] 4571 luminosity indicates $f \leq 0.01$.

\subsubsection{C-burning composition}
Introducing the full composition at $f=0.1$ (model C-burn30-0.1-2.0, panel G in Fig. \ref{fig:99d}) gives a temperature $T = 4350$ K (4330 K for OMg), $x_e = 0.14$ (0.15), and $x(\mbox{MgI}) = 0.003$ (0.004).
Thus, trace elements have a small effect on temperature and ionization at this filling factor. They provide about 10\% of the line cooling.
They have, however, led to the suppression of O I 1640 and Mg II 2795, 2802, and the formation of a more pronounced quasi-continuum (made up of many overlapping weak lines). This quasi-continuum is flat and is weak compared to the [O I] and Mg I lines, and does not resemble the 4000-5500 \AA\ component that is often seen in stripped-envelope SNe and may be present in the SLSNe as well. As the OMg model at this $f$-value, this model has too dim Mg I lines. The presence of trace metals has weakened Mg I] 4571.

At $f=0.01$ (model Cburn30-0.01-2.0, Panel H), the temperature is $T = 3950$ K (4530 K for OMg), $x_e = 0.078$ (0.076), and $x(\mbox{MgI}) = 0.14$ (0.072). Trace elements have a somewhat larger effect on temperature and ionization at this filling factor. They now provide 70\% of the cooling, mainly through Fe I (iron is 75\% neutral). Mg I] 4571 is significantly damped. Despite the high Mg I fraction, Mg I] 4571 cannot compete with iron for the cooling, and cools only 5\%. The line peaking at 5150 \AA\ is dominated by Fe I, and not Mg I 5180.
At this higher density, Na I D has emerged. The neutral fraction of sodium is $x(\mbox{Na I})=5\e{-3}$, and Na I provides 5\% of the cooling, which makes a luminosity in good agreement with the observed one.



At $f=0.001$ (model Cburn30-0.001-10, Panel I), the temperature is $T = 5490$ K (6470 K for OMg), $x_e = 0.082$ (0.11), and $x(\mbox{MgI}) = 0.002$ (0.001). Trace elements have now a significant effect on the temperature and ionization. 
The trace elements provide 85\% of the line cooling.
Mg I] 4571 cools only a few percent. At this high density there are also several optically thick iron-group lines in the vicinity of Mg I] 4571, and the line is further quenched by line blocking, as well as blended into the surrounding quasi-continuum. In fact, there are thousands of optically thick lines throughout the optical (due to the primordial iron-group abundances) and a pronounced quasi-continuum is formed; we are approaching a blackbody SED.

In summary, while there are only moderate differences between OMg and C-burn compositions at $f=0.1$, the large differences at $f \leq 0.01$ shows that at high density, the models are sensitive to the presence of trace elements and thereby to the metallicity of the progenitor. The metals
tend to damp the contrast between distinct lines and quasi-continuum, and makes in particular
Mg I] 4571 less pronounced.

\subsection{Varying mass} 
We now consider the models at lower masses, $M_{ej}=3$ and 10 \msun. Each of these models uses the full C burning composition, but vary in filling factor and energy deposition. Figure \ref{fig:varymass} 
shows a subset of 9 of these models, where for each $M$ (horisontal variation) and $f$ (vertical variation) combination, the model with energy deposition giving best overall reproduction of O and Mg lines to SN 2007bi is shown. 

%
%

\subsubsection{3 \msun~models} 
\label{sec:3Msun}
\begin{figure}
\includegraphics[width=1\linewidth]{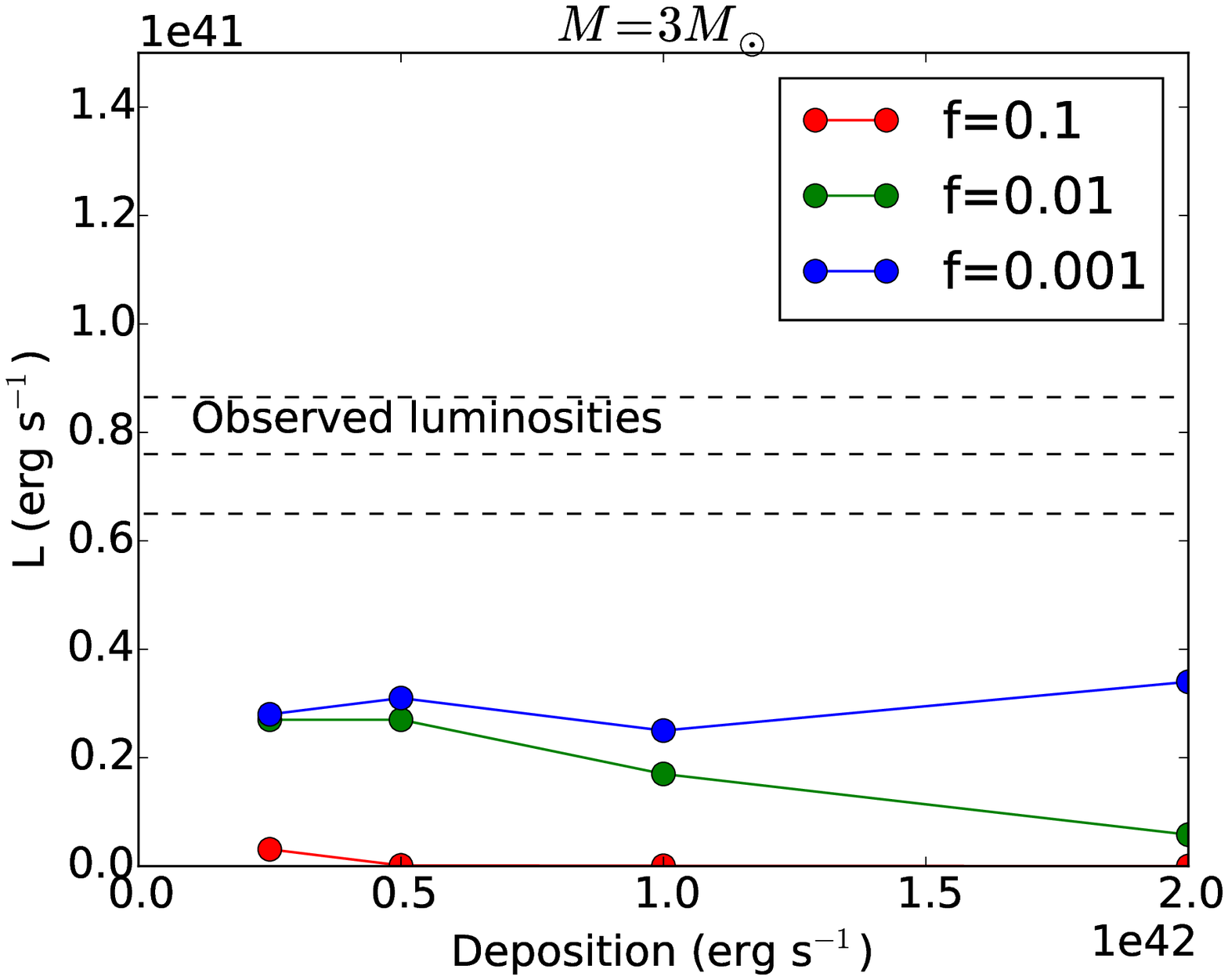}
\includegraphics[width=1\linewidth]{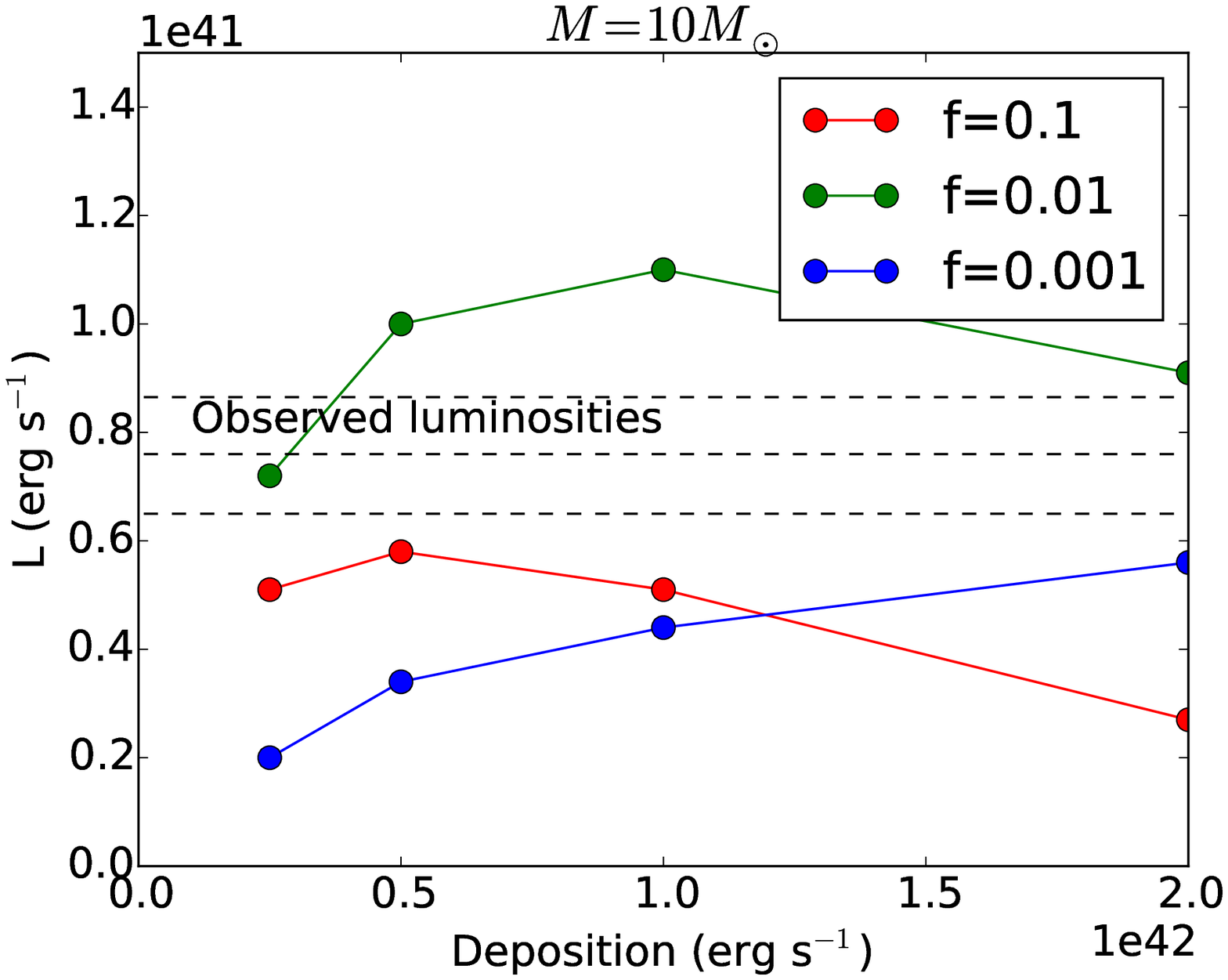}
\includegraphics[width=1\linewidth]{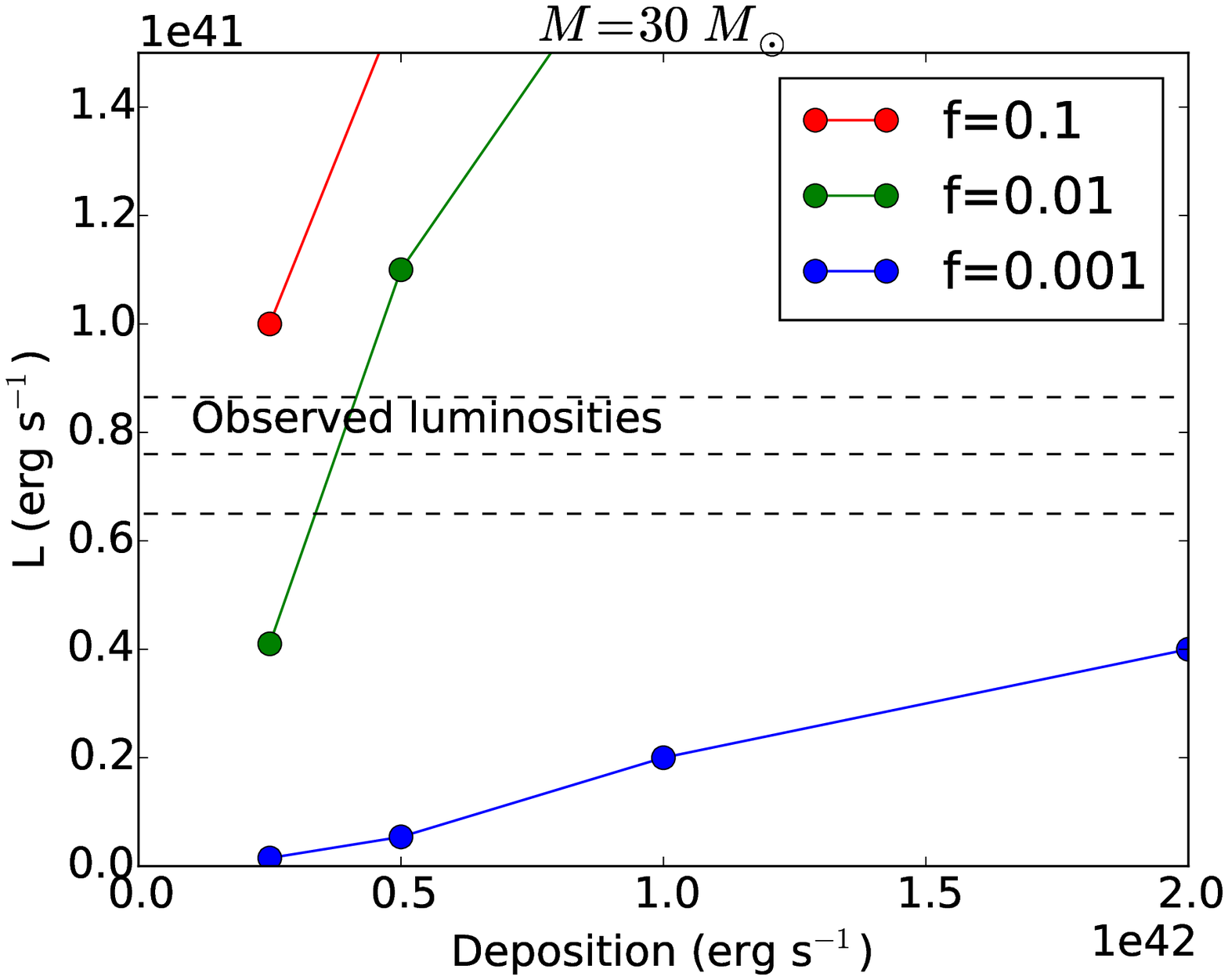}
\caption{The continuum-subtracted [O I] 6300, 6364 luminosity for the
3 \msun~(top), 10 \msun~(middle) and 30 \msun~(bottom) models. The observed
luminosities of SN 2007bi (+367d, lower line), LSQ14an (+410d, middle line) and SN 2015bn (+315d, top line) are marked as dashed lines.}
\label{fig:3O}
\end{figure}

The 3 \msun~models are shown in the left column of Fig. \ref{fig:varymass} (note that
the deposition energy is different for each $f$ value).
For no deposition value do the 3 \msun~models produce strong enough [O I] 6300, 6364
luminosity; increasing the deposition beyond a certain point reduces
the [O I] luminosity as O I becomes ionized to O II (see also Sect. \ref{sec:minmass} and Fig. \ref{fig:3O}). 

The low mass leads to high temperature and ionization, with
the spectra having a strong Mg II 2795, 2802 and a strong blue quasi-continuum between 1000-4000 \AA.
In a multi-component model this blue flux would probably be reprocessed to longer wavelengths by scattering and fluorescence,
and thus may not be relevant to compare with observations. The optical spectrum is at $f=0.1$ and $f=0.01$ dominated by [O II] 7320, 7330 and [O III] 4959, 5007.  
The O III fraction is 7\%, 2\% and 0.5\% in the best-fitting models at $f=0.1, 0.01$ and 0.001. Despite the much lower abundance than O II, O III provides a similar amount of cooling as O II because the [O III] 4959, 5007 transition is connected to the ground state, whereas [O II] 7320, 7330 connects two excited states. 
There is also some [O III] 4363, although a factor $\sim$10 weaker than [O III] 4959, 5007,
as this line arises from a higher state.

The $f=0.001$ model (panel C) fits some observed lines reasonably well, but the [O I] 6300, 6364 line is too weak by a factor 3, and in general the contrast between the emission lines and the quasi-continuum is too low in the 3 \msun~models. There is also very little Mg I emission in these models as the ionization state is high (even some Mg III is present).

The 3 \msun~models are discussed further in Sections \ref{sec:minmass} and \ref{sec:oIIIlines}.

\subsubsection{10 \msun~models} 
The 10 \msun\  models are shown in the middle panel of Fig. \ref{fig:varymass}. The models have $T=6000-6500$ K, $x_e \sim 0.5$, and small Mg I abundance. Increasing the mass from 3 to 10 \msun~allows reproduction of the [O I] 6300, 6364 luminosity as O I is more common.
The dramatic difference in [O I] 6300, 6364 strength between 3 and 10 \msun\ at $f=0.1$ illustrates the potentially strong non-linearity of the ionization solutions. In the 3 \msun\ models at $f=0.1$, O I has suffered runaway ionization, whereas in the 10 \msun\ models this has been avoided.

The [O II] 7320, 7330 and [O III] 4959, 5007 lines are less pronounced than in the 3 \msun~models as the ionization is lower. [O II] 7320, 7330 now tends to be too weak to explain the 7300 line, but [O III] 4959, 5007 makes a reasonable fit to SN 2007bi (while too weak for LSQ14an).

The $f=0.1$ and $f=0.01$ models have neglegible Mg I] 4571. The $f=0.001$ model brings out both Mg I] 4571 and Mg I 5180, and provides an overall reasonable fit to several parts of the spectrum.

\subsubsection{30 \msun~models} 
To illustrate some dependency on deposition, the models plotted in Fig. \ref{fig:varymass} use a different (higher) energy deposition than the ones plotted in Fig. \ref{fig:99d}. The $f=0.1$ and $f=0.01$ models have a deposition $5.0\e{41}$ \ergs\ instead of $2.0\e{41}$ \ergs, and the $f=0.001$ model has deposition $20\e{41}$ \ergs\ instead of $10\e{41}$ \ergs.

The best fit is produced for $f=0.01$. The higher mass
leads to lower temperatures and damping of the blue continuum seen in the 3 and 10 \msun\ models.
The more neutral gas state also leads to easier formation of Mg I lines, although it remains difficult to reproduce the full strength of Mg I] 4571.




\subsubsection{A minimum oxygen mass requirement}
\label{sec:minmass}
Figure \ref{fig:3O} shows observed and modelled [O I] 6300, 6364 line luminosities. The ``continuum'' is subtracted
in the following manner. Measure
\begin{eqnarray}
I_1 = \int_{\lambda_0-d\lambda_1}^{\lambda_0+d\lambda_1} F_\lambda d\lambda\\
I_2 = \int_{\lambda_0-d\lambda_2}^{\lambda_0+d\lambda_2} F_\lambda d\lambda
\end{eqnarray}
where $\lambda_0$ is the line center wavelength, $d\lambda_i = \lambda_0 dV_i/c$, and we use
$dV_1 = 8,000$ \kms and $dV_2 = 10,000$ \kms.
We should have, for a linear continuum of average strength $F_c$:
\begin{eqnarray}
I_1 = L_{line} + 2 F_c d \lambda_1\\
I_2 = L_{line} + 2 F_c d \lambda_2
\end{eqnarray}
Then
\begin{equation}
L_{line} = I_1 - \frac{I_2-I_1}{d\lambda_2/d\lambda_1 -1}
\end{equation}


Figure \ref{fig:3O} shows the [O I] 6300, 6364 luminosity in the model grids compared to the observed values.
As the top panel shows, the 3 \msun~models are always more than a factor 3 too dim, no matter
what density and amount of energy deposition we choose. From this
a minimum mass of $\sim$ 10 \msun~is needed. The 10 \msun\ and 30 \msun~model grids (middle and bottom panels) confirm that such O-zone masses can produce sufficient luminosity.

There is little quasi-continuum in [O I] 6300, 6364 in the observed spectra
(Fig. \ref{fig:allthree}), which makes errors due to the continuum subtraction limited. Also note that we have ignored
any correction due to extinction, which makes the observed luminosities somewhat
underestimated, going in the right direction for the mass limit. Finally, we expect little contamination of this feature by other lines.

\subsection{O III lines}
\label{sec:oIIIlines}
\citet{Lunnan2016} recently suggested the identification of broad [O III] 4363 and [O III] 4959, 5007 lines in PS1-14bj and LSQ14an at +200d. Figure \ref{fig:velspace} in the appendix shows a zoom-on in these lines in SN 2007bi, LSQ14an, and SN 2015bn.
SN 2007bi shows no clear broad [O III] lines, even though there may be hints of them. The narrow galaxy lines of H$\gamma$ 4341 and [O III] 4363 may merge together to form a line of some apparent broadening. LSQ14an has strong galaxy lines (clipped out in figures), but
there are also distinct broad [O III] components at +365d and +410d, with expansion velocities of 4000-5000 \kms.
SN 2015bn is similar to SN 2007bi, showing some hints of the [O III] lines, although they appear somewhat more pronounced in the spectra in \citet{Nicholl2016b}.

In the O-zone models, these [O III] lines emerge for many combinations of mass and density,
The [O III] 4959, 5007 lines have  lower excitation energy than [O III] 4363 and emerge more prominently. 
The emergence of [O III] 4959, 5007 is seen for most masses and filling factors, illustrating that [O I] 6300, 6364 and [O III] 4959, 5007 can be produced in the same physical region. However, [O III] 4363  is weak in the 10 and 30 \msun\ models, and only distinct in the 3 \msun~models. This line requires high temperature, which is more
easily obtained a low mass. 

\begin{figure}
\centering
\includegraphics[width=0.9\linewidth]{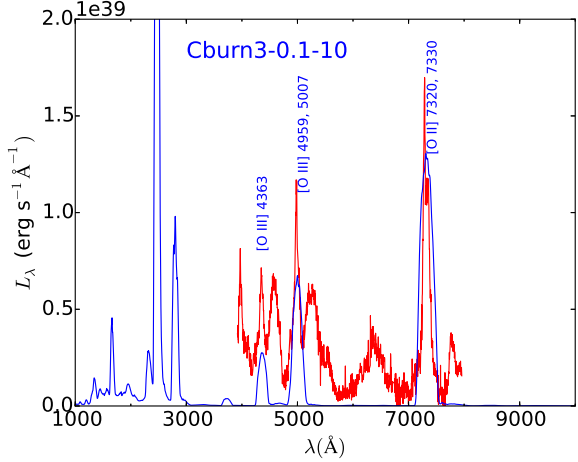}
\caption{A 3 \msun~model compared to LSQ14an. The model makes a good reproduction of 
O II and O III lines.}
\label{fig:Cburn3}
\end{figure}

As an illustration of this point, consider Fig. \ref{fig:Cburn3}, which compares the LSQ14an +410d spectrum with model Cburn3-0.1-10. This model produces [O II] 7320, 7330, [O III] 4363, and [O III] 4959, 5007 in good correspondance with the observed lines. The good agreement suggests that LSQ14an has a low-mass, dilute oxygen region reprocessing a large amount of energy. SN 2007bi and SN 2015bn show no evidence for
such a component, but the models show that the O II and O III lines are sensitive to
deposition, rapidly decreasing in strength if this is lower than $1\e{42}$ \ergs. Thus, it is possible
that LSQ14an displays these lines due to its higher energy deposition rather than any fundamental difference in ejecta properties.

It should be noted that the bulk of line flux seen in the O II and O III lines are at lower velocity (of order 4000 \kms, see also Fig. \ref{fig:velspace}) than the 8000 \kms\ used in the model, which was
chosen based mainly on the widths of neutral lines of O and Mg. It is interesting that the O II/O III velocity scale is the same as the observed O I 7774 one. Whereas production of O I 7774 is facilitated at high density, production of O II and O III lines
is facilitated at low density. 

The [O III] lines emerge despite there being a relatively low O III fraction in the models. They are efficient cooling lines, significantly more so than for example [O II] 7320, 7330, and can be prominent even with a low O III abundance. Still, the models with the strongest [O III] lines also have strong [O II] 7320, 7330. 

High-ionization lines can presumably arise both in an inner pulsar-wind nebula \citep{Chevalier1992,Gaensler2006,Metzger2014}, or in an outer interaction region \citep{Chevalier1994, Fransson2002}. Detailed modelling may be able to provide more constraints on this component.

\subsection{Oxygen recombination lines}
All three SNe show clear detection of O I 7774, and SN 2015bn additionally of O I 9263 and O I 1.13 $\mu$m. These lines are usually interpreted as recombination lines \citep{Maurer2010,J15a}. The O I 7774 luminosity is expected to obey, for $n_{e} = n_{\rm O II}$ \citep[][J15a hereafter]{J15a}
\begin{eqnarray}
\nonumber L _{7774}&=& \mathcal{V} \times  n_e^2 \alpha^{eff}(T) h\nu\\
\nonumber &=&4.5\e{41} \mbox{erg s}^{-1} \left(\frac{V}{8000~\mbox{km s}^{-1}}\right)^3f\\
\times  &&\left(\frac{t}{400d}\right)^3 \left(\frac{n_e}{10^8~\mbox{cm}^{-3}}\right)^2 \left(\frac{\alpha^{eff}(T)}{2\e{-13}~\mbox{cm}^3 \mbox{s}^{-1}}\right)
\end{eqnarray}xw
Here $\alpha^{eff}(T)$ is the effective recombination rate and $\mathcal{V}$ is the volume.
The observed luminosity in SN 2015bn (and similarly in the others) is $L_{7774}=3\e{40}$ \ergs\, which translates to $n_e = 3\e{7}f^{-1/2}$ cm$^{-3}$ (using $V=8000$ \kms), or $n_e=10^9$ cm$^{-3}$ for $f=0.001$. For $V=3000$ \kms\ a factor 4 higher value results. Because we have indications of $f< 0.1$ from the analysis in Sect. \ref{sec:varycomp}, the electron density becomes constrained to $n_e=10^8-10^9$ cm$^{-3}$. 

The O I 9263 and O I 1.13 $\mu$m recombination lines have theoretical luminosities about a factor 2 and 3, respectively,  lower than the O I 7774 line, in good agreement with the observed ratios. This strengthens the case
that they are indeed formed by recombination.

One limitation of the analytic treatment is the assumption that $n_e = n_{\rm O II}$. Charge transfer
reactions may cause significant deviation from this equality at high densities (J15a), and this could lead
to an underestimate of $n_e$. As discussed in Sect. \ref{sec:varycomp}, at very high density O I 7774 may also obtain
contributions by cooling, which would lead to an overestimate of $n_e$ by assuming pure recombination.

The zone mass is
\begin{eqnarray}
\nonumber M &=& \mathcal{V} \times n_e x_e^{-1} \bar{A}m_p = \frac{L_{7774}}{n_e \alpha^{eff} h\nu} x_e^{-1}\bar{A}m_p\\
\nonumber &=&80\ M_\odot \left(\frac{n_e}{10^8\ \mbox{cm}^{-3}}\right)^{-1} \left(\frac{x_e}{0.1}\right)^{-1}\left(\frac{\bar{A}}{16}\right)\\
&&\times \left(\frac{\alpha^{eff}(T)}{2\e{-13}\ \mbox{cm}^3 \mbox{s}^{-1}}\right)^{-1}
\end{eqnarray}
where $\bar{A}$ is the mean atomic weight. The uncertainties in $n_e$, $x_e$ and $\alpha^{eff}$ are somewhat too large to make a useful constraint, but if these can be constrained further a zone mass can be estimated.

\subsection{Mg I 1.50 $\mu$m recombination line}
J15a demonstrated that the Mg I 1.50 $\mu$m recombination line is a useful diagnostic, relatively free
from uncertainty in formation and blending. Its luminosity is
expected to be given by
\begin{eqnarray}
\nonumber L _{1.5 \mu m}= \mathcal{V} \times n_e n_{\rm MgII} \alpha^{eff}(T) h\nu = 6.6\e{38}\ \mbox{erg s}^{-1}\\
\times  \left(\frac{M_{\rm Mg}}{1~M_\odot}\right) \left(\frac{n_e}{10^8~\mbox{cm}^{-3}}\right) \left(\frac{\alpha^{eff}(T)}{1\e{-13}~\mbox{cm}^3 \mbox{s}^{-1}}\right)
\label{eq:mg}
\end{eqnarray}
where the second step assumes most magnesium to be singly ionized as typically obtained in the models.
The observed line in SN 2015bn has an expansion width of 5000 \kms. The measured luminosity within this velocity is $L_{1.5 \mu m}=1.7\e{40}$ \ergs, but about half is continuum so we take $L_{1.5 \mu m}=1\e{40}$ \ergs. For comparison, the measured luminosity in the Type IIb SN 2011dh at 200d was $1\e{38}$ \ergs, a factor 100 lower (J15a). From Eq. \ref{eq:mg}, $M_{\rm Mg} = 15\ M_\odot \left(n_e/10^8~\mbox{cm}^{-3}\right)^{-1}$. As any proposed model scenario would have $M_{\rm Mg} < 15$ \msun, this indicates $n_e > 10^8$ cm$^{-3}$. 
If we take $n_e=10^8-10^9$ cm$^{-3}$ as estimated from the O I recombination lines, the Mg mass is $1.5-15$ \msun.
The effective recombination rate depends on temperature as well as on whether Case B or Case C holds, but the variation between these cases and $T=2500-7500$ K is less than a factor 2 (see table C2 in J15a). 
As the Mg mass fraction is typically 10\% in C-burning ash, the inferred OMg zone mass is very large, $15-150$ \msun.

Model OMg30-0.001-10 reproduces the Mg I 1.50 $\mu$m line reasonably well (Fig. \ref{fig:OMg30PS15ae}). This model has
$M_{\rm Mg} = 3$ \msun, $n_e=1.2\e{9}$ cm$^{-3}$, and $T=6500$ K, giving a predicted luminosity of $2\e{40}$ \ergs\ using Eq. \ref{eq:mg}. All other models (lower $M_{ej}$ and/or larger $f$) produce weaker emission. The Mg I 1.50 $\mu$m emission is fainter by a factor of at least 10 in the 3 \msun~models, and the 10 \msun~models are on the low side as well.
The Mg I 1.50 $\mu$m line thus provides strong independent support for a large OMg zone mass ($M > 10 $ \msun) and a small filling factor ($f \sim 10^{-3}$).

\subsection{The Ca II lines}
\label{sec:caiilines}
The observed ratio between Ca II NIR and [Ca II] 7291, 7323 in SN 2015bn at +315d is 1.7, unusually high
for the nebular phase. There is some uncertainty in the true Ca II NIR luminosity, because
O I 8446 will scatter into the triplet, and the red side may be contaminated by [C I] 8727.
In addition, scattering in the triplet and fluorescence from the Ca II HK lines may contribute to the
luminosity.
However, O I 8446 is predicted to be only about half as strong as O I 7774, which would
make up only 10-20\% of the triplet luminosity. There is also no obvious sign of [C I] 8727
by an asymmetric red side. The absorption trough of Ca II HK also does not appear
to allow much of the observed luminosity in the branching lines.
Nevertheless, these contaminations mean an emissivity ratio
as low as 1 may be possible, and we take a ratio $1-2$ as being admissable.

\begin{figure*}
\includegraphics[width=1\linewidth]{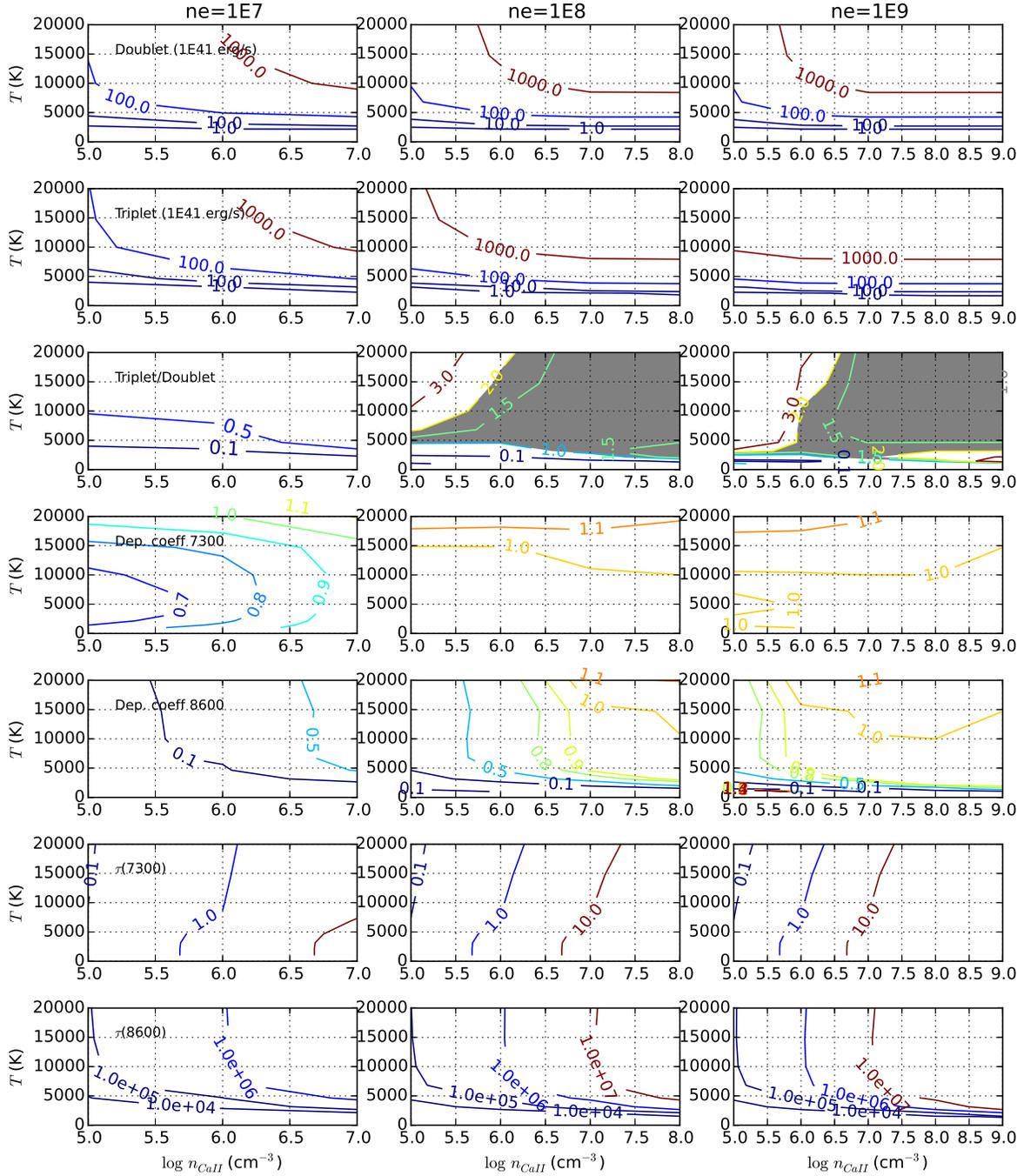} 
\caption{Formation of [Ca II] 7291, 7323 and the Ca II NIR triplet in a NLTE grid including collisional and
radiative processes. The three columns have $n_e=10^7$ cm$^{-3}$ (left), $n_e=10^8$ cm$^{-3}$ (middle) and $n_e=10^9$ cm$^{-3}$ (right). Note that the maximum $n(\mbox{CaII})$ values cannot exceed $n_e$. First row : [Ca II] 7291, 7323 luminosity in units of $10^{41}$ \ergs, for a volume corresponding to $V=8000$ \kms and $f=1$. Second row : Same for Ca II NIR triplet. Third row : The ratio Ca II NIR / [Ca II] 7291, 7323. The regions where the ratio=1-2 are marked gray. Fourth row: Departure coefficient for parent state of [Ca II] 7323. Fifth row : Departure coefficient
for parent state of Ca II 8662. Sixth row : Optical depth for [Ca II] 7323. Seventh row : Optical depth
for Ca II 8662.}
\label{fig:calcium}
\end{figure*}

Figure \ref{fig:calcium} (third row) shows the ratio of Ca II NIR/[Ca II] emission in on-the-spot NLTE calculations varying $n(\mbox{CaII})$, $n_e$, and $T$. These models show that a ratio $>1$ requires $n_e \gtrsim 10^8$ cm$^{-3}$.
The curves do not change much for $n_e$ higher than $10^8$ cm$^{-3}$. For $n_e \geq 10^8$ cm$^{-3}$ there are large plateau
regions where the ratio is 1-2. As the panels showing departure coefficients and optical depths illustrate, this is the optically thick LTE regime.

The regime producing the observed ratio allows for filling factors $f=0.001-1$ to give
the right luminosity (using the top panels), and temperatures
$T>4000$ K, thus no strong constraints on these parameters. The constraint $n_e \gtrsim 10^8$ cm$^{-3}$ translates to a total number of electrons
\begin{equation}
N_e > \mathcal{V} \times f \times 10^8 = 9\e{57}f
\end{equation}
The corresponding zone mass is
\begin{equation}
M = \frac{N_e}{x_e} \bar{A} m_p > 3000 M_\odot \left(\frac{x_e}{0.1}\right)^{-1} \left(\frac{\bar{A}}{40}\right) f
\end{equation}
To reach realistic masses this requires $f \ll 1$, independently confirming a small
filling factor as found from Mg and O lines. Although the Ca lines likely come from a O-burning
zone rather than a C-burning zone, a similar amount of clumping seems reasonable.

Several of the C-burn models in Section \ref{sec:omodels} produce Ca II NIR/[Ca II] ratios larger than unity. In these models the calcium number fraction is small, $10^{-5}$. For example, model C-burn10-0.001-5.0
has $T=6000 K$, $x_e=0.1$, $n_e = 10^8$ cm$^{-3}$, and $n(\mbox{CaII}) = 7\e{4}$ cm$^{-3}$. Its Ca II NIR/[Ca II] ratio
of a few agrees well with the NLTE grid prediction. While the luminosity of the Ca lines is an 
order of magnitude too faint compared to SN 2015bn, this illustrates the potential of small calcium abundances
to make strong lines.

The Ca II NIR triplet is observed also at two other nebular epochs of +343d and +392d in \citep{Nicholl2016b}. In these spectra the Ca II NIR/[Ca II] 7291, 7323 ratio is around 1, somewhat lower than seen here
at +315d. The ratio is expected to decrease with time as both electron density and temperature
decrease. The Ca II NIR line is also observed in LSQ14an (Fig. \ref{fig:PS15ae_optandnir}), but the likely dominance of [O II] 7320, 7330 over [Ca II] 7291, 7323 in this
SN prevents a determination of the Ca II NIR/[Ca II] 7291, 7323 ratio.


\section{Discussion}
\label{sec:discussion}
      %
     
The similarity of nebular-phase spectra of SLSNe to those of broad-lined Type Ic SNe established in Section \ref{sec:compobs} suggests the possibility
of a common origin for these explosions. This idea is also interesting in the context of the newly recognized class of ultra-long GRBs \citep{Levan2014}, the association between one of these and a bright Type Ic SN with early-time spectral appearance similar to SLSNe \citep{Greiner2015,Kann2016}, as well as progress in modelling rapidly rotating and magnetized cores \citep[e.g.][]{Mosta2015}.

The late-time spectra of SN 1998bw have been modelled as 5-7 \msun~of O-rich ejecta powered by 0.6-0.7 \msun~of \ni~\citep{Mazzali2001}.
While there is some debate whether broad-lined Type Ic SNe are powered by \ni~or a central
engine, existing \ni~models are successful enough to maintain this as the standard picture.
Thus, ``scaled-up'' \ni~powered explosion models must still be pursued for SLSNe, even though
the PISN variant of these have not compared well to observations so far \citep{Dessart2013,Jerkstrand2016}.

The late-time spectra of SN 2015bn and SN 2007bi are shown here to be almost identical (Fig. \ref{fig:allthree}). With the multi-epoch
coverage of SN 2015bn (and LSQ14an) we see significantly more rapid decline than $^{56}$Co decay (see also \citep{Nicholl2016b} and Inserra et al., in prep.). This cannot be achieved in PISN models, which have complete
gamma-trapping for at least 500d \citep{Jerkstrand2016}. 
Some transfer to the near-infrared may occur, but the NIR observations here indicate that this is a weak process.
Thus, if these are radioactivity-powered SNe,
they must be core-collapse explosions of less massive progenitors where gamma-ray escape can occur earlier.

Analytic light curve models can estimate ejecta masses, but are limited by assumed opacities
and explosion energies. The nebular phase can provide independent estimates in a phase
when the physics is different. 
Here we have found $M_{\rm O-zone} \approx 10-30$ \msun~for long-duration SLSNe from a variety of indicators.
An immediate consequence is a minimum progenitor mass of $M_{\rm ZAMS} \gtrsim 40$ \msun~
in a single-star progenitor scenario \citep[e.g.][]{Hirschi2004,Nomoto2006,Woosley2007,Ekstrom2012,Chieffi2013}.

Several independent observables indicate that this emitting mass occupies a small
fraction $f \lesssim 0.01$ of the expansion space given by the line widths, and this puts
important constraints on any model scenario. To some extent a value $f \ll 1$ would follow
if most luminosity is generated in a dense central region with $V \ll V_{max}$. Inspection
of the line profiles indicates, however, that much of the luminosity must come from regions with
$V \gtrsim 0.5 V_{max}$, suggesting true clumping or shell formation of high-velocity material.

Compression of the ejecta is conceivable in several scenarios. In one-dimensional magnetar models, the ejecta are pushed into a thin shell with $f \sim 0.001$ as they are accelerated by the pulsar wind pressure \citep{Kasen2010,Chen2016}.
In 2D models, instabilities lead to some degree of fragmentation of this shell, although
its main properties persist \citep{Chen2016}. The instabilities lead to a somewhat reduced
clumping level, and the mixing of material towards both lower and
higher velocity, thus giving line profiles deviating from the boxy profiles
which would be produced by a thin shell. Since none of the observed line profiles are boxy, the instabilities and fragmentation are one way to avoid a uniform shell. 

The \ni~bubble effect in radioactivity models \citep{Herant1991,Basko1994} provides also some degree of compression of the O zone, although with more modest factors. As the \ni~zone is hotter than the surrounding substrate due to radioactive decay in the first few days, it expands, an effect ignored in many explosion models. Kozyreva et al. (in prep) have investigated the $^{56}$Ni bubble effect for the PISN explosion of a 130 \msun\ He core, using the radiation hydrodynamics code STELLA. The $^{56}$Ni zone was diluted by a factor of $\sim$2, and the expansion created compression in the external Si/S shell, increasing its density by a factor of 2-3 (Kozyreva, priv. comm.).
Further out also the O zone showed some compression, but by a moderate factor of $\sim$ 20\% (Kozyreva, priv. comm.). In 3D simulations of core-collapse SNe Si and O
is somewhat more mixed, possibly allowing a compression of factor 2-3 also of oxygen. This modest effect still leaves the filling factor at values between $f=0.1-1$, higher than indicated by the modelling here.

In circumstellar interaction models, formation of a cool dense shell provides a way to compress the material \citep{Chevalier1994}. It may be difficult, however,
to obtain a high enough mass ($M \gtrsim 10$ \msun) in such a shell.

One model which cannot easily meet the requirements derived here is that of colliding shells in subsequent pulsational PISN eruptions. Models for these events predict O-shell masses of one or a few solar masses \citep{Woosley2007PP,Yoshida2016}, which would not provide enough total mass to explain the nebular spectra.
One should note that a star undergoing pulsational pair-instability eruptions may eventually explode as a core-collapse 
supernova, in which the mass can be much higher. Thus, we have no strong constraints on such a scenario, only
a scenario of collisions between pulsational (non-terminal) eruptions. 

The shell collision scenario is problematic also on velocity-grounds. \citet{Woosley2007PP} lists masses and energies for pulses on a grid of He stars between 48-60 \msun, and shell velocities are mostly below 1000-2000 \kms. The same holds for the 54-61 \msun\ CO core models presented by \citet{Yoshida2016}. It would clearly be difficult to reproduce the observed high bulk velocities in the SNe studied here. While photospheric spectra can trace velocities in small amounts of fast-moving ejecta, the nebular phase probes the bulk velocity which can be more directly linked to minimum E/M ratios. The light curves in \citet{Yoshida2016} also typically do not reach the SLSN range, unless the final CCSN is
involved in the interaction.

Most strong observed nebular emission lines have a secure identification. The largest uncertainty regards the line at 5250 \AA. In the models for SN 1998bw, \citet{Mazzali2001} obtains a strong line here due to  a blend of [Fe II] lines. The models in \citet{Sollerman2000} produce a blend of Mg I, Fe I] and [Fe II] lines, with Mg I 5180 appearing to dominate. Some further discussion of the 5250 \AA\ feature is given in \citet{Maeda2002}. Improved understanding of this line is an important goal for further modelling.

Several sources of uncertainty remain for the modelling of these spectra. One of the assumption in the SUMO models studied here is that the ejecta have reached a steady-state phase, when time-scales for the radiative transport and atomic processes are short compared to the dynamic time and the time-scale for energy input. The biggest concern is whether UV/blue photons are trapped in the transfer for long enough time that different parts of the spectrum are generated by power input at different times (i.e. energy is still diffusing out). Fluorescence should limit this time-dependence, but the exact magnitude of the effect is hard to judge.
Another issue is that the influence of the radiative transport is only qualitatively captured in the simplistic morphology assumed here.
While the ejecta are hot and the optical/NIR spectra are expected to be dominated by collisional cooling, the ionization balance of magnesium and similar elements with low ionization potential are governed by the radiation field. The models are also limited by the assumptions of a single density, the number of clumps, and composition. However, these uncertainties are mitigated by analysis with semi-analytic approaches which consistently point in the same direction of massive and dense ejecta.


The metallicity of the progenitor star has an influence on the emission from the OMg zone
through several mechanisms including cooling and line blocking. For the solar metallicity carbon burning composition explored
here, lines such as Mg I] 4571 tend to be quenched by the trace metals, worsening the fits
to observations compared to pure OMg compositions. However, there is dependency
on other model parameters such as the clump size, and a separate in-depth study would
be needed to see if any firm constraints on metallicity can be put from nebular spectra.
Running a single model at ten times reduced metallicity did not produce any great
changes.

The inferred electron densities of $n_e=10^8-10^9$ cm$^{-3}$ at $\sim$400d after explosion
are extraordinarily high. As a comparison, the electron density in the Type IIb SN 2011dh at +200d was inferred to
be $\sim10^6$ cm$^{-3}$ (J15a). Values of over $10^8$ cm$^{-3}$ were inferred for the Type IIn
SN 1995n \citep{Fransson2002}. \citet{BenAmi2014} inferred $n_e \sim 10^9$ cm$^{-3}$ for circumstellar matter emitting narrow lines in the (non-SLSN) Type Ic SN 2010mb. Because we still see mostly neutral species, the gas must have
$x_e < 1$. The inferred zone masses are large, at least 10 \msun, in agreement with modelling
of the thermal cooling lines.

\section{Summary and conclusions}
\label{sec:summary}
We have presented nebular-phase PESSTO and VLT spectra of the long-duration Type Ic SLSNe LSQ14an (+365d and +410d post-peak rest frame) and SN 2015bn (+250d and +315d), respectively, including the first nebular NIR spectra of SLSNe. We have carefully accounted for host galaxy contamination using galaxy models tied to the observed galaxy 
photometry and also applied this method to recalibrate the nebular phase
spectra of SN 2007bi.  We show that this makes a large difference to the spectral appearance, and 
after host correction the spectra of SN 2007bi, LSQ14an, and SN 2015bn show much similarity at about one year post peak, demonstrating homogeneity within the class. These spectra are in turn similar to those of broad-lined Type Ic SNe such as SN 1998bw at earlier phases (140-200 days), 
suggesting a possible common origin with GRB-SNe. 
The major difference seen in these spectral comparisons is that 
LSQ14an shows strong [O III] emission lines, and probably [O II] 7320, 7330, while they are 
significantly weaker in SN 2015bn and SN 2007bi.

SN 2015bn shows O I 9263, O I 1.13 $\mu$m and  Mg I 1.50 $\mu$m, but no strong emission by [Si I] and [S I] around 1.08 $\mu$m as predicted by PISN models. 
The NIR spectra of LSQ14an and SN 2015bn show strong Ca II NIR triplet emission at nebular times. NLTE models indicate that an electron density $n_e \gtrsim 10^8$ cm$^{-3}$ is needed to reproduce the observed Ca II NIR/[Ca II] 7291, 7373 ratio of $\sim$ 2 at +315d in SN 2015bn. Analytic formulae for the O I recombination lines yield similar values $n_e = 10^8-10^9$ cm$^{-3}$. These are extraordinarily high electron densities for the nebular phase.

Models for oxygen-zone emission show that
only models with $\gtrsim$10 \msun~of O can produce enough [O I] 6300, 6364 luminosity
to match observations, irrespective of the powering situation and the density. Too much energy
deposition in lower-mass models leads to ionization to O II before [O I] 6300, 6364 reaches the
observed brightness. Typical models produce detectable broad [O III] 4959, 5007 lines, strengthening the recent identification in observed spectra. The Mg I 1.50 $\mu$m recombination line in SN 2015bn indicates a large Mg mass of 1.5-15 \msun, which together with a typical mass fraction of 10\% gives a similar constraint of $M(\mbox{O-zone}) \gtrsim 15$ \msun. These masses require large progenitor ZAMS masses of at least 40 \msun.

To produce Mg I] 4571 and several other lines approaching their observed brightness, the O/Mg zone needs to be strongly clumped with a filling factor $f \lesssim 0.01$. A similar constraint is derived
for the zone emitting the Ca II lines. 

Overall we can view these results in the context of the three competing physical models which have been advanced to 
explain SLSNe : \ni~powered luminosity, engine driven explosions, and circumstellar interaction.

\begin{itemize}
\item SLSNe Type Ic show little resemblance with
PISN spectral models. In addition to optical comparisons published previously, two strong lines of Si I and S I at 1.08 $\mu$m are predicted \citep{Jerkstrand2016} but are not observed.
We cannot rule out that the luminosity is actually powered by \ni, however, and a massive core-collapse scenario is attractive given the strong similarities with SN 1998bw, assuming this was \ni\ powered. The requirement for compact regions of emission could possibly be explained by the \ni~bubble effect, but simulations need
to demonstrate the low value $f\lesssim 0.01$ for the oxygen material inferred here.

\item In the magnetar scenario, pulsar wind compression of the ejecta is one promising mechanism that can explain how the ejecta may be compressed into a dense region with a low filling factor. A thin shell would result in boxy line profiles, which are not observed, but instabilities and fragmentation as demonstrated in recent 2D simulations may produce profiles in better agreement with observed ones. The large ejecta masses and composition similar to type  Ic SNe would imply magnetar formation from massive Wolf-Rayet progenitors. 

\item Dense shell formation in circumstellar interaction models is also a conceivable scenario for forming the 
compact regions and producing an emitting zone with low filling factor.  The fact that we find some evidence
of a second, lower mass and lower density region producing [O II] and [O III] lines might also support an interaction scenario. However a mass of $\sim$10 \msun\ in the shell that produces the neutral oxygen and magnesium emission is  
required. The collision of shells from pulsational pair-instability eruptions is one variant of the CSI scenario,
but we find that the ejecta masses and velocities in published models are too low compared to those inferred from observations.


\end{itemize}

Finally, we note that our analysis is necessarily limited to the group of slowly evolving Type Ic SLSNe. There are no nebular spectra yet available for the faster declining, and more common SN 2005ap-like
events. This awaits the discovery of a low redshift ($z \lesssim 0.05$) event which is close enough to be observed in the nebular phase.

\appendix

\section{LSQ14an observations and data reductions}
\label{sec:14anobs}

\subsection{LSQ14an +365d spectrum (X-shooter)}
\label{sec:14anxshoot}

A spectrum was taken with the X-shooter spectrometer on the VLT on 2014 December 29 (MJD 57020, +425d post-peak observer frame, +365d rest frame). The total exposure time in each of the three arms was 1800s, and the spectrum was taken at
airmass = 1.487 with seeing of around 1\farcs4. The 1\farcs0 slit was used for the UVB arm and
the 0\farcs9 slits were employed in the VIS and NIR arms.  

We compared the reduced spectra from two different pipelines. Initially we took the ESO Advanced Data Products from 
the Science Archive Facility, which provide wavelength and flux calibrated spectra (with telluric corrections). 
The  ESO pipeline does not do optimal extraction in the last step of extracting the
1D spectrum, and hence the spectra are left with a series of positive and negative spikes due to cosmic rays 
and chip artefacts. We compared these to custom reduced spectra as described in 
\cite{2015A&A...581A.125K}. These reductions use the ESO pipeline to produce wavelength calibrated 
2D spectra for each of the three Xshooter arms and are then optimally extracted with a 
Moffat profile fit.  The reduction method of  \cite{2015A&A...581A.125K} produces cleaner spectra (less 
cosmic ray contamination) and better signal to noise in the UVB+VIS arms in particular (as the nodding sky reduction is not done which reduces noise by $\sqrt{2}$). We settled on the custom reductions from the \cite{2015A&A...581A.125K} pipeline, which produced spectra with dispersions of 0.4 \AA\,pix$^{-1}$ in the UVB+VIS arms and  0.6 \AA\,pix$^{-1}$ in the NIR arm. NIR telluric corrections were done using molecfit version 1.1.1 \citep{Smette2015,Kausch2015}.  The spectra at these pixel dispersions were of relatively low signal to noise, however this can be substantially improved upon with filtering and rebinning of the spectra.

The optical spectrum was calibrated to SN+host photometry on MJD 57021, which we measure as $B=20.81$, $V=20.20$, $R=20.01$ and $I=19.79$, giving a correction factor of 2.3. Lacking NIR photometry, the NIR
spectrum was calibrated with the same factor. 

The NIR spectrum suffer from positive and negative noise 
spikes at the positions of bright skylines. To improve 
signal to noise and reduced the appearance of the sky noise, 
we applied a median filter over a 10 \AA\ window. This simply replaces 
the pixel values with the median value over a binned window of 10 \AA. In the optical, the strong narrow nebular emission lines from the host galaxy were removed by interpolating across the base of the broad emission lines using 
a low order polynomial fit. In the red part of the X-shooter VIS arm, the strong sky lines leave
two residual noise features and these were removed using a simple interpolation.
The final spectra were rebinned to 2 \AA\ in the optical and 5 \AA\ in the NIR. A further 3-bin Gaussian smoothing was applied
to the optical spectrum.
 
\subsection{LSQ14an +410d spectrum (FORS2)}
\label{sec:14anfors2}
A second, later spectrum was obtained with the FORS2 spectrometer on the VLT over several nights
spanning 2015 February 16-23 (MJD 57069 to 57076). The  exposures comprised 
 $12\times1200$s  (4\,hrs total exposure time) with the 300V grism, blocking filter GG435 and a slit width of  1\farcs0. 
The spectra were reduced with standard procedures including bias subtraction, flat-fielding, wavelength and flux calibration. 
They were telluric line corrected using a sky model as described in the PESSTO pipeline in 
\cite{2015A&A...579A..40S}. The spectra were co-added to produce one deep spectrum, with an effective epoch 
of approximately +478d post-peak in the observer frame (+410d rest-frame). 

The spectrum was calibrated to SN+host photometry on MJD 57071, which we measure as $B=21.12$, $V=20.64$, $R=20.44$ and $I=20.12$, giving a correction factor of 1.03. 

For presentation in the paper, sky lines and narrow galaxy lines were removed as discussed above.

\section{SN 2015bn observations and data reduction}
\label{sec:2015bnobs}

\subsection{SN 2015bn +250d spectrum (EFOSC2)}
SN 2015bn was observed on 2016 January 1 with PESSTO (MJD 57388, +280d post-peak observer-frame, +250d rest-frame). The NTT and EFOSC2 were used with Gr\#13 and a 1\farcs0 slit width 
\citep[resolution and setup are presented in][]{2015A&A...579A..40S} and a series of 
3$\times$1800s separate exposures were taken. These data were reduced in standard fashion with the PESSTO 
pipeline and then co-added to produce one spectrum of 5400s duration. 

The optical spectrum was calibrated against photometry at at MJD 57373 (15d earlier observer frame) as presented in \citet{Nicholl2016}. Combining the host-subtracted SN photometry of $g=20.67$, $r=20.50$ and $i=20.18$
with the host photometry of $g=22.30$, $r=22.06$ and $i=22.06$\footnote{Petrosian AB magnitudes from the SDSS Data Release 12 \citep{Alam2015}} gives a SN+host photometry of $g=20.45$, $r=20.27$
and $i=20.00$. Scaling the spectrum to this photometry gave a correction factor of 0.96. 

\subsection{SN 2015bn +315d spectrum (X-shooter)}
SN 2015bn was observed again with VLT X-shooter over 6 days between MJD 57452 and 57459 (2016 March 5 to March 12), +350d post-peak observer frame (+315d rest-frame).  The exposures consisted of $4\times3200$s in UVB, $4\times3250$s in VIS  and $4\times3000$s in the NIR arm. As described in Section\,\ref{sec:14anxshoot}, we compared the ESO Advanced Data Products with the custom reduced spectra as produced by the \cite{2015A&A...581A.125K} pipeline and chose the latter method. These reductions produced spectra with dispersions of 0.4 \AA\,pix$^{-1}$ in the UVB and VIS arms and  0.6 \AA\,pix$^{-1}$ in the NIR. NIR telluric corrections were done using molecfit version 1.1.1.

The spectrum was calibrated against photometry at at MJD 57456 (contemporary) as presented in Nicholl et al. in prep. Combining the host-subtracted photometry of $g=22.02$, $r=21.51$ and $i=21.59$
with the host photometry of $g=22.30$, $r=22.06$ and $i=22.06$ gives a SN+host photometry of $g=21.40$, $r=21.00$
and $i=21.05$. Scaling the spectrum to this photometry gave a correction factor of 1.18.
The NIR spectrum was corrected with the same factor. As for LSQ14an, the NIR spectrum was median filtered with 10 \AA, and
the optical spectrum had narrow lines clipped out. 
The final spectra were rebinned to 2 \AA\ in the optical and 5 \AA\ in the NIR. A further 3-bin Gaussian smoothing was applied
to the optical spectrum.

\section{Observed line profiles}
\label{sec:linezoom}
\begin{figure*}
\centering
\includegraphics[width=0.4\linewidth]{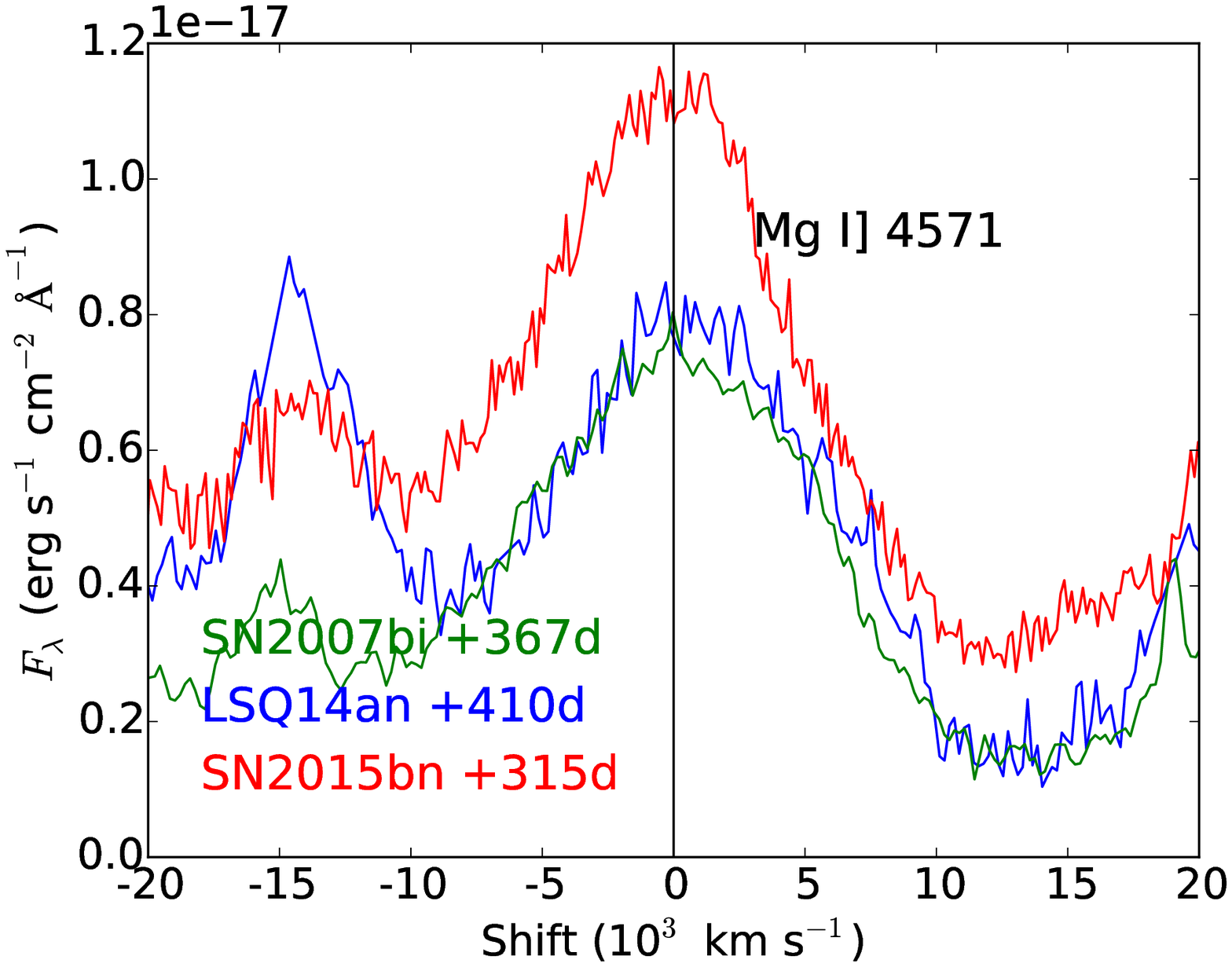} 
\includegraphics[width=0.4\linewidth]{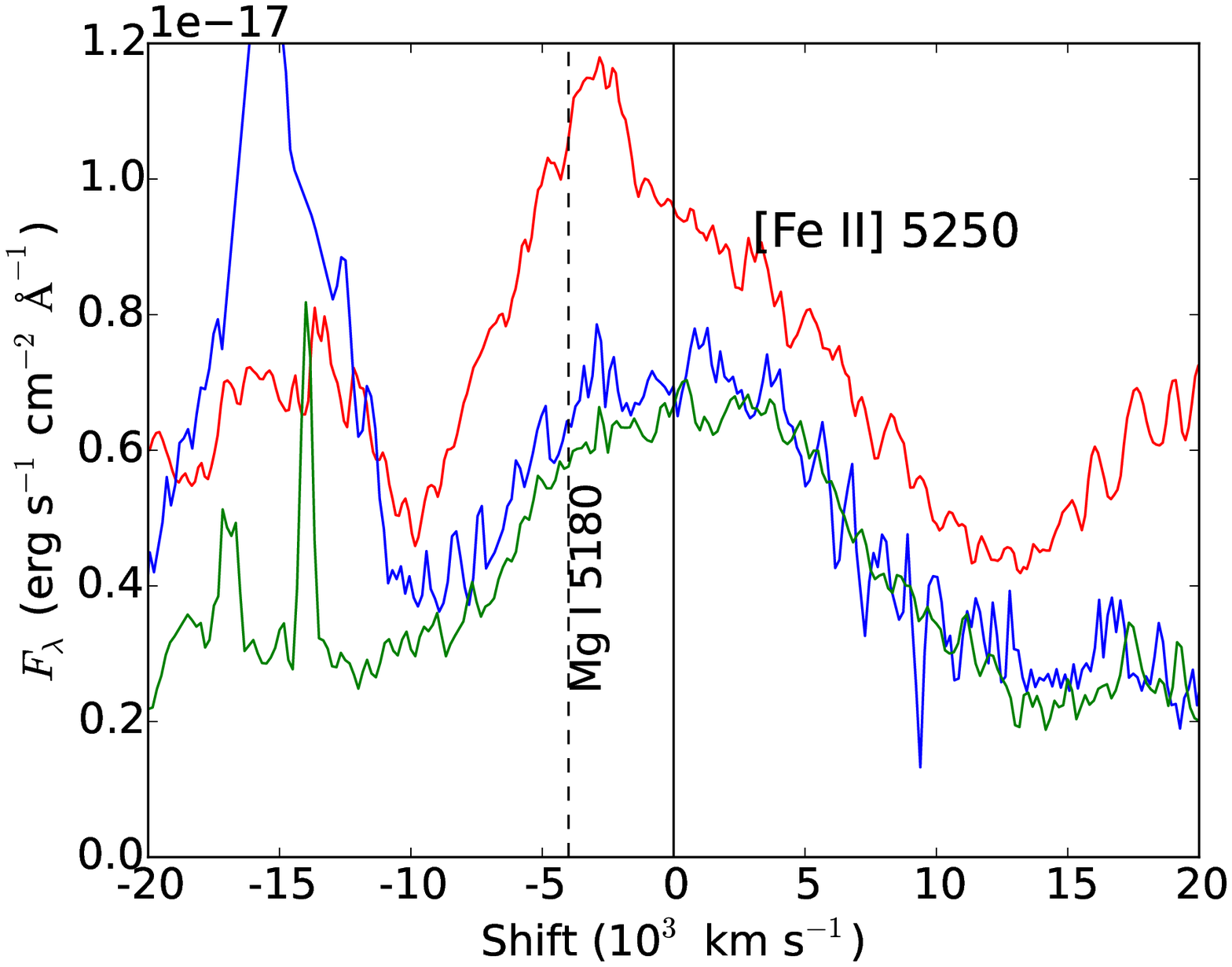} 
\includegraphics[width=0.4\linewidth]{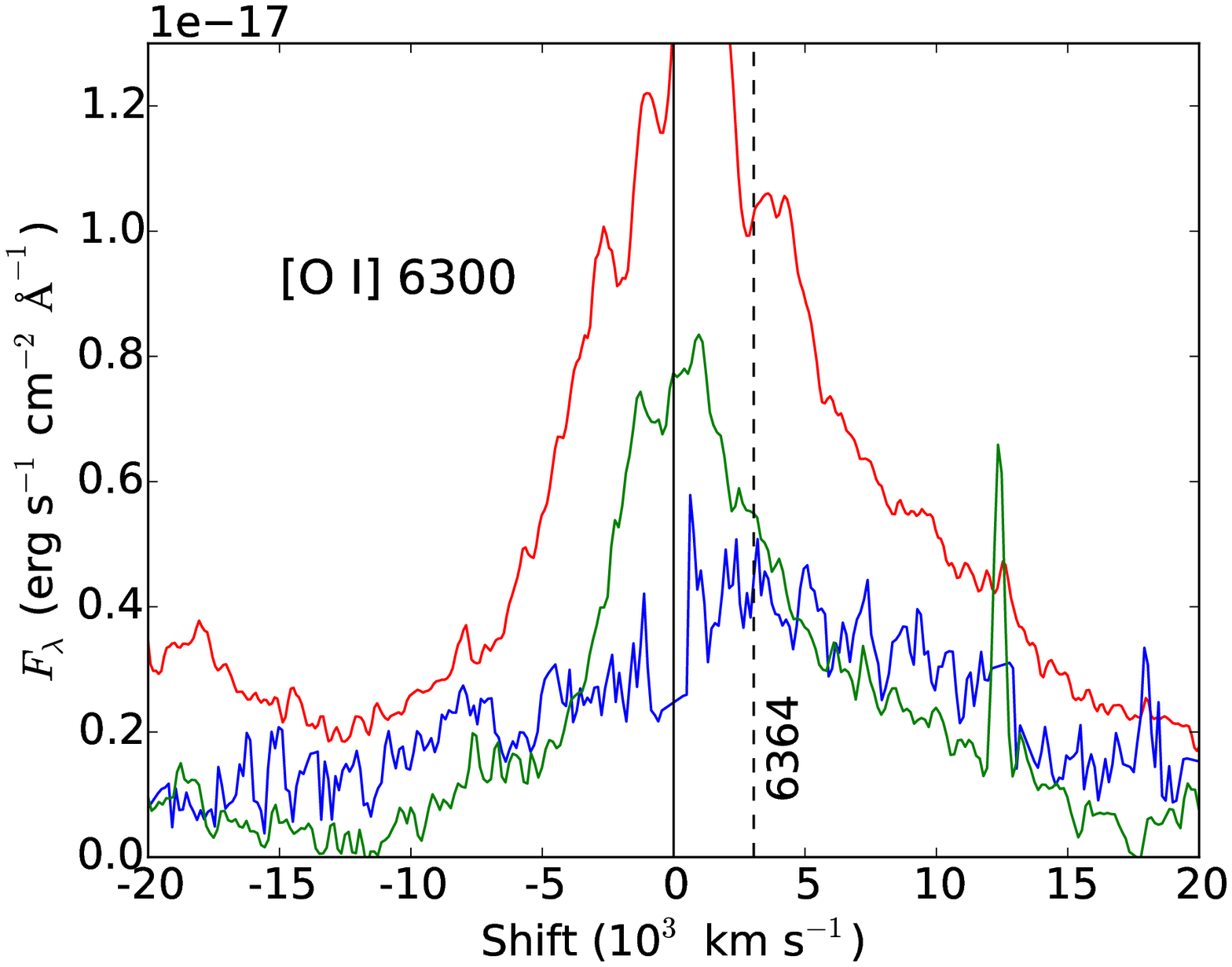}  
\includegraphics[width=0.4\linewidth]{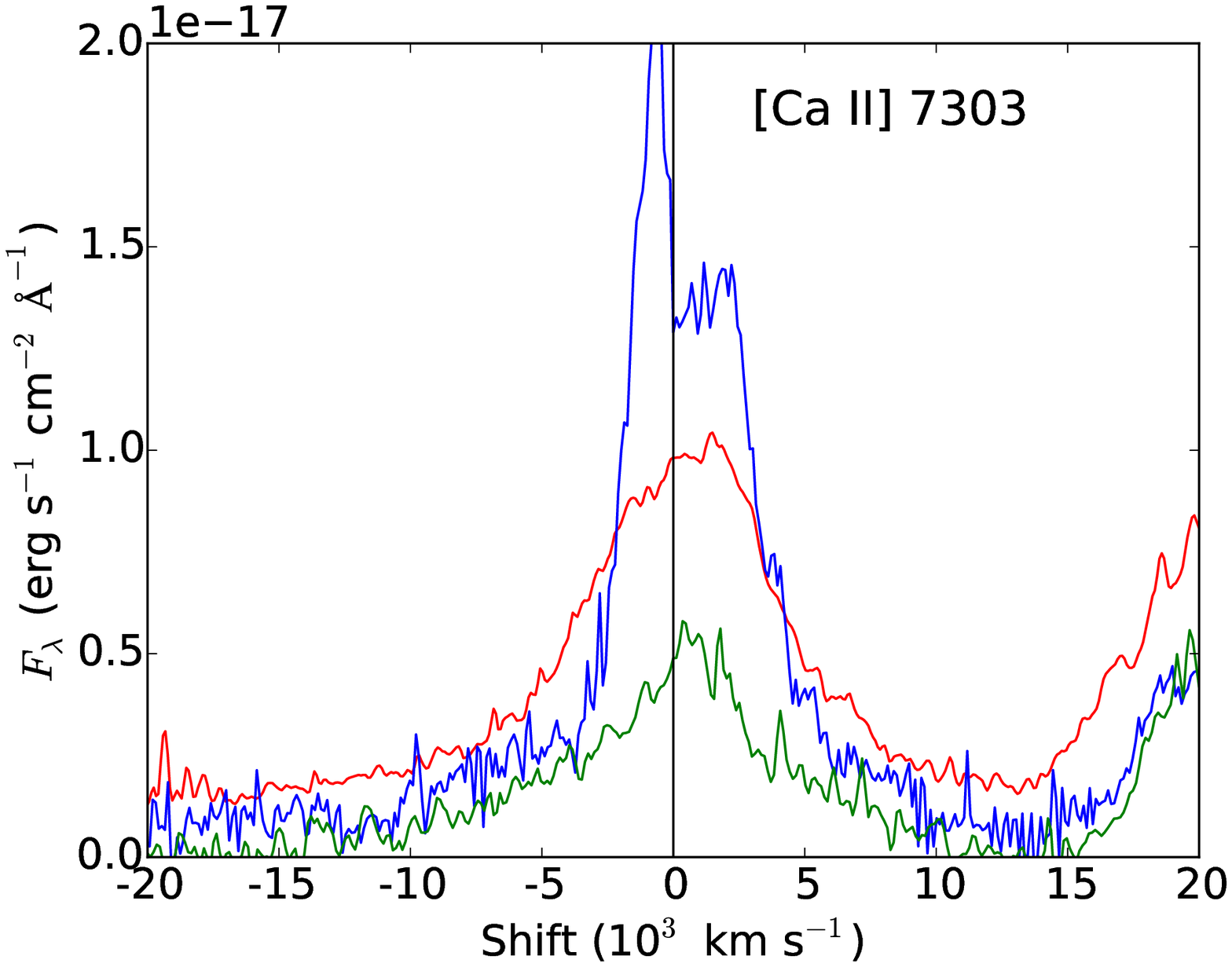} 
\includegraphics[width=0.4\linewidth]{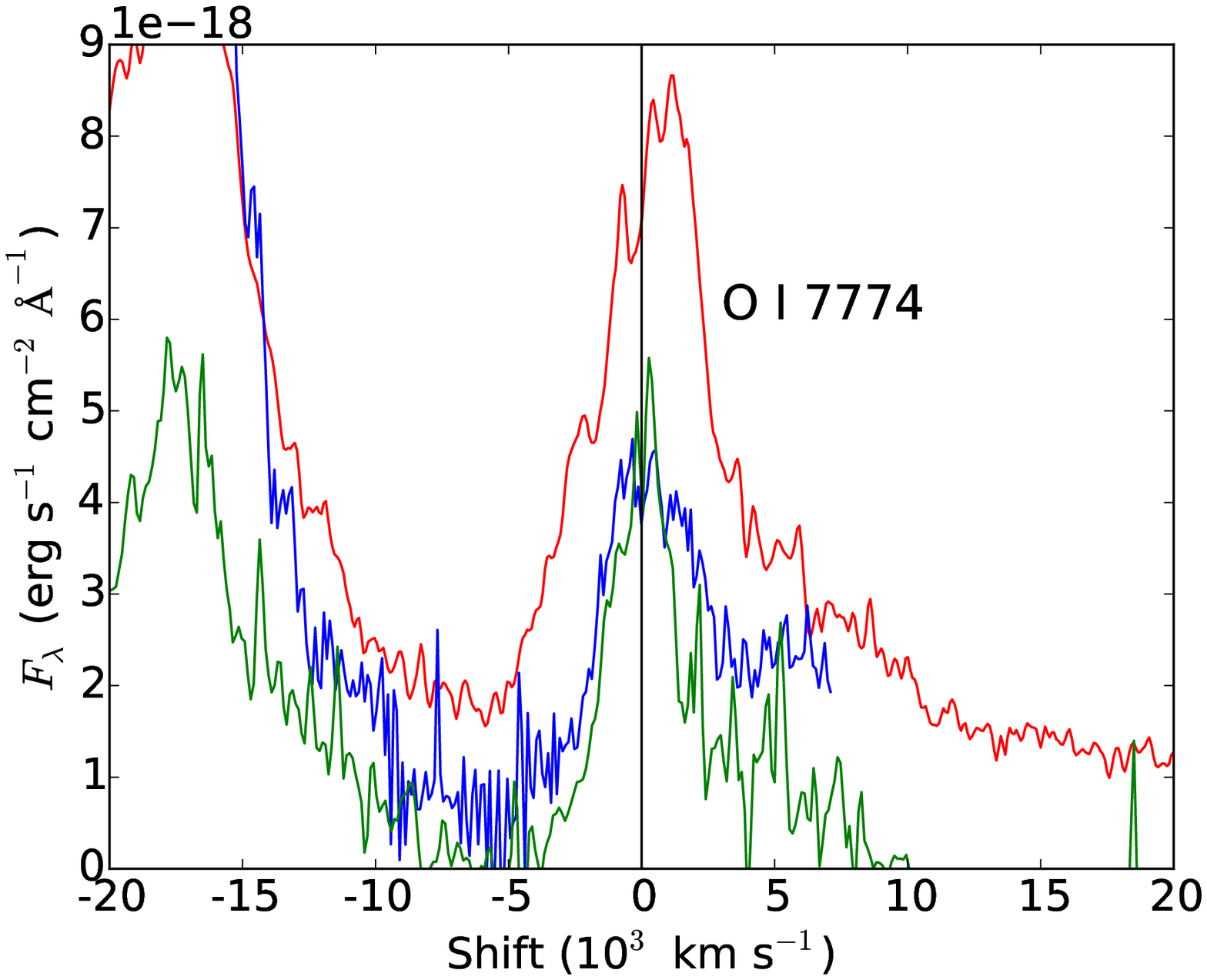} 
\includegraphics[width=0.4\linewidth]{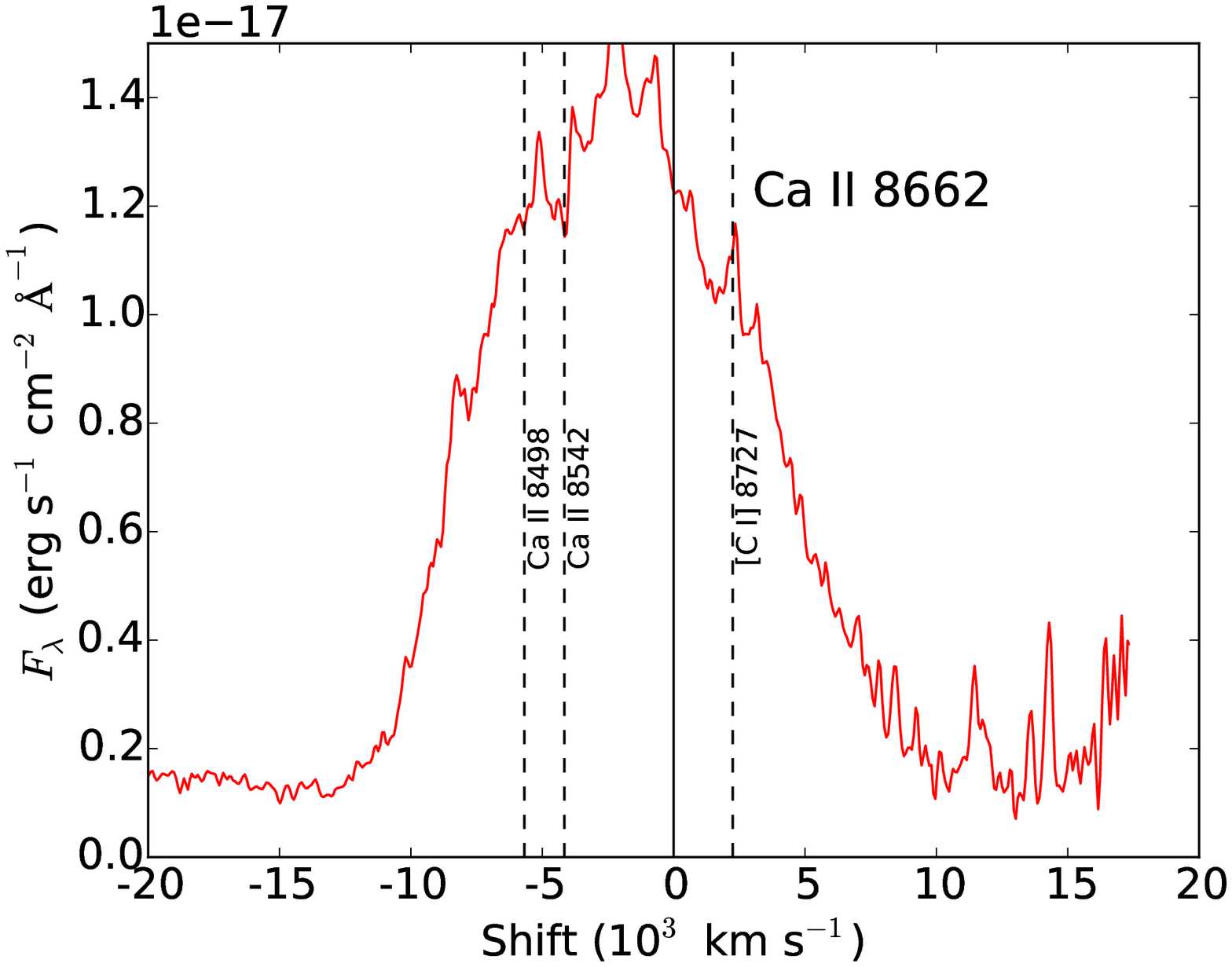}
\includegraphics[width=0.4\linewidth]{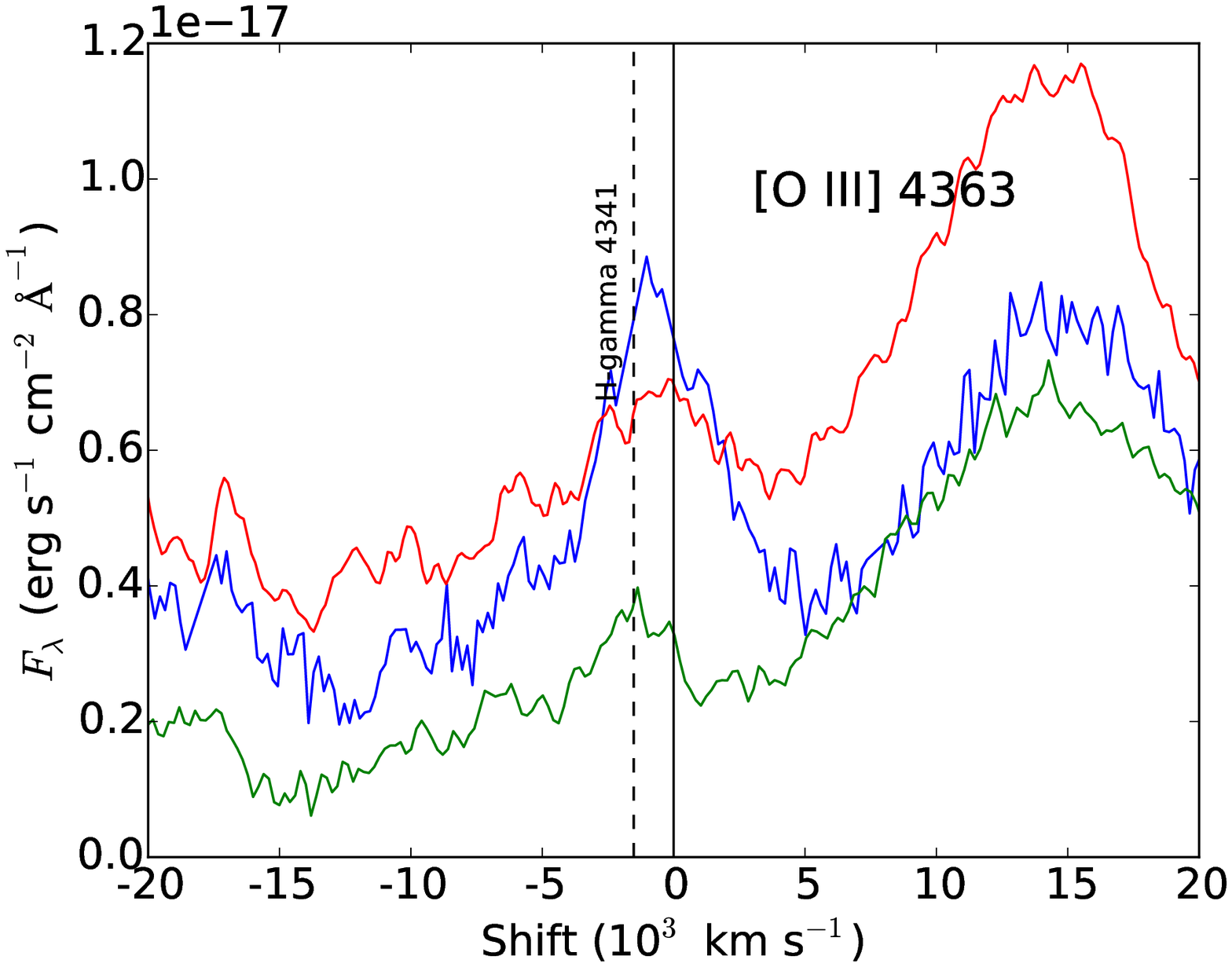} 
\includegraphics[width=0.4\linewidth]{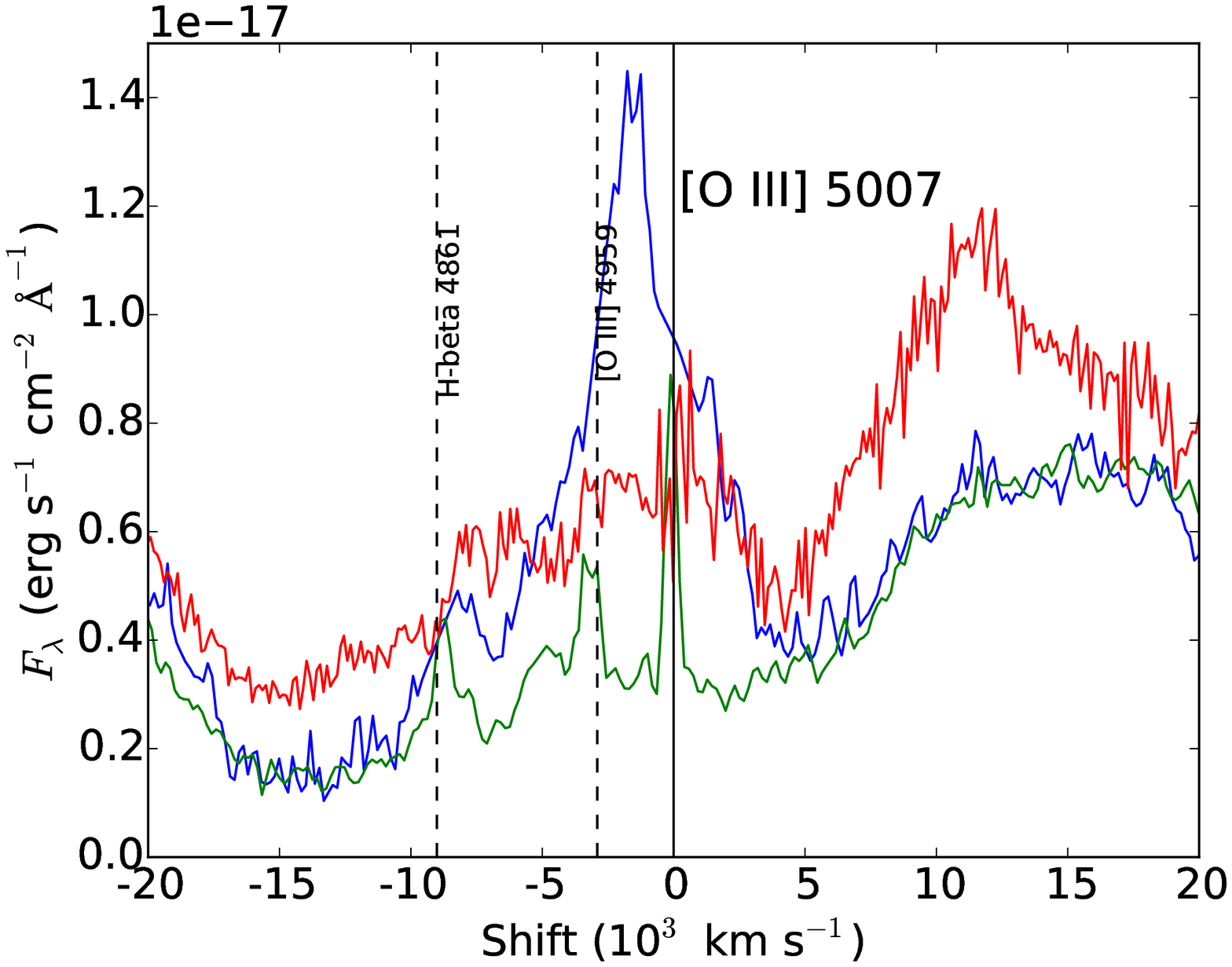} 
\caption{A comparison of key emission lines in SN 2007bi (green), LSQ14an (blue) and SN 2015bn (red).
Positive velocity is away from the observer.}
\label{fig:velspace}
\end{figure*}

Figure \ref{fig:velspace} compares the main emission lines in velocity space. There is emissivity in Mg I] 4571, [Fe II] 5250, and Ca II NIR out to $\sim$10,000 \kms.
[O I] 6300, 6364, [Ca II] 7291, 7323, and O I 7774 appear somewhat narrower, with most 
flux contained within $\sim$ 5000 \kms. In LSQ14an there is an unusual shape
of [Ca II] 7291, 7323 + [O II] 7320, 7330.

\section{Atomic data}

We have added O III to the spectral synthesis code, using 6 levels and 12 transitions. Energy levels
and A-values were taken from NIST\footnote{http://www.nist.gov/pml/data/asd.cfm} \citep{Reader1980, Moore1993, Wiese1996}. Collision strengths were taken from \citet{Lennon1994}.

\section{Runaway ionization}
\label{sec:icat}
Consider a trace element with total number abundance $n$, in a gas with electron number density $n_e$.
Let $N$ be the number of ionizing photons emitted per second, and assume only the trace element can absorb these. As long as all these are absorbed, the
ionization balance is
\begin{equation}
N = \alpha(T) V n_e n (1-x_n)
\end{equation}
where $x_n$ is the neutral fraction, and $V$ is the volume. The solution is
\begin{equation}
x_n = 1 - \frac{N}{\alpha(T)Vn_en}
\end{equation}
At full ionization ($x_n \rightarrow 0$), the number of recombinations is $\alpha(T) V n_e n$. If $N$ exceeds this, there is no solution. What happens then is
that depletion of the neutral abundance continues until escape of ionizing photons occur. The new
balance will be given by
\begin{equation}
N \left(1-e^{-\tau}\right) = \alpha(T) V n_e n 
\end{equation}
where $\tau = n x_n \sigma L = n x_n \sigma \left(3/4\pi\right)^{1/3}V^{1/3}$. Then, for $\tau \ll 1$, and ignoring the numeric constant
\begin{equation}
x_n = \alpha(T) V^{2/3} n N^{-1}\sigma^{-1}
\end{equation}
Thus, if the gas is still optically thick as $x_n \ll 1$, $x_n$ will jump down to a value
given by this equation; there is catastrophic runaway depletion.

\begin{figure}
\centering
\includegraphics[width=0.5\linewidth]{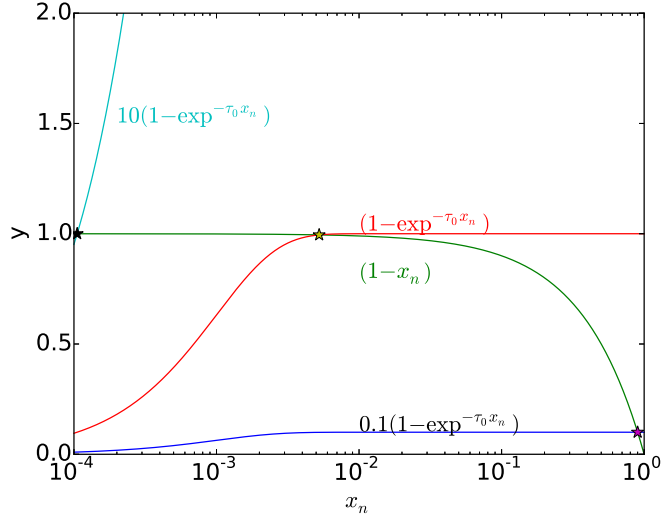}
\caption{Illustration of runaway ionization. The LHS (cyan, red, blue) and RHS (green) of Eq. \ref{eq:run} are plotted, for three different LHS functions. The $x_n$ solution is at the intersection of the curves.}
\label{fig:runaway}
\end{figure}

The generic solution is given by solving
\begin{equation}
\Gamma \left(1-e^{-\tau_0 x_n}\right) =  1-x_n
\label{eq:run}
\end{equation}
where $\Gamma = N/(\alpha V n_e n)$. Figure \ref{fig:runaway} shows illustrative solutions, for $\tau_0 = 10^{3}$ and three different $\Gamma$ values. As $\Gamma$ increases from zero, through 0.1 and towards 1, there is a gradual reduction in $x_n$. As $\Gamma > 1$, the solutions jump down to a value $x_n \lesssim 1/\tau_0$. Changing the ionization parameter by a factor of 10 from 0.1 to 1 decreases $x_n$ from 0.9 to 0.005, a factor 180. The decrease then slows down and follows $1/\Gamma$. Thus, the effect of changing $\Gamma$ by a factor 2 is small for $\Gamma \ll 1$ and for $\Gamma \gg 1$. But around $\Gamma \sim 1$, a factor 2 change in $\Gamma$ can give huge changes in $x_n$, of order $\tau_0$.
A similar situation can be shown to hold if $n_e$ is not assumed constant but $n_e = n(1-x_n)$.


\acknowledgements
We thank S. Blinnikov, C. Fransson, A. Kozyreva, P Mazzali, and C. Harris for discussion. 
AJ thanks the organisers and participants of MIAPP conference 'The Physics of Supernovae' and the workshop 'Supernovae: The Outliers' for stimulating discussions.
AJ acknowledges funding by the European Union's Framework
Programme for Research and Innovation Horizon 2020 under
Marie Sklodowska-Curie grant agreement No 702538, and Science and Technology Facilities Council DIRAC computing grants ACSP45 and ACSP74. The research leading to these results has received funding from the European Research Council under the European Union's Seventh Framework Programme (FP7/2007-2013)/ERfC Grant agreement n$^{\rm o}$ [291222]. Agreements 307260, 320360, and 615929 are also acknowledged. SJS acknowledges funding from STFC grants ST/I001123/1 and ST/L000709/1. 
This work is based in part on observations collected at the European Organisation for Astronomical Research in the Southern Hemisphere, Chile as part of PESSTO, (the Public ESO Spectroscopic Survey for Transient Objects Survey) ESO program 188.D-3003, 191.D-0935 and on the 
VLT ESO Programmes 094.A-0645, 096.D-0191and 296.D-5042.
KM acknowledges support from the STFC through an Ernest Rutherford Fellowship. ST acknowledges 
support by TRR 33 âThe Dark UniverseÔ of the German Research Foundation.
T.-W. Chen and TK acknowledge the support through the Sofia Kovalevskaja Award to P. Schady from the Alexander von Humboldt Foundation of Germany. This research has made use of the NASA/IPAC Extragalactic Database (NED) which is operated by the Jet Propulsion Laboratory, California Institute of Technology, under contract with the National Aeronautics and Space Administration.
We have made use of the Weizmann interactive supernova data repository - http://wiserep.weizmann.ac.il.
The Pan-STARRS1 Surveys (PS1) have been made possible through contributions of the Institute for Astronomy, the University of Hawaii, the Pan-STARRS Project Office, the Max-Planck Society and its participating institutes, the Max Planck Institute for Astronomy, Heidelberg and the Max Planck Institute for Extraterrestrial Physics, Garching, The Johns Hopkins University, Durham University, the University of Edinburgh, Queen's University Belfast, the Harvard-Smithsonian Center for Astrophysics, the Las Cumbres Observatory Global Telescope Network Incorporated, the National Central University of Taiwan, the Space Telescope Science Institute, the National Aeronautics and Space Administration under Grant No. NNX08AR22G issued through the Planetary Science Division of the NASA Science Mission Directorate, the National Science Foundation under Grant No. AST-1238877, the University of Maryland, and Eotvos Lorand University (ELTE) and the Los Alamos National Laboratory.

\bibliographystyle{apj}
\bibliography{bibl}


\end{document}